\documentclass[superscriptaddress,aps,reprint,pra,floatfix,longbibliography]{revtex4-1}

\usepackage{graphicx}
\usepackage{dcolumn}
\usepackage{bm}
\usepackage{amsmath,amssymb,amsfonts,amsthm}
\usepackage{physics}
\usepackage{color}

\definecolor{darkblue}{rgb}{0,0,.65}
\definecolor{darkgreen}{rgb}{0.3,0.6,0.3}
\definecolor{cyan1}{rgb}{0.0, 0.6, 0.6}
\usepackage[%
pdfstartview=FitH,%
breaklinks=true,%
bookmarks=true,%
colorlinks=true,%
anchorcolor=black,%
citecolor=red,
filecolor=black,%
menucolor=black,%
urlcolor=darkblue,%
linkcolor=blue,%
]{hyperref}

\providecommand{\ket}[1]{\ensuremath{\left|{#1}\right.\rangle}}
\providecommand{\bra}[1]{\ensuremath{\langle\left.{#1}\right|}}

\providecommand{\floor}[1]{\left \lfloor{#1}\right \rfloor }

\graphicspath{{.}{./Figures/}}

\begin{document}
	
	\title{Non-Hermitian scattering on a tight-binding lattice}
	
	
	\author{Phillip C. Burke}
	\affiliation{ Department of Theoretical Physics, Maynooth University, Maynooth, Kildare, Ireland
	}
	\author{Jan Wiersig}
	\affiliation{Institut f\"ur Physik, Otto-von-Guericke-Universit\"at Magdeburg, Postfach 4120, D-39016 Magdeburg, Germany}
	\author{Masudul Haque}
	\affiliation{ Department of Theoretical Physics, Maynooth University, Maynooth, Kildare, Ireland
	}
	\affiliation{Max-Planck Institute for the Physics of Complex Systems, Dresden, Germany}
	\date{\today}
	
	\begin{abstract}
		
		We analyze the scattering dynamics and spectrum of a quantum particle on a tight-binding lattice
		subject to a non-Hermitian (purely imaginary) local potential.  The reflection, transmission and
		absorption coefficients are studied as a function of the strength of this absorbing potential.
		The system is found to have an exceptional point at a certain strength of the potential.
		Unusually, all (or nearly all) of the spectrum pairs up into mutually coalescing eigenstate pairs
		at this exceptional point.  At large potential strengths, the absorption coefficient decreases and
		the effect of the imaginary potential is similar to that of a real potential.  We quantify this
		similarity by utilizing properties of a localized eigenstate.
		
	\end{abstract}
	\maketitle
	
	
	\section{Introduction}\label{sec:Introduction}
	
	In recent years there has been a surge of interest in quantum systems that are described by
	non-Hermitian Hamiltonians.  Although Hermiticity is regarded as a postulate of standard quantum
	mechanics, non-Hermitian Hamiltonians are useful as effective descriptions of systems where loss or
	gain plays an important role, such as open quantum systems \cite{IRotter_JPA2009_opensystems} and
	optical systems described by wave equations formally analogous to a Schr\"odinger equation
	\cite{Cao_Wiersig_RMP2015, Ozdemir_Rotter_Nori_NatMaterials2019_review,
		El-Ganainy_etal_CommunicPhys2019_review_nonHermOptics}.  By now, a number of experimental
	platforms for the study of non-Hermitian quantum mechanics are available.  These include lasers or
	optical resonators \cite{Tureci_Rotter_PRL2012_lasers, Brandstetter_NatCom2014_laser,
		Wiersig_Rotter_PNAS2016_chiralmodes, Miao_Longhi_Liang_Science2016}, coupled optical waveguides
	\cite{Guo_etal_PRL2009_transparency, Segev_NatPhys2010_PTsym, Segev_PRL2011_photoniclattice,
		Szameit_Rudner_PRL2015_topological_transition, Cerjan_etal_NatPhot2019_Weyl}, microwave resonators
	\cite{Persson_IRotter_Barth_PRL2000_microwave, Dembowski_etal_PRL2001, Dembowski_etal_PRL2003,
		Doppler_Nature2016_encircling} and arrays thereof \cite{Schomerus_Nature2015_resonatorchain},
	optical microcavities \cite{Cao_Wiersig_RMP2015, Chen_etal_Nature2017_microcavity,
		Yi_Kullig_Wiersig_PRL2018_EPs_MicrodiskCavity}, optomechanical systems
	\cite{Xu_etal_Nature2016_optomechanical}, photonic crystals
	\cite{Zhen_etal_Nature2015_EPring_from_DiracCone, Zhou_etal_Science2018_FermiArc}, acoustics
	\cite{Zhu_etal_JournAcousticalSoc2015, Fleury_etal_NatComm2015_accoustic_invisible,
		Shi_etal_NatComm2016_acoustics_unidirectional, Ding_PRX2016_acoustics_multipleEPs} atom-cavity
	composites \cite{Choi_etal_PRL2010_AtomCavity}, exciton-polariton systems in semiconductor
	microcavities \cite{Gao_etal_Nature2015_excitonpolariton,Snoke_Truscott_Ostrovskaya_PRL2018_chiral},
	and various other arrangements \cite{PRL2007_hydrogenatom,
		Kottos_Christodoulides_PRL2011_unidirectional, Regensburger_etal_NatPhys2012, Ott_PRL2013,
		Lu_Wiersig_etal_ScienceBulletin2018_PT_cavityQED,   Li_Harter_Joglekar_NatComm2019_Floquet_coldatom}.
	
	Non-Hermitian Hamiltonians lead to various phenomena not present in Hermitian systems. In general,
	the eigenvalues of non-Hermitian Hamiltonians are complex. 
	The left and right eigenstates of a non-Hermitian Hamiltonian are generally not equal --- we confine our discussion to right eigenstates. The eigenstates are in general not mutually orthogonal.
	This non-orthogonality becomes extreme at points in the parameter space
	referred to as \emph{exceptional points} \cite{Kato_book1995, Heiss2004, Berry2004,
		Mueller_IRotter_JPA2008_exceptional, Heiss_JPA2012}. At an exceptional point, the eigenvalues
	appear to become degenerate.  However, it is not a genuine degeneracy as the corresponding
	eigenvectors coalesce as well.  This results in our eigenstates no longer providing a basis spanning
	the entire Hilbert space.  The Hamiltonian matrix is therefore non-diagonalizable and is a defective
	matrix \cite{Golub_VanLoan_MatrixComputations_book1996, Watkins_MatrixComputations_book2004} at
	these exceptional points.
	The surviving eigenstate at an exceptional point is always chiral \cite{Heiss_Harney_EPJD2001_chirality}; this chirality has been
	observed experimentally \cite{Dembowski_etal_PRL2003, Wiersig_Rotter_PNAS2016_chiralmodes,
		Miao_Longhi_Liang_Science2016, Liu_Wiersig_etal_LaserPhotRev2018_chirality, Snoke_Truscott_Ostrovskaya_PRL2018_chiral}.
	Other phenomena associated with exceptional points include loss-induced transparency
	\cite{Guo_etal_PRL2009_transparency}, unidirectional transmission
	\cite{Kottos_Christodoulides_PRL2011_unidirectional,
		Regensburger_etal_NatPhys2012,Shi_etal_NatComm2016_acoustics_unidirectional}, lasers with
	non-monotonic pump-dependence \cite{Tureci_Rotter_PRL2012_lasers}, enhanced sensing
	\cite{Wiersig_PRL2014_sensitivity, Wiersig_PRA2016_Sensors,
		Hodei_etal_Nature2017_EnhancedSensitivity}, etc.
	Exceptional points are also associated with the real-to-complex spectral transition for parity-time
	({$\cal PT$}) symmetric Hamiltonians \cite{Heiss_JPA2012}.   
	
	In this work, we are concerned with the non-Hermitian physics of a quantum particle on a
	tight-binding lattice.  Previous studies of non-Hermitian effects for a lattice particle include
	Anderson localization \cite{Hatano_Nelson_PRL1996_localization, Goldsheid_PRL1998_Andersonmodel,
		Heinrichs_PRB2001_Andersonmodel, Barontini_PhysRevA.91.032114_Anderson} and localization in quasiperiodic potentials
	\cite{Longhi_PRL2019_quasicrystals, Longhi_PRB2019_AubryAndre}, invisibility (reflectionless
	scattering) due to non-Hermitian hopping \cite{Longhi_PRA2010_invisibility} or oscillating imaginary
	scatterer \cite{Longhi_EPL2017_Floquet}, flat-band physics \cite{Flach_PRB2017_flatband}, Bloch
	oscillations \cite{Longhi_PRL2009_BlochOscil}, {$\cal PT$} symmetry obtained by combining an
	absorbing potential on one site with an emitting potential on another
	\cite{Bendix_Kottos_Shapiro_PRL2009_latticePT, Jin_Song_PRA2009_latticePT,
		Jin_Song_PRA2010_latticePT, Joglekar_Scott_Babbey_Saxena_PRA2010_tightbindingPT,
		Jin_Song_PRA2016_PTflux, Zhu_etal_PRA2016_latticePT_scattering, Ortega_etal_arXiV2019_latticePT},
	etc.  In addition, non-Hermitian tight-binding lattices form the basis of the study of non-Hermitian
	topological many-body systems, a topic of rapidly growing interest
	\cite{NGoldman_Zilberberg_Carusotto_RMP2018_topologicalphotonics,
		MartinezAlvarez_etal_EPJST2018_topological_review, torres_arXiv2019_review_topological}.  A few
	studies have also addressed \emph{interacting} many-body systems in non-Hermitian lattice systems
	\cite{McClarty_Rau_PRB2019_magnon, LuitzPiazza_2019_manybody}.
	
	We will consider an imaginary potential on one site of the lattice, serving as an absorbing
	scattering potential.  This can be regarded as a lattice analog of a delta-function scattering
	potential in the continuum which is purely imaginary. An imaginary scattering potential is linked to
	measurement \cite{Allcock_AnnPhys1969_1,Allcock_AnnPhys1969_2,Allcock_AnnPhys1969_3}, and is thus
	related to quantum first-passage time problems and the quantum Zeno effect
	\cite{Misra_Sudarshan_Zeno_JMathPhys1977, Facchi_Pascazio_JPA2008,
		Zezyulin_Barontini_Ott_MacroscopicZeno_PRL2012, Krapivsky_Luck_Mallick_JSP2014_survival,
		Dhar_PRA2015_measurements, Dhar_JPA2015_arrival,Kozlowski_Mekhov_PRA2016_Zeno,
		Chiocchetta_Kollath_Diehl_PRL2019_QuantumZeno}.
	In analogy to the quantum Zeno effect, it is expected that an imaginary potential will have
	suppressed absorption when the strength of the potential is large.  This suggests that the
	absorption might be non-monotonic as a function of the strength of the dissipative potential.  In
	this work, we explicitly show non-monotonic dependence of the amount of absorption on the potential
	strength, in the context of a simple lattice model.  The Hamiltonian is
	\begin{equation}
	\label{eq:ham}
	H \ = \ -J\sum_{j=1}^{L-1} \Big(\ket{j}\bra{j+1} + \ket{j+1}\bra{j} \Big) -i\gamma \ket{q}\bra{q},
	\end{equation}
	with $1\leq{q}\leq{L}$.  Here $\gamma$ is a positive constant, so that the imaginary potential is absorbing.  The labels for
	the bra's and ket's here are site labels: The particle lives on an $L$-site chain with open boundary
	conditions.  The hopping strength will henceforth be set to $J=1$, i.e., energies and times are
	measured in units of $J$ and $1/J$ respectively, and are therefore presented without units.  Also,
	the spacing between sites is set to unity, so that lengths and wavenumbers are dimensionless as
	well. 
	
\begin{figure}[tb]
		\includegraphics*[width=\linewidth]{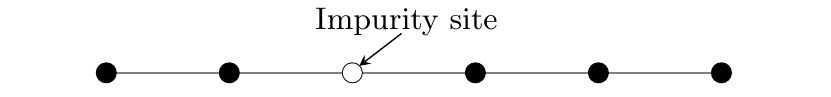}
		\caption{The impurity is placed at one of the central sites of the lattice, as shown here
			for $L=6$.  In this case, it could equivalently be placed on the fourth instead of the third site.
			For odd $L$, there is a definite central site.}
		\label{fig:lattice}
	\end{figure}

	The site $q$ is the location of the dissipative impurity.  Since we want to study reflection and
	transmission, it is convenient to place the particle at the center of the lattice, at either site
	$\floor{\frac{L}{2}}$ or $\floor{\frac{L}{2}}+1$ (Fig.~\ref{fig:lattice}).
	
	We present a study of the dynamics and eigenspectrum of the system \eqref{eq:ham}.  By scattering
	wavepackets numerically off the dissipative impurity, we show how the reflection, transmission and
	absorption fractions depend on the strength $\gamma$ of the impurity.  These results are compared
	with the continuum problem, which is a variant of the standard textbook problem of quantum
	scattering off a Hermitian delta-function potential.  In both cases the absorption coefficient is
	found to be a non-monotonic function of $\gamma$, having a maximum at a point that depends on the
	momentum of the incident particle or wavepacket.  In addition, we present the spectrum of the
	Hamiltonian, which shows an unusual exceptional point at $\gamma=2$ at which all (or nearly all,
	depending on $L$) of the eigenvalues pair up.  The absorption coefficient is non-monotonic and has a
	maximum near, but not necessary at, the exceptional point.  At large $\gamma$, the absorption is
	vanishingly small, and the system behaves as if the impurity were a real potential $V$.  In
	particular the system has a (anti-)bound eigenstate, which allows us to draw a correspondence
	between values of $\gamma$ and $V$.  The localized eigenstate is purely a lattice phenomenon with no
	analogue in the continuum.
	
	In Section \ref{sec:RTA} we present the scattering results and comparisons with the continuum case.
	Section \ref{sec:eps} discusses the spectrum and exceptional points.  In Section \ref{sec:Large} we
	investigate the system at large $\gamma$ values, and draw a comparison between real and imaginary
	potentials via their bound states.  In Section \ref{sec:concl} we present some discussion and
	concluding remarks.  The Appendixes present some further details on the eigenvalues and eigenstates.

	\section{Scattering at an absorbing potential --- Reflection, Transmission,   Absorption }\label{sec:RTA}
	
	In this section, we examine the scattering of a quantum particle by the dissipative impurity.
	To make a comparison with the corresponding continuum system, we first work out the results for the
	continuum system in \ref{subsec:scattering_continuum}, before turning back to our lattice problem in 
	\ref{subsec:scattering_lattice}.

	\subsection{Continuum scattering by imaginary delta-potential} \label{subsec:scattering_continuum}
	
	Complex scattering potentials in the continuum have been considered generally in the literature \cite{MUGA2004357, Allcock_AnnPhys1969_2,Allcock_AnnPhys1969_3}. We are specifically interested in the case of an imaginary potential of delta-function shape, which is the analog of the single-site potential on a lattice.
	
	In the continuum, the wavefunction $\psi(x)$ satisfies the time-independent Schr\"{o}dinger
	Equation:
	\begin{gather}
	-\frac{\hbar^{2}}{2m}\frac{d^{2}\psi(x)}{dx^2} + V(x)\psi(x) \ = \ E\psi(x).
	\end{gather}
	(We will eventually set $\hbar=1$ but retain it for now.)  We take $V(x)$ to be a negative imaginary
	delta potential: $V(x) \ = \ -i\gamma\delta(x)$.

	\begin{figure}[tb]
		\includegraphics[width=\linewidth]{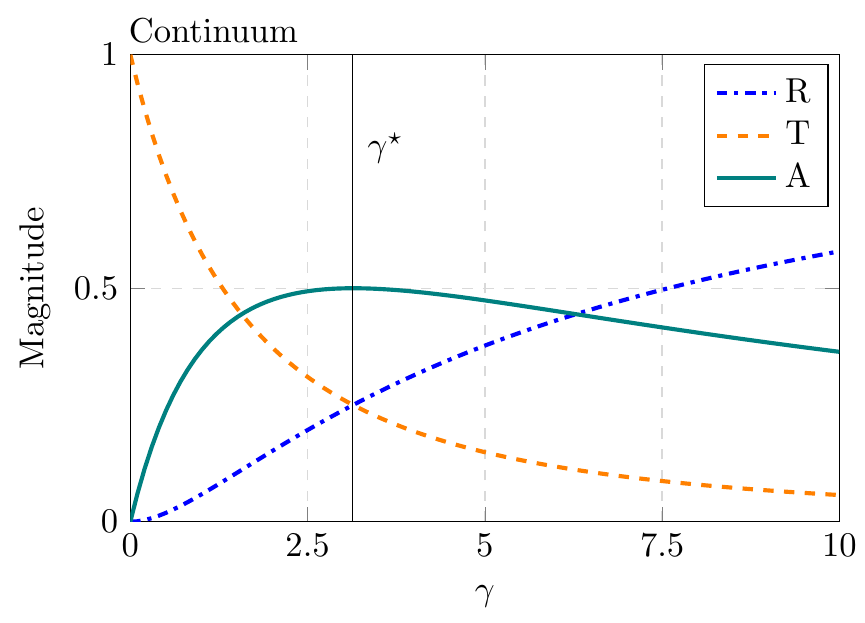}
		\caption{Continuum scattering.  The reflection, transmission and absorption probabilities
			($R$, $T$, $A$), plotted against the strength $\gamma$ of the dissipative delta-potential.
			Here $k = \pi/2$, $\hbar = 1$, and $m=0.5$.}
		\label{fig:RTAcontinuum}
	\end{figure}

	Solving the scattering problem is a variation of the standard textbook scattering problem with
	a real delta-function potential \cite{Griffiths_QM_book2017}.  We take the wavefunction to be of the
	form $e^{ikx}+re^{-ikx}$ on the left half-line ($x<0$) and of the form $te^{ikx}$ on the right
	half-line ($x>0$), with wavenumber $k>0$.  We then use the appropriate (dis)continuity conditions at
	$x=0$ to solve for the reflection and transmission amplitudes ($r$, $t$).  This yields
	\begin{align} \label{eq:rt}
	r \ = \ \frac{-1}{1 + \dfrac{k\hbar^{2}}{m\gamma}} \quad , \quad
	t \ = \ \frac{1}{1 + \dfrac{m\gamma}{k\hbar^{2}}} \, .
	\end{align}
	Using \eqref{eq:rt} we can obtain the reflection, transmission, and now also absorption probability
	as functions of the parameter $\gamma$: 
	\begin{equation}
	R=|r|^{2}, \quad T=|t|^{2}, \quad A = 1-R-T \label{eq:RTA} \ .
	\end{equation}
	We see in Fig.~\ref{fig:RTAcontinuum} that $R = T$ for a particular value of $\gamma$, and that $A$
	is maximized by some value of $\gamma$. Using equations \eqref{eq:rt} and \eqref{eq:RTA}, we find that these
	points are both equal to
	\begin{equation}
	\gamma^{\star} = \frac{k\hbar^2}{m}. \label{maxA_cont}
	\end{equation}
	These expressions depend on $\hbar$ and the mass $m$.  We set $\hbar=1$.  To facilitate comparison
	with the lattice situation, we choose $m=1/2$ so that the quadratic dispersion ($\hbar^2k^2/2m$) on
	the continuum matches the low-energy part of the cosine dispersion ($-2\cos{k}$) on the lattice
	without impurity.  Thus
	\begin{align} \label{eq:rt_unitless}
	r \ = \ \frac{-\gamma}{\gamma + 2k} \, , \quad t \ = \ \frac{2k}{\gamma + 2k} \, , \quad \gamma^{\star} = 2k \, .
	\end{align}

	\subsection{Lattice}\label{subsec:scattering_lattice}
	
	We now turn to the lattice problem.  Through numerical time evolution we will calculate the
	reflection and transmission fractions, $R$ and $T$, and obtain the absorption fraction using $A = 1
	- R-T$. 
	
	We initialize our particle as a (discrete version of) a Gaussian wavepacket,  localized around the
	site $j_0$ and carrying lattice momentum $k$:
	\begin{equation} 
	\ket{\psi(0)} = \sum_j \psi_j(0)\ket{j}  =~ \mathcal{N}^{-1} \sum_j   e^{ \frac{-(j-j_{0})^2}{2\sigma^{2}} } \, e^{ikj}  \, \ket{j}
	\end{equation}
	where $\mathcal{N}$ is a normalization constant.  A positive $k$ ensures that the wavepacket will
	propagate rightwards initially. The position $j_0$ is chosen such that the wavepacket starts on the
	left side of the lattice, and does not initially overlap significantly with either the lattice edges
	or the impurity.  The width $\sigma$ is chosen to be significantly larger than $1$, but significantly smaller than $L/2$.
	The wavepacket is evolved using the Hamiltonian: $\ket{\psi(t)} = e^{-iHt}\ket{\psi(0)}$.
	Expressing the wavefunction at time $t$ in the site basis, $\ket{\psi(t)} = \sum_j
	\psi_j(t)\ket{j}$, the coefficients $\psi_j(t)$ provide the occupancies,   $|\psi_j(t)|^2$.

	\begin{figure}
		\includegraphics*[width=1.0\linewidth]{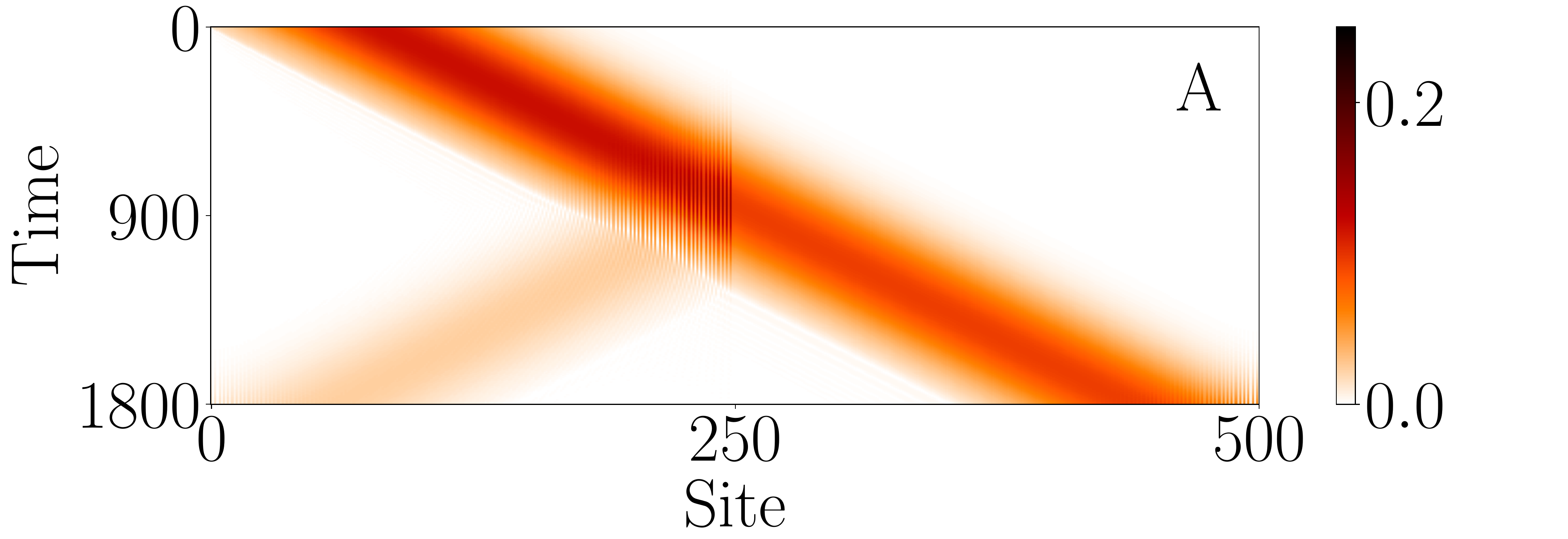}\\
		\includegraphics*[width=1.0\linewidth]{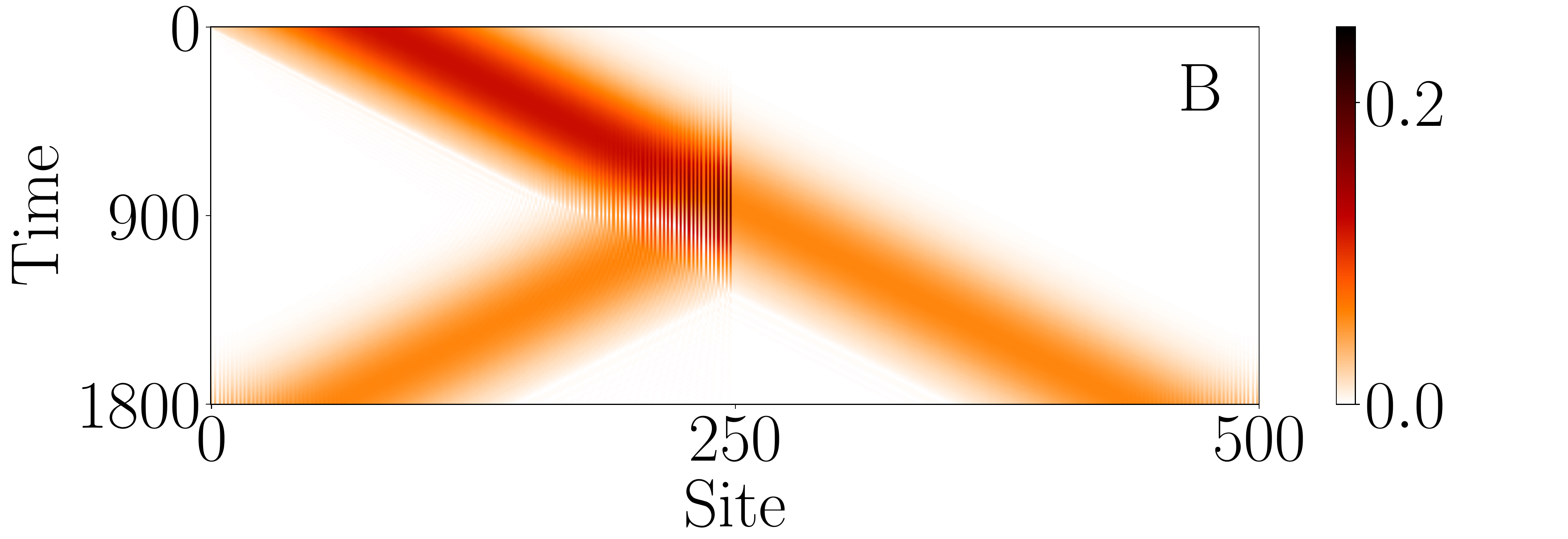}\\
		\includegraphics*[width=1.0\linewidth]{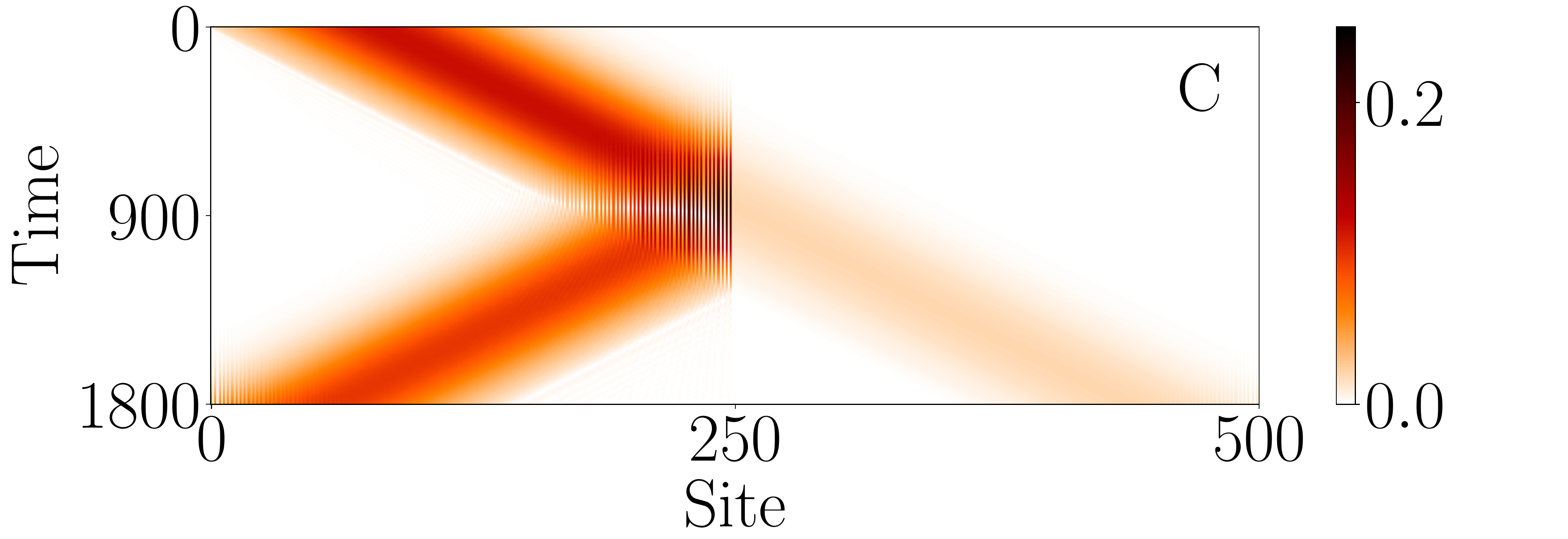}
		\caption{Wavepacket evolution illustrated by a density plot of site occupancies $|\psi_{j}|^2$. Here $L$ = 500, $\sigma$ = 40, $k$ = $\pi/2$. \textbf{A}: $\gamma = 0.5$ - Shows less of the wavepacket being reflected than transmitted. \textbf{B}: $\gamma = 2$ - Shows roughly similar amounts of the wavepacket being
			reflected/transmitted. \textbf{C}: $\gamma = 10$ - Shows less of the wavepacket being
			transmitted than reflected.}
		\label{fig:wavepacket}
	\end{figure}
	
	\begin{figure}
		\includegraphics[width=\linewidth]{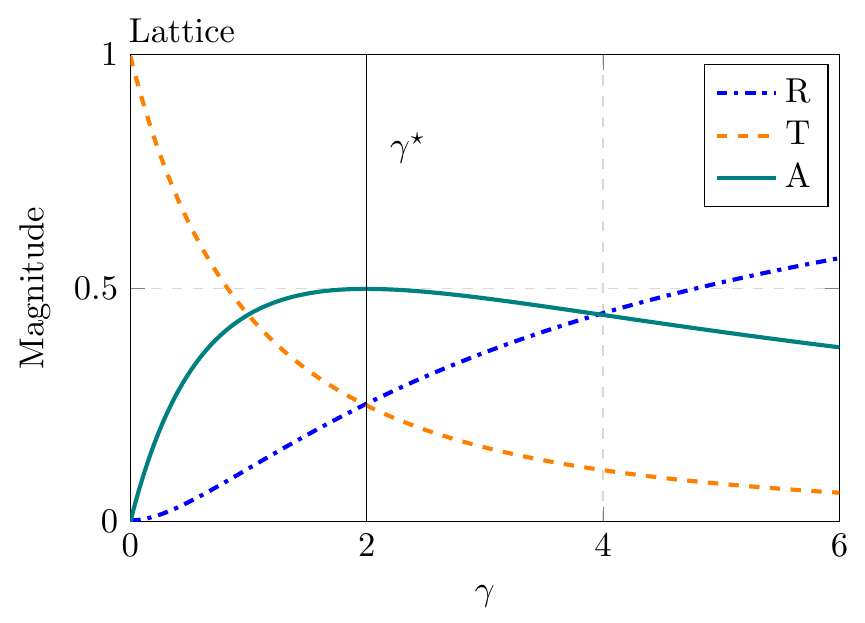}
		\caption{Reflection, transmission and absorption probabilities calcuated using wavepacket evolution on the lattice. ($R,T,A$ plotted against  $\gamma$.) Here $L=500$, $\sigma$ = 40, $k$ = $\pi/2$.}
		\label{fig:RTA_lattice}
	\end{figure}

	Fig.~\ref{fig:wavepacket} shows the evolution of a wavepacket for three different values of
	$\gamma$, initially localized near the left end of a $500$-site lattice.  After the particle is
	incident on the impurity, we see different portions being reflected and transmitted.
	Choosing a time after the collision has occurred, such that the reflected and transmitted packets are
	well-separated from the impurity, one can define the coefficients based on the wavefunction
	coefficients at this time.  The reflected (transmitted) fraction is the weight to the left (right)
	of the impurity.  Denoting the impurity site as $q$,
	\begin{align}
	R = \sum_{j=1}^{q} |\psi_j|^{2}, \quad T = \sum_{j=q+1}^{L} |\psi_j|^{2}, \quad A = 1-R-T \ .
	\end{align}
	Fig.~\ref{fig:RTA_lattice} shows the results of calculating the coefficients for a lattice with 500
	sites, with the impurity at site 250, for a range of values for $\gamma$.  The coefficients are
	extracted from time evolution with a $\sigma=40$ wavepacket. 
	The observation of weights on the left and right parts of the lattice is performed at a time well after the wavepacket has scattered off the impurity, but well before either the reflected or the transmitted wavepacket reaches one of the boundaries.  For Fig.~\ref{fig:RTA_lattice}, this time was $t = 160$.  For other values of $k$ (Fig.~\ref{fig:MaxA}), the times are different as the speed of the wavepacket depends on $k$.
	We have checked that the dependence on $\sigma$ is negligible provided $1\ll\sigma\ll{L/2}$.  For
	both Fig.~\ref{fig:wavepacket} and Fig.~\ref{fig:RTA_lattice}, the wavepacket momentum is
	$k=\pi/2$, for which the dispersion of the wavepacket is least severe.

	\subsection{Comparison between Continuum and Lattice} \label{subsec:continuum_lattice_comparison}
	
	Comparing Figures \ref{fig:RTAcontinuum} and \ref{fig:RTA_lattice}, we see that our lattice results
	are very similar to the continuum results, except for a rescaling of $\gamma$.  In the continuum
	case, we have found that the main feature [maximum of $A(\gamma)$, or crossing point of $R(\gamma)$
	and $T(\gamma)$] occurs at a value of $\gamma$ that is proportional to the momentum, $\gamma^{\star}
	= 2k$.  One therefore expects that in the lattice case $\gamma^{\star}$ should also depend on the
	momentum of the scattered particle.  More specifically, since the single-particle dispersion changes
	as $k^2\to-2\cos{k}$ in going from the continuum to lattice, one expects from the dependence of
	$\gamma^{\star}=2k$ in the continuum that the dependence might be $\gamma^{\star}=2\sin{k}$ on the
	lattice.
	
\begin{figure}
		\includegraphics*[width=\linewidth]{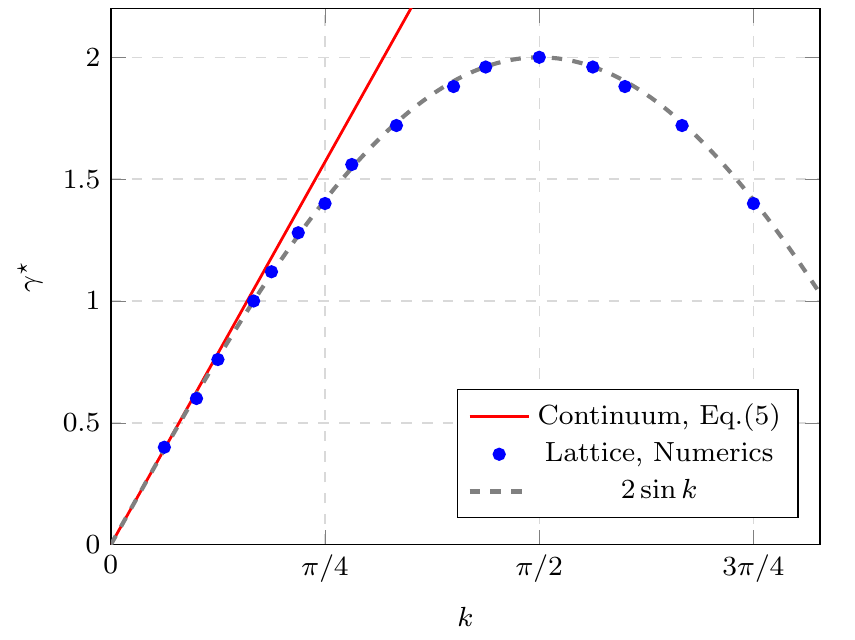}
		\caption{Comparing results of the value of $\gamma$ for which absorption is maximised in the
			continuum ($\gamma^{\star}=2k$) and on the lattice (obtained from numerical wavepacket
			evolution).  Lattice results are obtained with $L = 250$ and $\sigma = 15$.  For
			comparison, the function $2\sin{k}$ is plotted (dashed curve).}
		\label{fig:MaxA}
	\end{figure}

	We can extract $\gamma^{\star}$ for various momenta by running our numerical time evolution of
	wavepacket scattering for various momenta and identifying the maximum of
	$A(\gamma)$. 
	The results are shown in Fig.~\ref{fig:MaxA}, comparing the continuum and lattice case.  Indeed
	the momentum dependence of the $\gamma^{\star}$ appears to be $\approx2\sin{k}$ on the lattice, with
	a maximum of $\gamma^{\star}\approx2$ for $k=\pi/2$.
	
	\begin{figure}
		\includegraphics[width=1.0\linewidth]{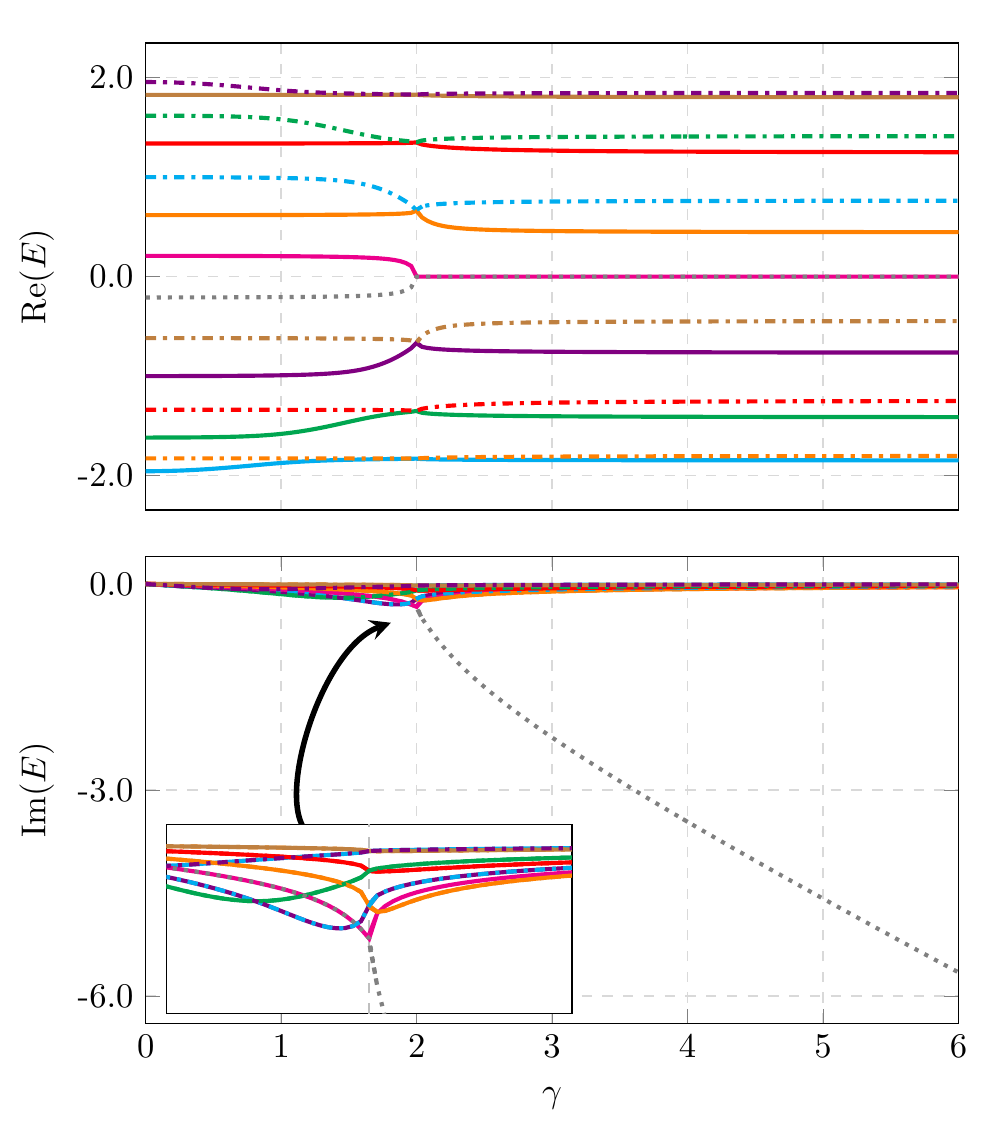}\\
		\caption{Energy spectrum of the Hamiltonian \eqref{eq:ham}, for $L = 14$, as function of the
			potential strength $\gamma$.  Real and imaginary parts of the eigenvalues are plotted
			separately.}
		\label{fig:Spectrum}
	\end{figure}
	
	\begin{figure}
		\includegraphics[width=1.0\linewidth]{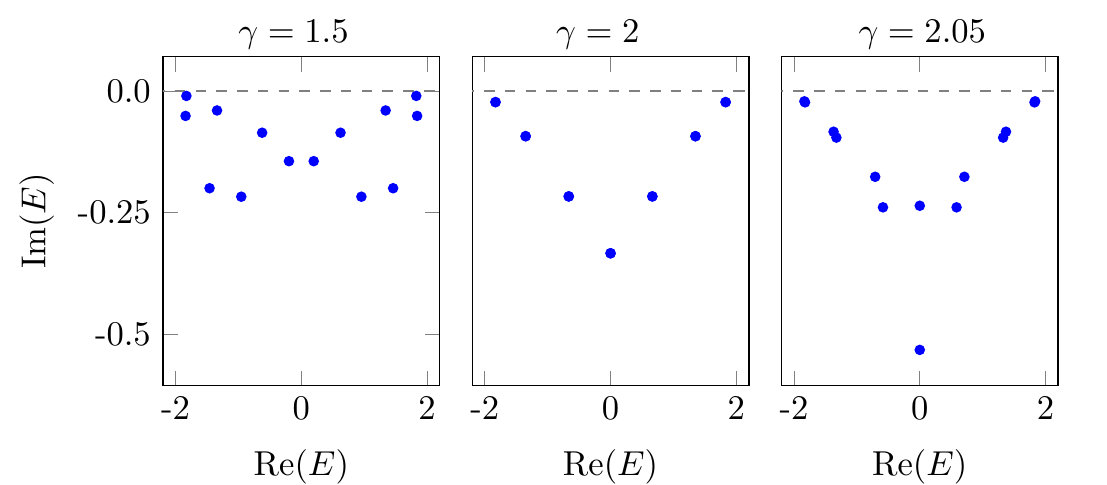}\\
		\caption{Eigenvalues of the Hamiltonian \eqref{eq:ham}, for $L = 14$, for three values of
			$\gamma$, below, at and above the exceptional point.  In each case, the $L$ eigenvalues
			are plotted on the complex plane.  For $\gamma=2$, only $L/2$ points are visible because
			the eigenvalues are paired.}
		\label{fig:Spectrum_complexplane}
	\end{figure}
	
	\section{Spectrum and exceptional points}\label{sec:eps}
	
	It turns out that the value  $\gamma\approx2$ also plays a  special role in the spectrum of our
	non-Hermitian lattice Hamiltonian. 
	
	Previously we presented data for systems with 500 and 250 sites.  For clarity, we now show the
	spectrum of smaller systems.  Fig.~\ref{fig:Spectrum} presents the eigenvalues for a system with 14
	sites as a function of $\gamma$.  As the eigenvalues are complex, the real and imaginary components
	are shown separately.  We also show the 14 eigenvalues in the complex plane, for
	three different values of $\gamma$, in Fig.~\ref{fig:Spectrum_complexplane}. 
	For any value of $\gamma$, the real part of the eigenvalues are generally spaced between $-2$ and
	$+2$, as one expects from a tight-binding one-dimensional lattice.  The most visibly striking
	feature in the spectrum is that, at $\gamma=2$, the eigenvalues coalesce in pairs.  (The coalescence
	is visible in the real parts --- the imaginary parts are already paired up even at $\gamma<2$.)
	
	This is not a higher-order exceptional point \cite{GraefeKorsch_JPA2008_higherorder,
		Heiss_JPA2008_higherorder_chirality, Demange_Graefe_JPA2011_higherorder}, but rather an
	exceptional point where all eigenvalues pair up as second-order exceptional points, not just two
	eigenvalues.  Of course, observing the eigenvalues is not sufficient to say that this is an
	exceptional point --- the eigenfunctions also need to coalesce.  Indeed, considering the pair of
	eigenstates whose eigenvalues become equal at $\gamma=2$, we find numerically that one of the
	eigenstates becomes equal to $-i$ times the other eigenstate.
	
	In Appendix \ref{app_degen} we show analytically that the eigenvalues always group into degenerate
	pairs at $\gamma=2$, for an even-$L$ lattice with the impurity at one of the central sites.   One
	can also show that the corresponding eigenstates for every such pair are linearly dependent.  
	
	Unlike exceptional points which separate a {$\cal PT$}-symmetric phase from a {$\cal PT$}-symmetry-broken phase, the eigenvalues of our system are complex on both sides of the
	exceptional point.  The imaginary parts on average have larger magnitude near the exceptional point, and generally decrease as one moves away from $\gamma=2$, with one striking exception.
	The exception corresponds to one of the two eigenvalues whose real part becomes zero.  The imaginary
	part becomes large and negative as $\gamma$ increases, and eventually becomes $\approx-\gamma$.
	This eigenvalue corresponds to a bound state localized at the dissipative impurity,
	which we will analyze in the next section.  
	
	The structure of the spectrum discussed here for $L=14$ is true for $L\mod4=2$.  For other values of
	$L$, there are variations, which we detail in Appendix \ref{app_size}. 
	In particular, for odd values of $L$, there is only a single pair of eigenvalues coalescing ($L\mod4=3$), or none at all ($L\mod4=1$). However, even with an odd number of sites the localized eigenstate still exists for large values of $\gamma$.
	In Appendix \ref{app_location} we also discuss the dependence of the location of the impurity site.
	
	\section{Large $\mathbf{\gamma}$} \label{sec:Large}
	
	At large $\gamma$ the absorption decreases, suggesting that the effect of the imaginary potential is
	similar to that of a real potential.  In this section we draw a comparison between the effects of
	real and imaginary on-site potentials. 
	
\begin{figure}
		\includegraphics*[width=\linewidth]{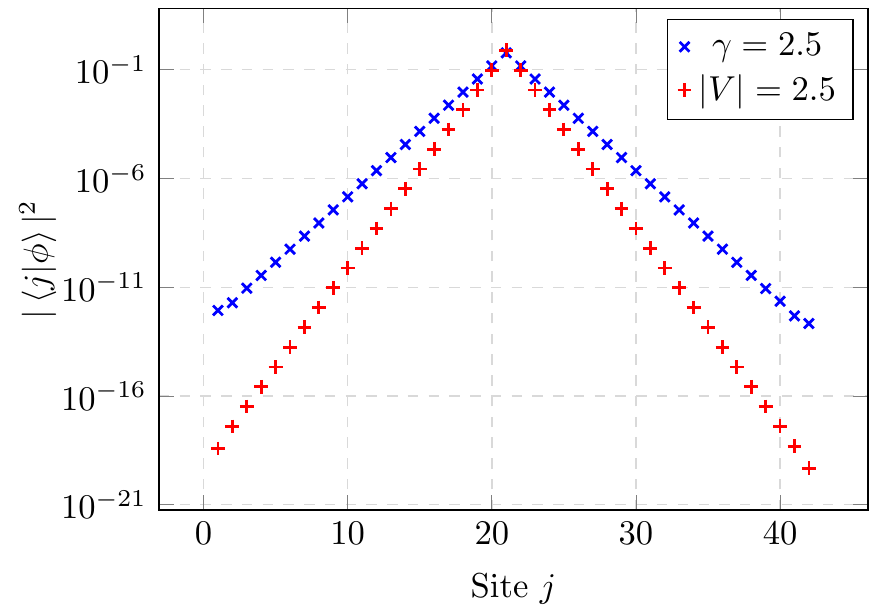}
		\caption{Site occupancies of the localized eigenstate, for both a real ($V$) and an imaginary
			($-i\gamma$) potential of magnitude 2.5, and $L = 42$ sites.  The scale is log-linear.}
		\label{fig:bound}
	\end{figure}

	In Section \ref{sec:eps} we saw there was a single eigenvalue, with a corresponding eigenstate,
	which had a purely imaginary negative component.  At large $\gamma$ the eigenenergy approaches
	$-i\gamma$, for which a plausible explanation would be that the eigenstate is localized at or around the
	impurity site $q$ and hence its energy is primarily determined by the $-i\gamma \ket{q}\bra{q}$ term
	in the Hamiltonian \eqref{eq:ham}.  Indeed the corresponding eigenstate is numerically found to be
	exponentially localized around the impurity site  (Fig.~\ref{fig:bound}). 
	
	For comparison, we also consider the effect of a real potential, i.e., the Hermitian Hamiltonian 
	\begin{equation}
	\label{eq:ham_real}
	H \ = \ -J\sum_{j=1}^{L-1} \Big(\ket{j}\bra{j+1} + \ket{j+1}\bra{j} \Big) + V \ket{q}\bra{q}
	\end{equation}
	Here $V$ is a real parameter which could be either positive or negative.  It is known that this
	Hamiltonian supports a bound state for negative $V$ and an anti-bound state for positive $V$.  (The
	spectrum, which is real, contains one state which separates from the band and at large $|V|$
	approaches $V$.)  This eigenstate is exponentially localized around site $q$.
	
\begin{figure}
		\includegraphics*[width=0.9\linewidth]{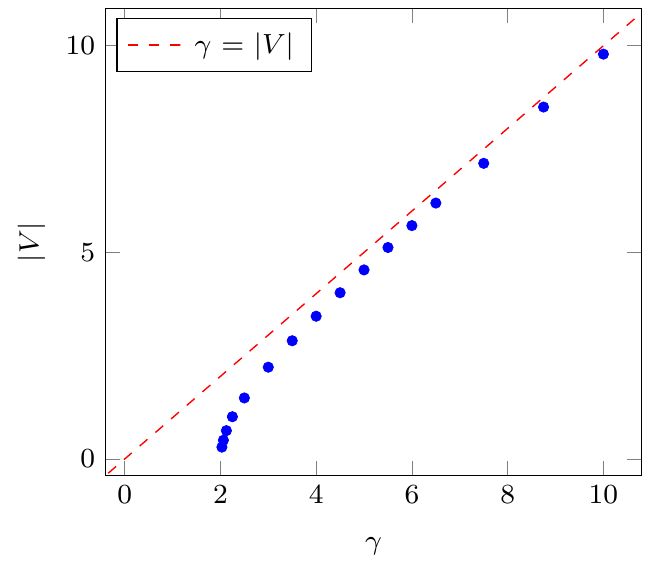}
		\caption{A correspondence between the parameters of the real and imaginary potentials, using
			the localization length of the bound states, for $L = 42$. }
		\label{fig:gversusv}
	\end{figure}

	In Fig.~\ref{fig:bound} we show the exponential localization of the eigenstate both for the real
	potential ($|V|=2.5$) and for the dissipative impurity $\gamma=2.5$.  At these values, the
	eigenstate is more strongly localized (has smaller localization length) for the case of the real
	potential, Eq.~\eqref{eq:ham_real}.
	Approximating the occupancies at site $j$ by the form $\propto e^{(j-q)/\alpha}$, where $q$ is the
	impurity position, one can extract the localization length $\alpha$.  By extracting $\alpha$ for the
	localized eigenstate for various values of $\gamma$ in the case of our non-Hermitian system
	\eqref{eq:ham}, and for various values of $V$ in the case of the system \eqref{eq:ham_real}, we can
	assign to each $\gamma>2$ a value of $V$, for which the same localization length is obtained.
	Results of this calculation are shown in Fig.~\ref{fig:gversusv}, for a system with $L=42$ sites.
	This quantifies the idea that, at large $\gamma$, an absorbing impurity behaves like a real-valued
	impurity.
	
	For values of $\gamma < 2$, there is no bound state.  For $\gamma$ slightly larger than $2$, the
	localization length corresponds to the bound state of a very weak real potential (very small $|V|$).
	As $\gamma$ grows, the corresponding $|V|$ increases and asymptotically approaches $|V|=\gamma$.  In
	other words, the effect of an absorbing impurity of large strength $\gamma\gg2$ is similar to that of a
	real-valued impurity of the same strength.  
	
	One can ask whether in the continuum case there is a similar correspondence --- in that case also
	the absorption is low for the non-Hermitian model at large $\gamma$.  It is well-known that the
	negative real delta-potential has a single bound (localized) state.  However, neither the positive
	real potential, nor the imaginary potential, have bound states.  (If one assumes that there is a
	bound state for some potential, $\lambda \delta({x})$, one finds that $\lambda$ must have a real
	component, which is negative.)  Hence no quantitative correspondence can be drawn in terms of the
	localization length, as we have done for the lattice.
	
	The existence of a strongly localized eigenstate provides a simple `spectral' interpretation of the suppression of absorption at large $\gamma$ that we have presented in Subsection \ref{subsec:scattering_lattice}.  For large $\gamma$, the localized eigenstate has near-zero overlap with the incident wavepacket, because in the initial state the wavepacket is far from the impurity site.  Thus, the wavepacket is `shielded' from the impurity, because its dynamics is confined to the subspace of all the other eigenstates which have near-zero weight at the impurity site.  Therefore the wavepacket undergoes almost no absorption.  Curiously, for the suppression of absorption in the continuum case (Subsection \ref{subsec:scattering_continuum}), the same interpretation cannot be used, as there is no localized eigenstate in that case.
	
	In Appendix \ref{app_states} we show site occupancy profiles for a sample of some of the
	eigenstates.  Other than the special (localized) eigenstate, the other eigenstates resemble those
	for a real potential --- the eigenstates at the bottom and top of the band have few nodes, while
	those near the center of the band have many nodes.

	\section{Discussion and Context} \label{sec:concl}
	
	We have studied the scattering dynamics and the spectrum of a tight-binding single-particle system
	with a non-Hermitian absorbing impurity at one site, focusing on the case where the impurity is near
	the center of the lattice.

	Setups loosely similar to ours have been explored in a few other recent works.  In
	Ref.~\cite{Longhi_EPL2017_Floquet}, scattering off a localized lattice impurity is studied, in the
	case where the strength and phase of the impurity are oscillating.  Scattering was studied using
	Gaussian wavepackets, as in the present work.  For certain parameters, the oscillatory non-Hermitian
	impurity was reported to allow perfect transmission (`Floquet invisibility').  In
	Ref.~\cite{Wiersig_PRA2018}, the lattice impurity was placed at the lattice edge and the role of the
	non-orthogonality of the eigenstates on the non-unitary time evolution was explored.  In addition,
	some related issues have been discussed in the context of {$\cal PT$}-symmetric lattice systems
	formed by having imaginary potentials on multiple sites
	\cite{Bendix_Kottos_Shapiro_PRL2009_latticePT, Jin_Song_PRA2009_latticePT,
		Jin_Song_PRA2010_latticePT, Joglekar_Scott_Babbey_Saxena_PRA2010_tightbindingPT,
		Jin_Song_PRA2016_PTflux, Zhu_etal_PRA2016_latticePT_scattering, Ortega_etal_arXiV2019_latticePT}.
	The spectrum of lattices with two impurities has been studied in
	Refs.~\cite{Jin_Song_PRA2009_latticePT, Ortega_etal_arXiV2019_latticePT}.
	Ref.~\cite{Bendix_Kottos_Shapiro_PRL2009_latticePT} reported an eigenstate which is localized on the
	two impurity sites --- this may be considered a {$\cal PT$}-symmetric version of the localized
	eigenstate we have studied.  Refs.~\cite{Jin_Song_PRA2009_latticePT, Jin_Song_PRA2010_latticePT}
	have made comparisons between the non-Hermitian system and corresponding Hermitian system, as we
	have done.  After the appearance of our preprint, our single-particle non-Hermitian Hamiltonian has appeared in Ref.~\cite{Chiocchetta_Kollath_Diehl_arXiv2019_manybodyQuantumZeno} as an effective Hamiltonian.
	
	\textit{Experimentally}, lattice systems with localized losses have
	been studied in several contexts.  In the setup of
	Ref.~\cite{Labouvie_PhysRevLett.115.050601,Labouvie_PhysRevLett.116.235302},
	a Bose-Einstein condensate is realized in a one dimensional optical
	lattice, with engineered losses on a single site acting as a local
	dissipative potential. Connecting single-particle results such as ours
	to many-boson physics in such a setup remains an interesting challenge
	for future work.

	A realization more similar to the single particle tight-binding system
	considered in this work is that with photonic lattice systems, such as
	those in
	Refs.~\cite{Szameit_nature_2014_susy_mode_conv,Eichelkraut2018}. In
	this setup, photonic lattices are realized using femtosecond laser
	writing to inscribe waveguide arrays with appropriate index profiles
	in fused silica. The physics of photons in such an architecture can be
	well-described by a tight-binding model, with an additional spatial
	direction taking the role of time.  This setup, or its variants, has
	been used to demonstrate a number of paradigmatic tight-binding
	phenomena, including Bloch oscillations
	\cite{Silberberg_photoniclattice_BlochOscil_PRL1999} and Anderson
	localization \cite{Segev_photoniclattice_andersonloc_Nature2007,
		Szameit_dynamicloc_Andersonloc_OptExpr2020}.  Both one-dimensional
	and two-dimensional lattices have been realized, and lossy sites and
	other types of non-Hermiticity have been explored
	\cite{Szameit_Segev_PRA2011,
		Szameit_Rudner_PRL2015_topological_transition, Eichelkraut2018}.  It
	is possible to create localized excitations (wavepackets) and observe
	their propagation \cite {Szameit_supersymmetric_OptLett2014,
		Szameit_dynamicloc_Andersonloc_OptExpr2020}.  Thus, studies of
	scattering off lossy sites should be possible in such a setup.
	
	Another possible experimental setting for observing scattering off
	non-Hermitian potentials in a tight-binding lattice might be microwave
	realizations using coupled dielectric resonators, such as that
	discussed in \cite{Schomerus_Nature2015_resonatorchain}.  This setup is well approximated by a nearest-neighbour tight-binding
	Hamiltonian. The resonance frequency of an isolated resonator, and the
	coupling strength between two resonators (due to the evanescent
	electromagnetic field), correspond to the on-site energy  and
	to the hopping term, respectively.  A
	controllable on-site loss is created by placing an absorbing material
	on a particular resonator.

	In the present work, by explicit time evolution starting from initial states which are
	momentum-carrying wavepackets, we found the reflection, transmission and absorption coefficients
	($R$, $T$, $A$) as a function of the impurity strength $\gamma$ and of the incident momentum $k$.
	The absorption was shown to first increase and then decrease as the strength $\gamma$ is increased.
	It can be argued that this non-monotonic behavior is related to the quantum Zeno effect.  The
	experimental non-monotonic behavior of Ref.~\cite{Ott_PRL2013} can be interpreted in the same light.
	We have demonstrated and analyzed the effect in a simple lattice setting.  We have also compared
	with the scattering of a single particle in a continuum from an absorptive delta-potential.
	
	We have also presented the spectrum of the non-Hermitian system.  The system we focus on --- even
	number of sites, impurity at one of the central sites --- has an unusual exceptional point
	structure.  At the same value of $\gamma$, \emph{all} the eigenstates of the systems coalesce in
	pairs.  This is not a higher-order of exceptional point \cite{GraefeKorsch_JPA2008_higherorder,
		Heiss_JPA2008_higherorder_chirality, Demange_Graefe_JPA2011_higherorder}, rather, it is a
	collection of many second-order coalescences at the same point in parameter space.  At larger
	$\gamma$, the spectrum contains one localized eigenstate.  This is another way in which a strong
	absorptive impurity acts like a real-valued impurity potential.  This feature is particular to the
	lattice as there are no bound states in the corresponding continuum problem.  The eigenvalue
	corresponding to the localized eigenstate has a \emph{purely imaginary} value.
	
	Our work opens up several avenues of research.  We have explored scattering dynamics. A detailed study of other types of dynamics remains to be done, not only for tight-binding lattices, but also for continuum particles subjected to localized absorbers. Extending such dynamical considerations to nonlinear cases \cite{Konotop_Ott_PRL2009,Zezyulin_Barontini_Ott_MacroscopicZeno_PRL2012, Labouvie_PhysRevLett.116.235302} also deserves further exploration.
	The spectral part of the present study provides motivation for a more thorough investigation of the
	spectrum of relatively simple non-Hermitian models.  The structure we have found --- many pairs
	coalescing at the same point --- suggests that non-Hermitian spectra may hold more surprises not yet
	known in the literature.

	\begin{acknowledgments}
		We thank S.~Nulty (Maynooth) for analytically demonstrating the degeneracy at $\gamma=2$ (Appendix
		\ref{app_degen}).  PCB thanks Maynooth University (National University of Ireland, Maynooth) for funding provided via the John \& Pat Hume
		Scholarship.
	\end{acknowledgments}
	
	\appendix
	
	\section{Analytical expressions for spectrum}
	\label{app_degen}
	In the main text, we have shown numerically that the eigenvalues of our system coalesce in pairs at
	$\gamma=2$, for even $L$, when the impurity site $q$ is one of the central sites, i.e., when $q=L/2$
	or $q=(L/2)+1$.  In this Appendix, we analyze the eigenvalues analytically.  We express the
	characteristic polynomial (whose roots are the eigenvalues) in a form which allows us to predict,
	first, that all the eigenvalues pair up when $q$ is one of the central sites, and second, that this
	is a multiple exceptional point because each eigenstate pair is linearly dependent.  The
	characteristic polynomial is treated in Section \ref{subsec_general_q} and the case of $q=L/2$
	(or $q=L/2+1$) is considered in Section \ref{subsec_central_q}.

	\subsection{General location, $q$} \label{subsec_general_q}

	We want to find the eigenvalues of the  $L\times L$ matrix 
	\begin{equation}
	\label{matrix_equation}
	\left[H_q\right]_{jk}=-\delta_{j,k+1}-\delta_{j+1,k}-i\gamma\delta_{jq}\delta_{jk} \, .
	\end{equation}
	Here $1\leq q\leq L$. The characteristic polynomial of this matrix up to a minus sign is the
	determinant of the tridiagonal matrix 
	
	\begin{equation}
	\pmqty{\lambda & 1 & 0 & \ldots &\ldots &\ldots &\ldots &\ldots\\ 
		1 & \lambda & 1 &\ldots &\ldots &\ldots &\ldots &\ldots\\ 
		\vdots & \vdots & \ddots &\ddots &\ddots &\ldots &\ldots &\ldots\\ 
		\vdots & \vdots & \vdots &1 &\lambda +i\gamma &1 &\ldots &\ldots\\ 
		\vdots & \vdots & \vdots &\vdots &\ddots &\ddots &\ddots &\ldots\\
		\vdots & \vdots & \vdots &\vdots &\vdots &1 &\lambda &1\\
		\vdots & \vdots & \vdots &\vdots &\vdots &\vdots &1 &\lambda\\} . 
	\label{char_M}
	\end{equation}
	
	Now the determinants of tridiagonal matrices satisfy a recurrence relation. If  $P_{n}$ is the
	determinant of the $n\times{n}$ matrix with elements
	\begin{equation}
	A_{ij}=b_i\delta_{i,j+1}+c_{j}\delta_{i+1,j}+a_i\delta_{ij}, 
	\end{equation}
	then 
	\begin{equation}
	P_n=a_n P_{n-1}-c_{n-1}b_{n-1}P_{n-2} \, . 
	\end{equation}
	This recurrence relation can be verified by determinant expansion and appears in numerous sources,
	e.g., is mentioned in Section 8.4 of Ref.~\cite{Golub_VanLoan_MatrixComputations_book1996}.
	The characteristic polynomial  of  $H$ \eqref{matrix_equation}, i.e., the determinant of
	the matrix \eqref{char_M}, therefore satisfies 
	\begin{align} \label{Pn_recur}
	P_n&=\lambda P_{n-1}-P_{n-2}, \quad \text{if } n\neq q \notag	\\
	P_q&=(\lambda+i\gamma) P_{q-1}-P_{q-2}, \quad \text{if } q>1  \notag\\
	P_0&=1 \notag \\
	P_1&=\lambda+i\gamma \delta_{q1} .
	\end{align}
	A standard method of solving such linear recurrence relations is to use the $Z$ transform.  Ignoring
	the second line in Eq.~\eqref{Pn_recur}, i.e., ignoring the impurity, we can get an expression for
	$P_n$ in terms of $P_0$, $P_1$ and $\lambda$.  Defining $F(z)=Z\{P_n\}$ and using a shift theorem,
	we get 
	\begin{gather}
	z^2F(z)-z^2 P_0-zP_1=\lambda (zF(z)-zP_0)-F(z) \, .
	\end{gather}
	After solving for $F(z)$ and decomposing into partial fractions, one can take the inverse $Z$
	transform, yielding
	\begin{align}\label{Pn_result_from_Z}
	P_n=\frac{P_0}{\sqrt{\lambda^2-4}}\left[(x_+)^{n+1}-(x_-)^{n+1}\right] \notag\\
	+\frac{P_1-\lambda P_0 }{\sqrt{\lambda^2-4}}\left[(x_+)^n-(x_-)^n\right]
	\end{align}
	where $x_{\pm}(\lambda)=\frac{1}{2}\left[\lambda \pm \sqrt{\lambda^2-4}\right]$.
	
	Defining:
	\begin{equation} \label{Kn1}
	K_n(\lambda):=\begin{cases}\dfrac{1}{\sqrt{\lambda^2-4}}\left[(x_+)^{n+1}-(x_-)^{n+1}\right] &
	\text{for}\; n\geq 0 \\ 0 & \text{for}\; n<0
	\end{cases}
	\end{equation}
	we can rewrite Eq.~\eqref{Pn_result_from_Z} as
	\begin{equation}\label{Pn_intermsof_Kn}
	P_n=P_0 K_{n}+(P_1-\lambda P_0) K_{n-1} . 
	\end{equation}
	Since we have derived this ignoring the impurity, Eqs.~\eqref{Pn_result_from_Z}, 
	\eqref{Pn_intermsof_Kn} are valid either for $q=1$, in which case $P_1=\lambda+i\gamma$, or for
	values of $n$ less than $q$.
	
	For $q=1$, we have $P_0=1$ and $P_1=\lambda+i\gamma$ so that $P_n=K_{n}+i\gamma K_{n-1}$, and therefore:
	\begin{equation} \label{P_L_q1}
	P_L = K_{L}+i\gamma K_{L-1}\quad \text{for } q=1.
	\end{equation}
	
	We now turn to $q>1$.  For $n<q$, Eqs.~\eqref{Pn_result_from_Z} and
	\eqref{Pn_intermsof_Kn} are valid directly with  $P_0=1$ and $P_1=\lambda$, i.e., with $P_1-\lambda
	P_0=0$, so that
	\begin{equation}
	P_n=K_{n}\quad \text{for } q>1 \text{ and } n<q.
	\end{equation}
	We have expressions for $P_n$ up to $n=q-1$, but we want $P_L$ and $L\geq{q}$.  To go beyond $q$, we
	define a new sequence of functions $Q_n(\lambda)$, satisfying the same recurrence relation as $P_n$
	(\ref{Pn_recur}), except with new initial conditions: $Q_0=P_{q-1}$ and
	$Q_1=P_q=(\lambda+i\gamma) P_{q-1}-P_{q-2}$.  Thus we need to solve
	\begin{align}
	Q_n&= \lambda Q_{q-1}-Q_{q-2},\quad Q_0=K_{q-1}, \notag\\ Q_1&=(\lambda+i\gamma) K_{q-1}-K_{q-2}.
	\end{align}
	Now we have already solved the same recurrence relation for $P_n$, using the $Z$ transform.  The
	solution is $Q_n=Q_0 K_{n}+(Q_1-\lambda Q_0) K_{n-1}$.  Therefore
	\begin{align}
	Q_n= K_{q-1}K_{n}+(i\gamma K_{q-1}-K_{q-2}) K_{n-1} \, .
	\end{align}
	Noting that $Q_n(\lambda)=P_{n+q-1}(\lambda)$, the determinant of the full matrix can be found as
	$P_{L}(\lambda)=Q_{L-q+1}(\lambda)$. Thus
	\begin{equation} \label{P_L_anyq}
	P_{L}(\lambda)=K_{q-1}K_{L-q+1}+(i\gamma K_{q-1}-K_{q-2}) K_{L-q} \, .
	\end{equation}
	
	We now introduce a slight change of notation: We refer to this polynomial as $P_{L,q}$.  In other
	words, the characteristic polynomial of the Hamiltonian matrix of a lattice of size $L$ and having
	the impurity at position $q$ will be called $P_{L,q}$.  Note that Eq.~\eqref{P_L_anyq} reduces to
	Eq.~\eqref{P_L_q1} for $q=1$; thus
	\begin{align}
	P_{L,q}=K_{q-1}K_{L-q+1}+(i\gamma K_{q-1}&-K_{q-2}) K_{L-q} 
	\end{align}
	for all positions of the impurity, $1\leq q\leq L$.
	
	By binomial-expanding $(x_{\pm})^{n+1}$, one can show that 
	\begin{align} \label{sym1}
	P_{L,q}(-\lambda^*)=(-1)^L &P_{L,q}(\lambda)^*\  .
	\end{align}
	This shows that the zeros of $P_{L,q}$ (eigenvalues of $H$) are symmetric by reflection through the
	imaginary axis in the complex plane, since if $\lambda=a+ib$ is a zero then $-\lambda^*=-a+ib$ is
	also a zero.  This symmetry is obvious from the spectra shown in
	Fig.~\ref{fig:Spectrum_complexplane}.

	\subsection{Impurity at center} \label{subsec_central_q}
	
	We now turn to the case we have focused on in this paper: when $L$ is even and $q=L/2$ or
	$q=\frac{L}{2}+1$.  In this case,
	\begin{align*}
	P_{L,\frac{L}{2}}&=K_{\frac{L}{2}-1}K_{\frac{L}{2}+1}+(i\gamma K_{\frac{L}{2}-1}-K_{\frac{L}{2}-2}) K_{\frac{L}{2}}\\
	&=K_{\frac{L}{2}-1}(\lambda K_{\frac{L}{2}}-K_{\frac{L}{2}-1})+(i\gamma K_{\frac{L}{2}-1}-K_{\frac{L}{2}-2}) K_{\frac{L}{2}}\\
	&=-(K_{\frac{L}{2}-1})^2+K_{\frac{L}{2}}\left(\lambda K_{\frac{L}{2}-1}-K_{\frac{L}{2}-2}\right)\\
	& \qquad\qquad\qquad\qquad\qquad\qquad\qquad\qquad +i\gamma K_{\frac{L}{2}-1} K_{\frac{L}{2}}\\
	&=(K_{\frac{L}{2}})^2-(K_{\frac{L}{2}-1})^2+i\gamma K_{\frac{L}{2}-1} K_{\frac{L}{2}} \,. 
	\end{align*}
	Now precisely when $\gamma=2$, this can be written as
	\begin{equation}
	P_{L,L/2}=\left(K_{\frac{L}{2}}+iK_{\frac{L}{2}-1}\right)^2 \,. 
	\end{equation} 
	This means that every root of the polynomial is a zero of order at least $2$, i.e., the
	eigenspectrum is doubly degenerate at $\gamma=2$.  We have thus analytically derived the most
	prominent feature of the spectrum presented in the main text.
	
	We now argue that, for a tridiagonal system such has ours, a coalescence of eigenvalues implies a
	coalescence of eigenstates, i.e., that the eigenstates corresponding to the equal eigenvalues are
	always linearly dependent.  Consider some eigenvalue $\lambda$ and corresponding eigenvector $X =
	(x_1,x_1,\ldots,x_L)^T$.  Due to the form of the matrix, all the components $x_i$ can be written as
	a function of $\lambda$ and the terms on the diagonals, times the first component $x_1$.  If we have
	any two eigenvectors with the same eigenvalue $\lambda$, the functions in the eigenvectors are the
	same functions, and hence the eigenvectors only differ in the choice of $x_1$, i.e., they are
	linearly dependent. Thus, if there is a degeneracy at some point, the eigenvectors are linearly
	dependent, and hence we have an exceptional point.


	\section{Eigenstates} \label{app_states}
	
	We show some eigenstates of the system, through their occupancy profiles.  
	
\begin{figure}
		\includegraphics*[width=\linewidth]{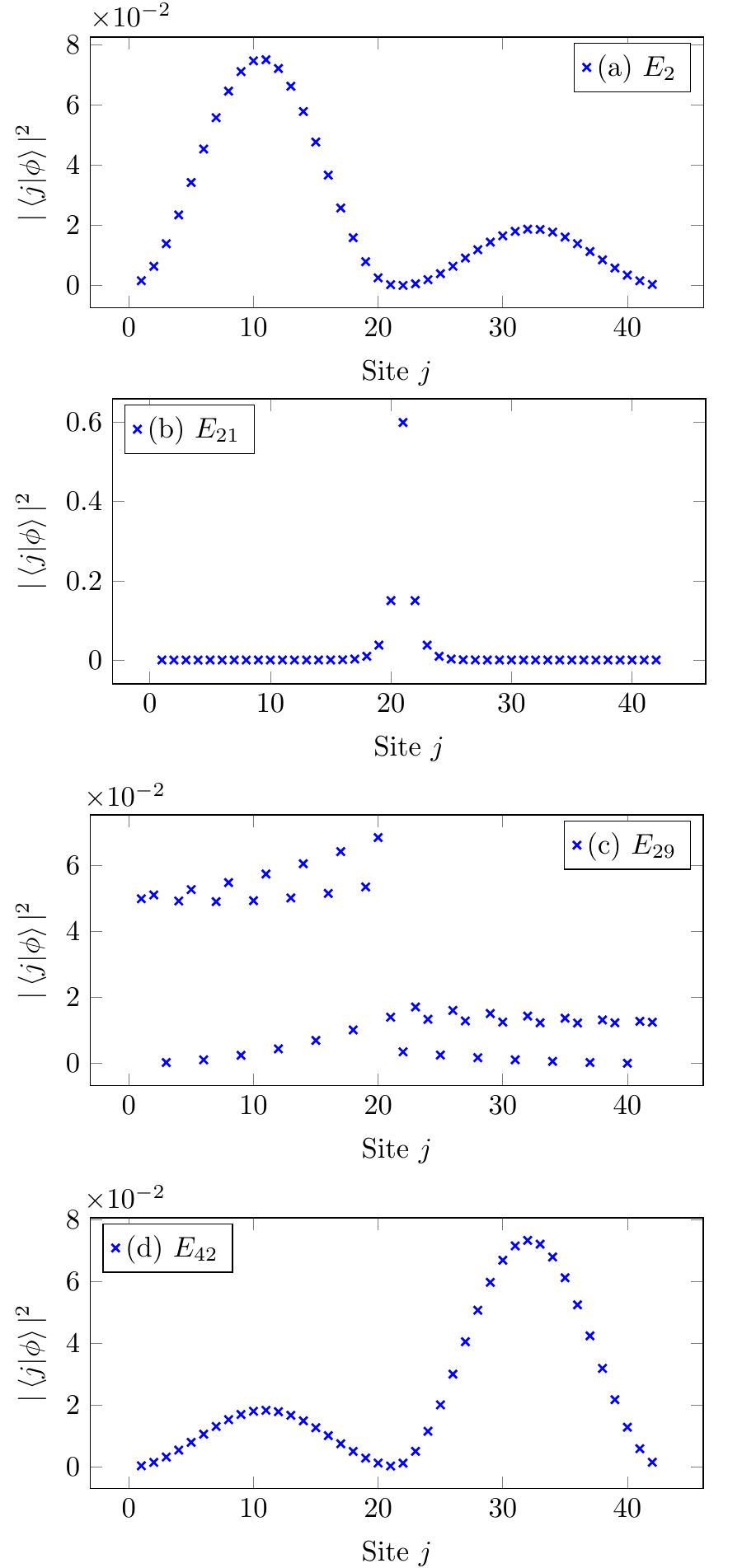}
		\caption{Occupancy profiles of a sample of the eigenstates, for a $L = 42$ system and
			$\gamma = 2.5$.  The second shown eigenstate from the top is the localized eigenstate.}
		\label{fig:eigstates}
	\end{figure}

	Since the eigenvalues are complex, there is no particularly natural way to order them. Here we order
	the eigenstates based on their real component, and then by their imaginary component, from smallest
	to largest, i.e., $1-2i$ comes before $1+2i$. Fig.~\ref{fig:eigstates} illustrates a selection of
	the eigenstates of a system with $L=42$ sites. They are labeled as `$E_i$', i.e. the eigenstate
	presented is the state corresponding to the $i^{th}$ eigenvalue, when ordered in the described
	manner.
	
	Note that the eigenvectors coefficients $\innerproduct{j}{\phi}$ are themselves complex; we only show the
	occupancies $\left|\innerproduct{j}{\phi}\right|^2$ and not the real and imaginary parts
	separately.  (Here $\ket{\phi}$ is the eigenvector in question and $j$ is the site index.)


	\section{Size dependence of the spectrum}\label{app_size}
	
	In Fig.~\ref{fig:Spectrum} we saw coalescence of every pair of eigenvalues at $\gamma=2$.  This
	was for a system with $L=14$ sites, and the impurity at site $q=7$.  We now outline the
	$L$-dependence of the spectrum.  The pattern is different for odd $L$.  For even $L$, there is a
	difference between $L$ values satisfying $L=4n+2$ and those satisfying $L=4n$, where $n$ is a
	non-negative integer.
	
	\begin{figure}
		\includegraphics[width=\linewidth]{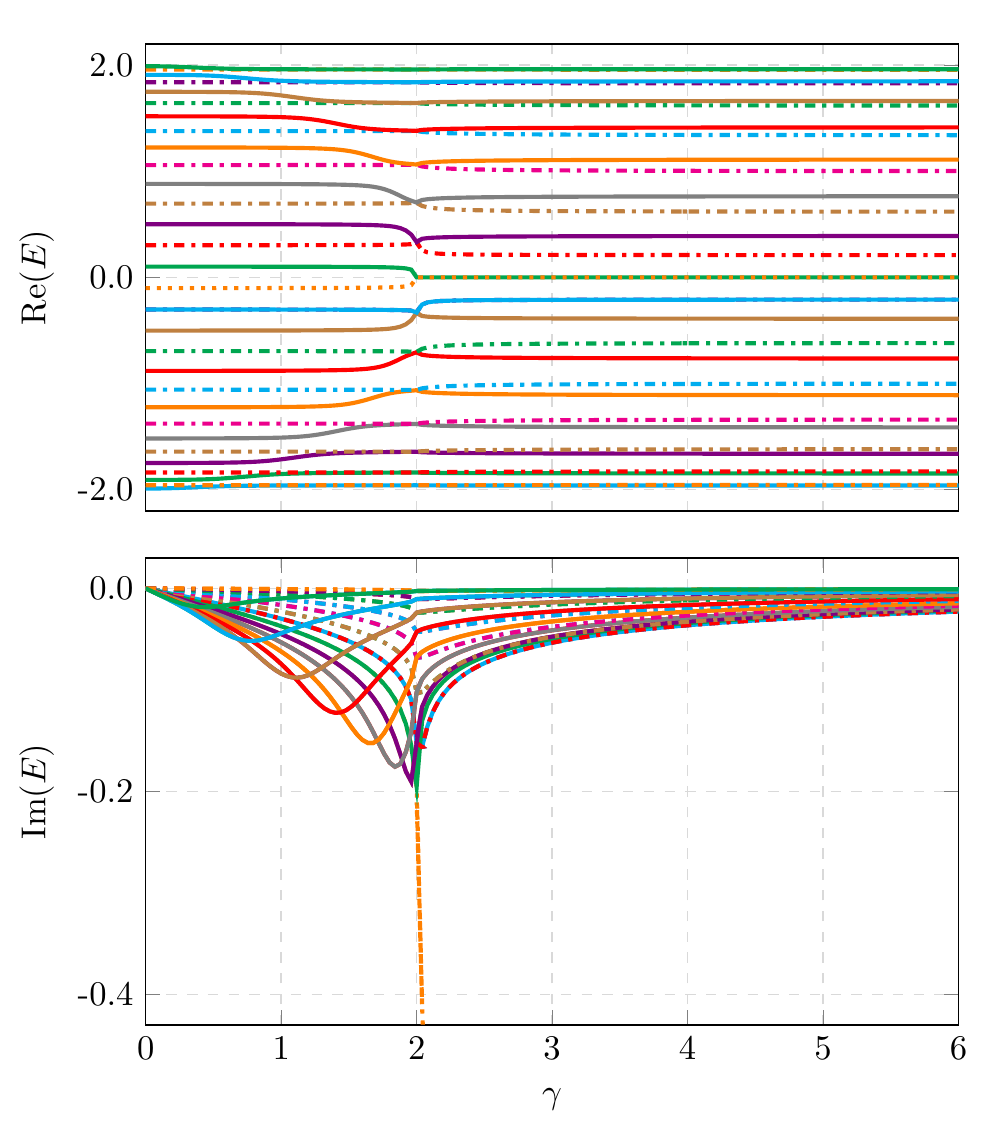}
		\caption{Energy spectrum of a system with $L = 30$.  As this value is in the $L=4n+2$
			sequence, the features are the same as those described in the main text for $L=14$.}
		\label{fig:Spectrum30}
	\end{figure}

	The case $L=14$, presented in the main text, belongs to the $L=4n+2$ sequence ($6$, $10$, $14$,
	$18$, \ldots).  In Fig.~\ref{fig:Spectrum30} we show the case of $L=30$, showing exactly the same
	pattern: all eigenvalues pair up in a multiple exceptional point exactly at $\gamma=2$.  There are
	an odd number of pairs, and the eigenvalues with central real values have zero eigenvalue after the
	coalescence, i.e., for $\gamma>2$.  One of these two eigenvalues correspond to the localized
	eigenstate, and has imaginary part growing with $\gamma$.

	\begin{figure}
		\includegraphics[width=\linewidth]{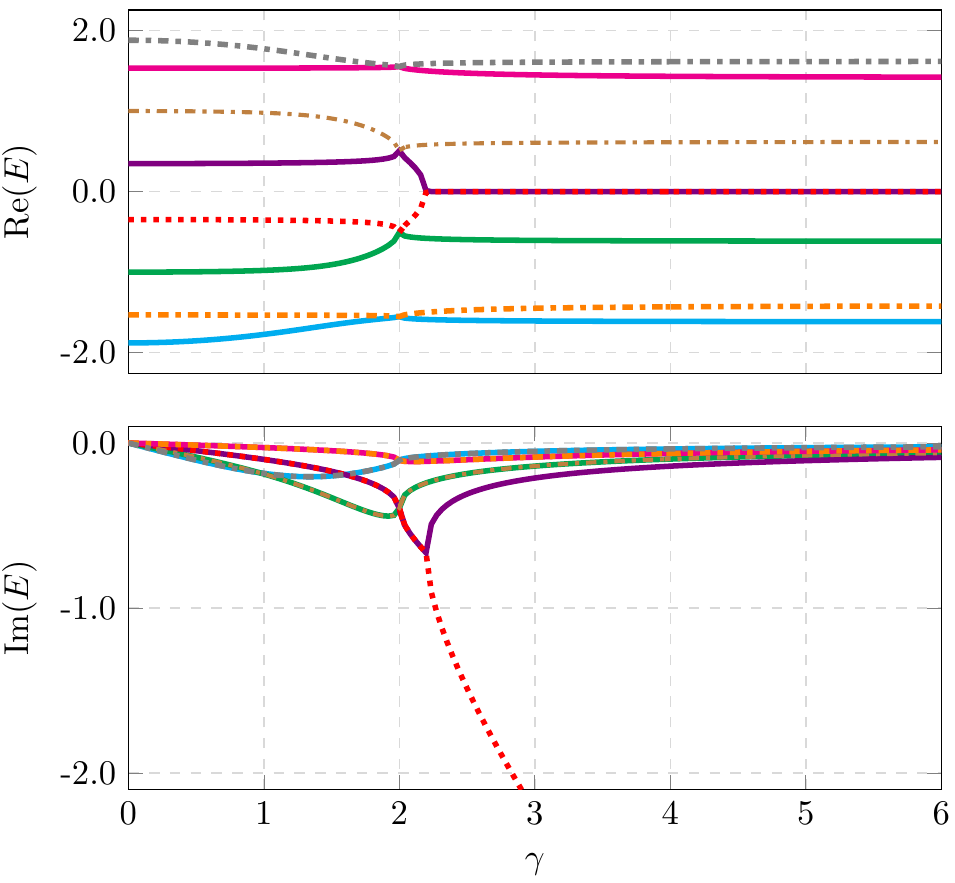}
		\caption{Energy spectrum of a system with $L = 8$.  For values of $L$ in the sequence
			$L=4n$, there is an extra exceptional point slightly above $\gamma=2$.  The localized
			eigenstate appears beyond this new exceptional point.} 
		\label{fig:Spectrum8}
	\end{figure}

	For even $L$ values satisfying $L=4n$, the situation is very similar, with one additional structure.
	As proved in Appendix \ref{app_degen} for even $L$, at exactly $\gamma=2$, all eigenvalues pair up;
	this is true for both $L=4n+2$ and $L=4n$.  In addition, for $L=4n$, at a value slightly above
	$\gamma=2$, the two eigenvalues with real values nearest to zero coalesce in an additional
	exceptional point, as seen in Fig.~\ref{fig:Spectrum8} for $L=8$.  It is at this point,
	$\gamma=\gamma_1>2$, that the localized state appears and the imaginary part of the corresponding
	eigenvalue separates off and starts to increase unboundedly in the negative direction.  With
	increasing $L$ in the sequence $L=4n$, the location of the new exceptional point, $\gamma_1$,
	approaches $2$.
	
	\begin{figure}
		\includegraphics*[width=\linewidth]{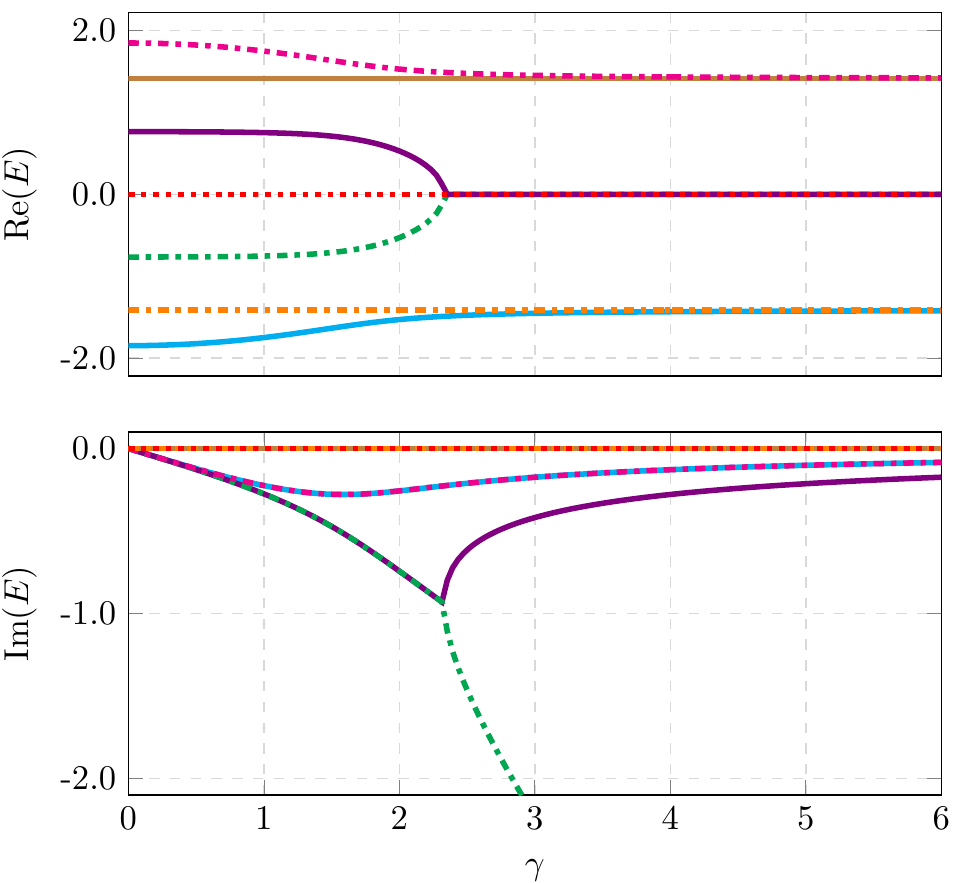}
		\caption{Energy spectrum for $L = 7$.  The impurity is on the central site, $q=4$.}
		\label{fig:Spectrum7}
	\end{figure}
	
	We now turn to odd $L$, with the impurity placed on the central site, $q=(L+1)/2$.  For $L=4n+3$,
	there is only a single exceptional point.  This appears to be a third-order exceptional point, and
	appears at a value $\gamma>2$.  An example is shown in Fig.~\ref{fig:Spectrum7}, for $L=7$.  As
	the system size tends to infinity, the location of the point tends to $\gamma\to2$.  There is always
	a single eigenvalue that has a zero real component --- the two other eigenvalues with real parts
	closest to zero merge with this at the exceptional point.

	\begin{figure}
		\includegraphics*[width=\linewidth]{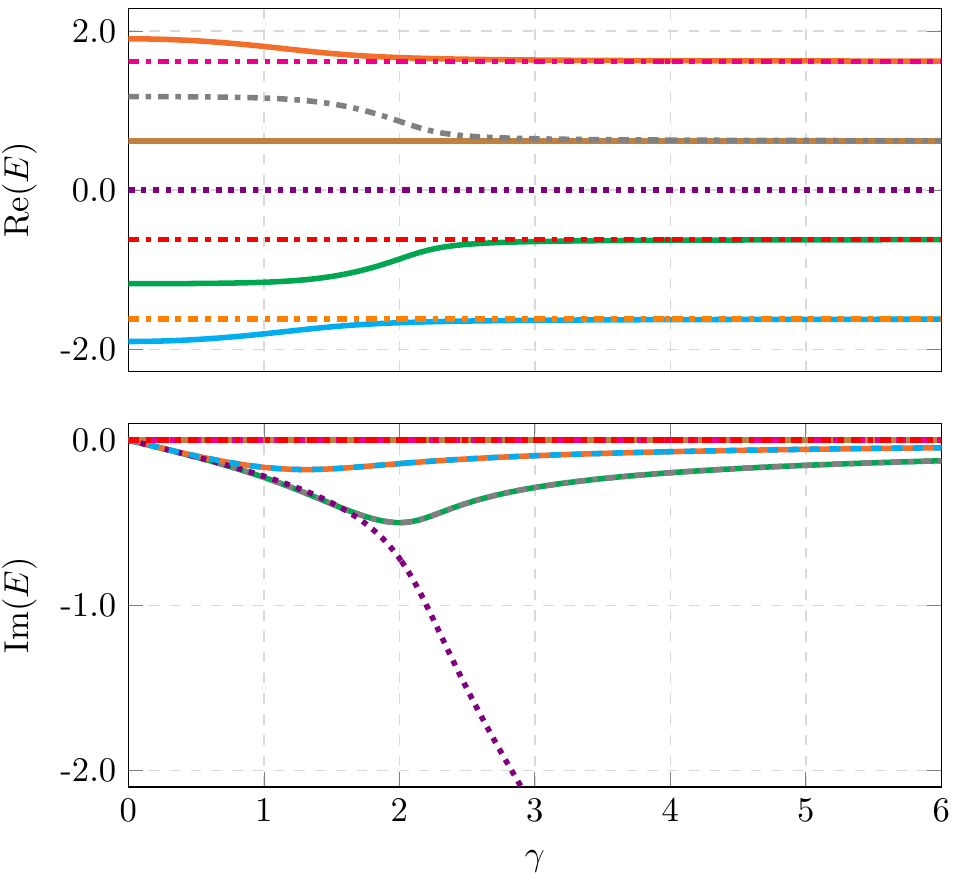}
		\caption{Energy spectrum for $L = 9$.  The impurity is on the central site, $q=5$.}
		\label{fig:Spectrum9}
	\end{figure}

	Finally, for $L=4n+1$, there appears to be no exceptional points; nevertheless, at large $\gamma$
	the eigenvalues pair up gradually.  An example is shown in Fig.~\ref{fig:Spectrum9} for $L=9$.  A
	single eigenvalue remains unpaired with zero real component.  Although this does not merge with any
	other eigenvalue, around $\gamma\approx2$ the imaginary component of this eigenvalue starts
	increasing unboundedly with $\gamma$, indicating that the corresponding eigenstate becomes
	localized.
	
	In summary, although there are differences in detail between the four cases, there is always a bound
	state at large $\gamma$, and around $\gamma=2$ there is always some reorganization of the spectrum.
	With increasing $L$, the location of these features converge toward $\gamma=2$.

	
	\section{Effect of impurity location}\label{app_location}
	
	In Appendix \ref{app_size}, we illustrated the dependence on the lattice size $L$, focusing on the
	case where the impurity is located at the center of the lattice, $q=L/2$ or $q=(L+1)/2$.  In this
	Appendix we briefly discuss the dependence of the location $q$ of the impurity, focusing on the
	case $L=4n+2$. 
	
	When the impurity is not on one of the central sites, the eigenvalues do not all coalesce as pairs
	at $\gamma=2$.  As the impurity is moved from the edge site towards the center ($q=1$, $q=2$,...)
	there is an exceptional point at a value of $\gamma$ which is less than $2$ for odd $q$ and larger
	than $2$ for even $q$.  At this exceptional point, the two eigenvalues with real parts closest to
	zero coalesce.  As in the case of a centrally located impurity, when $\gamma$ is raised further
	beyond this value, the real parts of these two eigenvalues are locked at zero, and the imaginary
	part of one of this pair starts to increase in magnitude.  This indicates an eigenstate localized at
	the impurity.  (E.g., for $q=1$ this state is localized at the edge of the lattice.)
	
	
	%


\begin{thebibliography}{98}%
		\makeatletter
		\providecommand \@ifxundefined [1]{%
			\@ifx{#1\undefined}
		}%
		\providecommand \@ifnum [1]{%
			\ifnum #1\expandafter \@firstoftwo
			\else \expandafter \@secondoftwo
			\fi
		}%
		\providecommand \@ifx [1]{%
			\ifx #1\expandafter \@firstoftwo
			\else \expandafter \@secondoftwo
			\fi
		}%
		\providecommand \natexlab [1]{#1}%
		\providecommand \enquote  [1]{``#1''}%
		\providecommand \bibnamefont  [1]{#1}%
		\providecommand \bibfnamefont [1]{#1}%
		\providecommand \citenamefont [1]{#1}%
		\providecommand \href@noop [0]{\@secondoftwo}%
		\providecommand \href [0]{\begingroup \@sanitize@url \@href}%
		\providecommand \@href[1]{\@@startlink{#1}\@@href}%
		\providecommand \@@href[1]{\endgroup#1\@@endlink}%
		\providecommand \@sanitize@url [0]{\catcode `\\12\catcode `\$12\catcode
			`\&12\catcode `\#12\catcode `\^12\catcode `\_12\catcode `\%12\relax}%
		\providecommand \@@startlink[1]{}%
		\providecommand \@@endlink[0]{}%
		\providecommand \url  [0]{\begingroup\@sanitize@url \@url }%
		\providecommand \@url [1]{\endgroup\@href {#1}{\urlprefix }}%
		\providecommand \urlprefix  [0]{URL }%
		\providecommand \Eprint [0]{\href }%
		\providecommand \doibase [0]{http://dx.doi.org/}%
		\providecommand \selectlanguage [0]{\@gobble}%
		\providecommand \bibinfo  [0]{\@secondoftwo}%
		\providecommand \bibfield  [0]{\@secondoftwo}%
		\providecommand \translation [1]{[#1]}%
		\providecommand \BibitemOpen [0]{}%
		\providecommand \bibitemStop [0]{}%
		\providecommand \bibitemNoStop [0]{.\EOS\space}%
		\providecommand \EOS [0]{\spacefactor3000\relax}%
		\providecommand \BibitemShut  [1]{\csname bibitem#1\endcsname}%
		\let\auto@bib@innerbib\@empty
		\bibitem [{\citenamefont {Rotter}(2009)}]{IRotter_JPA2009_opensystems}%
		\BibitemOpen
		\bibfield  {author} {\bibinfo {author} {\bibfnamefont {I.}~\bibnamefont
				{Rotter}},\ }\bibfield  {title} {\enquote {\bibinfo {title} {A
					non-{H}ermitian {H}amilton operator and the physics of open quantum
					systems},}\ }\href {\doibase 10.1088/1751-8113/42/15/153001} {\bibfield
			{journal} {\bibinfo  {journal} {Journal of Physics A: Mathematical and
					Theoretical}\ }\textbf {\bibinfo {volume} {42}},\ \bibinfo {pages} {153001}
			(\bibinfo {year} {2009})}\BibitemShut {NoStop}%
		\bibitem [{\citenamefont {Cao}\ and\ \citenamefont
			{Wiersig}(2015)}]{Cao_Wiersig_RMP2015}%
		\BibitemOpen
		\bibfield  {author} {\bibinfo {author} {\bibfnamefont {H.}~\bibnamefont
				{Cao}}\ and\ \bibinfo {author} {\bibfnamefont {J.}~\bibnamefont {Wiersig}},\
		}\bibfield  {title} {\enquote {\bibinfo {title} {Dielectric microcavities:
					Model systems for wave chaos and non-{H}ermitian physics},}\ }\href {\doibase
			10.1103/RevModPhys.87.61} {\bibfield  {journal} {\bibinfo  {journal} {Rev.
					Mod. Phys.}\ }\textbf {\bibinfo {volume} {87}},\ \bibinfo {pages} {61--111}
			(\bibinfo {year} {2015})}\BibitemShut {NoStop}%
		\bibitem [{\citenamefont {{\"O}zdemir}\ \emph {et~al.}(2019)\citenamefont
			{{\"O}zdemir}, \citenamefont {Rotter}, \citenamefont {Nori},\ and\
			\citenamefont {Yang}}]{Ozdemir_Rotter_Nori_NatMaterials2019_review}%
		\BibitemOpen
		\bibfield  {author} {\bibinfo {author} {\bibfnamefont {S.K.}\ \bibnamefont
				{{\"O}zdemir}}, \bibinfo {author} {\bibfnamefont {S.}~\bibnamefont {Rotter}},
			\bibinfo {author} {\bibfnamefont {F.}~\bibnamefont {Nori}}, \ and\ \bibinfo
			{author} {\bibfnamefont {L.}~\bibnamefont {Yang}},\ }\bibfield  {title}
		{\enquote {\bibinfo {title} {Parity-time symmetry and exceptional points in
					photonics},}\ }\href {\doibase 10.1038/s41563-019-0304-9} {\bibfield
			{journal} {\bibinfo  {journal} {Nature Materials}\ }\textbf {\bibinfo
				{volume} {18}},\ \bibinfo {pages} {783--798} (\bibinfo {year}
			{2019})}\BibitemShut {NoStop}%
		\bibitem [{\citenamefont {El-Ganainy}\ \emph {et~al.}(2019)\citenamefont
			{El-Ganainy}, \citenamefont {Khajavikhan}, \citenamefont {Christodoulides},\
			and\ \citenamefont
			{{\"O}zdemir}}]{El-Ganainy_etal_CommunicPhys2019_review_nonHermOptics}%
		\BibitemOpen
		\bibfield  {author} {\bibinfo {author} {\bibfnamefont {R.}~\bibnamefont
				{El-Ganainy}}, \bibinfo {author} {\bibfnamefont {M.}~\bibnamefont
				{Khajavikhan}}, \bibinfo {author} {\bibfnamefont {D.N.}\ \bibnamefont
				{Christodoulides}}, \ and\ \bibinfo {author} {\bibfnamefont {S.K.}\
				\bibnamefont {{\"O}zdemir}},\ }\bibfield  {title} {\enquote {\bibinfo {title}
				{The dawn of non-{H}ermitian optics},}\ }\href {\doibase
			10.1038/s42005-019-0130-z} {\bibfield  {journal} {\bibinfo  {journal}
				{Communications Physics}\ }\textbf {\bibinfo {volume} {2}},\ \bibinfo {pages}
			{37} (\bibinfo {year} {2019})}\BibitemShut {NoStop}%
		\bibitem [{\citenamefont {Liertzer}\ \emph {et~al.}(2012)\citenamefont
			{Liertzer}, \citenamefont {Ge}, \citenamefont {Cerjan}, \citenamefont
			{Stone}, \citenamefont {T\"ureci},\ and\ \citenamefont
			{Rotter}}]{Tureci_Rotter_PRL2012_lasers}%
		\BibitemOpen
		\bibfield  {author} {\bibinfo {author} {\bibfnamefont {M.}~\bibnamefont
				{Liertzer}}, \bibinfo {author} {\bibfnamefont {Li}~\bibnamefont {Ge}},
			\bibinfo {author} {\bibfnamefont {A.}~\bibnamefont {Cerjan}}, \bibinfo
			{author} {\bibfnamefont {A.~D.}\ \bibnamefont {Stone}}, \bibinfo {author}
			{\bibfnamefont {H.~E.}\ \bibnamefont {T\"ureci}}, \ and\ \bibinfo {author}
			{\bibfnamefont {S.}~\bibnamefont {Rotter}},\ }\bibfield  {title} {\enquote
			{\bibinfo {title} {Pump-induced exceptional points in lasers},}\ }\href
		{\doibase 10.1103/PhysRevLett.108.173901} {\bibfield  {journal} {\bibinfo
				{journal} {Phys. Rev. Lett.}\ }\textbf {\bibinfo {volume} {108}},\ \bibinfo
			{pages} {173901} (\bibinfo {year} {2012})}\BibitemShut {NoStop}%
		\bibitem [{\citenamefont {Brandstetter}\ \emph {et~al.}(2014)\citenamefont
			{Brandstetter}, \citenamefont {Liertzer}, \citenamefont {Deutsch},
			\citenamefont {Klang}, \citenamefont {Sch{\"o}berl}, \citenamefont
			{T{\"u}reci}, \citenamefont {Strasser}, \citenamefont {Unterrainer},\ and\
			\citenamefont {Rotter}}]{Brandstetter_NatCom2014_laser}%
		\BibitemOpen
		\bibfield  {author} {\bibinfo {author} {\bibfnamefont {M.}~\bibnamefont
				{Brandstetter}}, \bibinfo {author} {\bibfnamefont {M.}~\bibnamefont
				{Liertzer}}, \bibinfo {author} {\bibfnamefont {C.}~\bibnamefont {Deutsch}},
			\bibinfo {author} {\bibfnamefont {P.}~\bibnamefont {Klang}}, \bibinfo
			{author} {\bibfnamefont {J.}~\bibnamefont {Sch{\"o}berl}}, \bibinfo {author}
			{\bibfnamefont {H.E.}\ \bibnamefont {T{\"u}reci}}, \bibinfo {author}
			{\bibfnamefont {G.}~\bibnamefont {Strasser}}, \bibinfo {author}
			{\bibfnamefont {K.}~\bibnamefont {Unterrainer}}, \ and\ \bibinfo {author}
			{\bibfnamefont {S.}~\bibnamefont {Rotter}},\ }\bibfield  {title} {\enquote
			{\bibinfo {title} {Reversing the pump dependence of a laser at an exceptional
					point},}\ }\href {\doibase 10.1038/ncomms5034} {\bibfield  {journal}
			{\bibinfo  {journal} {Nature Comm.}\ }\textbf {\bibinfo {volume} {5}},\
			\bibinfo {pages} {4034} (\bibinfo {year} {2014})}\BibitemShut {NoStop}%
		\bibitem [{\citenamefont {Peng}\ \emph {et~al.}(2016)\citenamefont {Peng},
			\citenamefont {{\"O}zdemir}, \citenamefont {Liertzer}, \citenamefont {Chen},
			\citenamefont {Kramer}, \citenamefont {Y{\i}lmaz}, \citenamefont {Wiersig},
			\citenamefont {Rotter},\ and\ \citenamefont
			{Yang}}]{Wiersig_Rotter_PNAS2016_chiralmodes}%
		\BibitemOpen
		\bibfield  {author} {\bibinfo {author} {\bibfnamefont {B.}~\bibnamefont
				{Peng}}, \bibinfo {author} {\bibfnamefont {{\c S}.K.}\ \bibnamefont
				{{\"O}zdemir}}, \bibinfo {author} {\bibfnamefont {M.}~\bibnamefont
				{Liertzer}}, \bibinfo {author} {\bibfnamefont {W.}~\bibnamefont {Chen}},
			\bibinfo {author} {\bibfnamefont {J.}~\bibnamefont {Kramer}}, \bibinfo
			{author} {\bibfnamefont {H.}~\bibnamefont {Y{\i}lmaz}}, \bibinfo {author}
			{\bibfnamefont {J.}~\bibnamefont {Wiersig}}, \bibinfo {author} {\bibfnamefont
				{S.}~\bibnamefont {Rotter}}, \ and\ \bibinfo {author} {\bibfnamefont
				{L.}~\bibnamefont {Yang}},\ }\bibfield  {title} {\enquote {\bibinfo {title}
				{Chiral modes and directional lasing at exceptional points},}\ }\href
		{\doibase 10.1073/pnas.1603318113} {\bibfield  {journal} {\bibinfo  {journal}
				{Proceedings of the National Academy of Sciences}\ }\textbf {\bibinfo
				{volume} {113}},\ \bibinfo {pages} {6845--6850} (\bibinfo {year}
			{2016})}\BibitemShut {NoStop}%
		\bibitem [{\citenamefont {Miao}\ \emph {et~al.}(2016)\citenamefont {Miao},
			\citenamefont {Zhang}, \citenamefont {Sun}, \citenamefont {Walasik},
			\citenamefont {Longhi}, \citenamefont {Litchinitser},\ and\ \citenamefont
			{Feng}}]{Miao_Longhi_Liang_Science2016}%
		\BibitemOpen
		\bibfield  {author} {\bibinfo {author} {\bibfnamefont {P.}~\bibnamefont
				{Miao}}, \bibinfo {author} {\bibfnamefont {Z.}~\bibnamefont {Zhang}},
			\bibinfo {author} {\bibfnamefont {J.}~\bibnamefont {Sun}}, \bibinfo {author}
			{\bibfnamefont {W.}~\bibnamefont {Walasik}}, \bibinfo {author} {\bibfnamefont
				{S.}~\bibnamefont {Longhi}}, \bibinfo {author} {\bibfnamefont {N.M.}\
				\bibnamefont {Litchinitser}}, \ and\ \bibinfo {author} {\bibfnamefont
				{L.}~\bibnamefont {Feng}},\ }\bibfield  {title} {\enquote {\bibinfo {title}
				{Orbital angular momentum microlaser},}\ }\href {\doibase
			10.1126/science.aaf8533} {\bibfield  {journal} {\bibinfo  {journal}
				{Science}\ }\textbf {\bibinfo {volume} {353}},\ \bibinfo {pages} {464--467}
			(\bibinfo {year} {2016})}\BibitemShut {NoStop}%
		\bibitem [{\citenamefont {Guo}\ \emph {et~al.}(2009)\citenamefont {Guo},
			\citenamefont {Salamo}, \citenamefont {Duchesne}, \citenamefont {Morandotti},
			\citenamefont {Volatier-Ravat}, \citenamefont {Aimez}, \citenamefont
			{Siviloglou},\ and\ \citenamefont
			{Christodoulides}}]{Guo_etal_PRL2009_transparency}%
		\BibitemOpen
		\bibfield  {author} {\bibinfo {author} {\bibfnamefont {A.}~\bibnamefont
				{Guo}}, \bibinfo {author} {\bibfnamefont {G.~J.}\ \bibnamefont {Salamo}},
			\bibinfo {author} {\bibfnamefont {D.}~\bibnamefont {Duchesne}}, \bibinfo
			{author} {\bibfnamefont {R.}~\bibnamefont {Morandotti}}, \bibinfo {author}
			{\bibfnamefont {M.}~\bibnamefont {Volatier-Ravat}}, \bibinfo {author}
			{\bibfnamefont {V.}~\bibnamefont {Aimez}}, \bibinfo {author} {\bibfnamefont
				{G.~A.}\ \bibnamefont {Siviloglou}}, \ and\ \bibinfo {author} {\bibfnamefont
				{D.~N.}\ \bibnamefont {Christodoulides}},\ }\bibfield  {title} {\enquote
			{\bibinfo {title} {Observation of $\mathcal{P}\mathcal{T}$-symmetry breaking
					in complex optical potentials},}\ }\href {\doibase
			10.1103/PhysRevLett.103.093902} {\bibfield  {journal} {\bibinfo  {journal}
				{Phys. Rev. Lett.}\ }\textbf {\bibinfo {volume} {103}},\ \bibinfo {pages}
			{093902} (\bibinfo {year} {2009})}\BibitemShut {NoStop}%
		\bibitem [{\citenamefont {R\"uter}\ \emph {et~al.}(2010)\citenamefont
			{R\"uter}, \citenamefont {Makris}, \citenamefont {El-Ganainy}, \citenamefont
			{Christodoulides}, \citenamefont {Segev},\ and\ \citenamefont
			{Kip}}]{Segev_NatPhys2010_PTsym}%
		\BibitemOpen
		\bibfield  {author} {\bibinfo {author} {\bibfnamefont {C.~E.}\ \bibnamefont
				{R\"uter}}, \bibinfo {author} {\bibfnamefont {K.~G.}\ \bibnamefont {Makris}},
			\bibinfo {author} {\bibfnamefont {R.}~\bibnamefont {El-Ganainy}}, \bibinfo
			{author} {\bibfnamefont {D.~N.}\ \bibnamefont {Christodoulides}}, \bibinfo
			{author} {\bibfnamefont {M.}~\bibnamefont {Segev}}, \ and\ \bibinfo {author}
			{\bibfnamefont {D.}~\bibnamefont {Kip}},\ }\bibfield  {title} {\enquote
			{\bibinfo {title} {Observation of parity–time symmetry in optics},}\ }\href
		{\doibase 10.1038/nphys1515} {\bibfield  {journal} {\bibinfo  {journal}
				{Nature Physics}\ }\textbf {\bibinfo {volume} {6}} (\bibinfo {year} {2010}),\
			10.1038/nphys1515}\BibitemShut {NoStop}%
		\bibitem [{\citenamefont {Alfassi}\ \emph {et~al.}(2011)\citenamefont
			{Alfassi}, \citenamefont {Peleg}, \citenamefont {Moiseyev},\ and\
			\citenamefont {Segev}}]{Segev_PRL2011_photoniclattice}%
		\BibitemOpen
		\bibfield  {author} {\bibinfo {author} {\bibfnamefont {B.}~\bibnamefont
				{Alfassi}}, \bibinfo {author} {\bibfnamefont {O.}~\bibnamefont {Peleg}},
			\bibinfo {author} {\bibfnamefont {N.}~\bibnamefont {Moiseyev}}, \ and\
			\bibinfo {author} {\bibfnamefont {M.}~\bibnamefont {Segev}},\ }\bibfield
		{title} {\enquote {\bibinfo {title} {Diverging rabi oscillations in
					subwavelength photonic lattices},}\ }\href {\doibase
			10.1103/PhysRevLett.106.073901} {\bibfield  {journal} {\bibinfo  {journal}
				{Phys. Rev. Lett.}\ }\textbf {\bibinfo {volume} {106}},\ \bibinfo {pages}
			{073901} (\bibinfo {year} {2011})}\BibitemShut {NoStop}%
		\bibitem [{\citenamefont {Zeuner}\ \emph {et~al.}(2015)\citenamefont {Zeuner},
			\citenamefont {Rechtsman}, \citenamefont {Plotnik}, \citenamefont {Lumer},
			\citenamefont {Nolte}, \citenamefont {Rudner}, \citenamefont {Segev},\ and\
			\citenamefont {Szameit}}]{Szameit_Rudner_PRL2015_topological_transition}%
		\BibitemOpen
		\bibfield  {author} {\bibinfo {author} {\bibfnamefont {J.M.}\ \bibnamefont
				{Zeuner}}, \bibinfo {author} {\bibfnamefont {M.C.}\ \bibnamefont
				{Rechtsman}}, \bibinfo {author} {\bibfnamefont {Y.}~\bibnamefont {Plotnik}},
			\bibinfo {author} {\bibfnamefont {Y.}~\bibnamefont {Lumer}}, \bibinfo
			{author} {\bibfnamefont {S.}~\bibnamefont {Nolte}}, \bibinfo {author}
			{\bibfnamefont {M.S.}\ \bibnamefont {Rudner}}, \bibinfo {author}
			{\bibfnamefont {M.}~\bibnamefont {Segev}}, \ and\ \bibinfo {author}
			{\bibfnamefont {A.}~\bibnamefont {Szameit}},\ }\bibfield  {title} {\enquote
			{\bibinfo {title} {Observation of a topological transition in the bulk of a
					non-{H}ermitian system},}\ }\href {\doibase 10.1103/PhysRevLett.115.040402}
		{\bibfield  {journal} {\bibinfo  {journal} {Phys. Rev. Lett.}\ }\textbf
			{\bibinfo {volume} {115}},\ \bibinfo {pages} {040402} (\bibinfo {year}
			{2015})}\BibitemShut {NoStop}%
		\bibitem [{\citenamefont {Cerjan}\ \emph {et~al.}(2019)\citenamefont {Cerjan},
			\citenamefont {Huang}, \citenamefont {Wang}, \citenamefont {Chen},
			\citenamefont {Chong},\ and\ \citenamefont
			{Rechtsman}}]{Cerjan_etal_NatPhot2019_Weyl}%
		\BibitemOpen
		\bibfield  {author} {\bibinfo {author} {\bibfnamefont {A.}~\bibnamefont
				{Cerjan}}, \bibinfo {author} {\bibfnamefont {S.}~\bibnamefont {Huang}},
			\bibinfo {author} {\bibfnamefont {M.}~\bibnamefont {Wang}}, \bibinfo {author}
			{\bibfnamefont {Kevin~P.}\ \bibnamefont {Chen}}, \bibinfo {author}
			{\bibfnamefont {Y.}~\bibnamefont {Chong}}, \ and\ \bibinfo {author}
			{\bibfnamefont {M.C.}\ \bibnamefont {Rechtsman}},\ }\bibfield  {title}
		{\enquote {\bibinfo {title} {Experimental realization of a weyl exceptional
					ring},}\ }\href {\doibase 10.1038/s41566-019-0453-z} {\bibfield  {journal}
			{\bibinfo  {journal} {Nature Photonics}\ }\textbf {\bibinfo {volume} {13}},\
			\bibinfo {pages} {623–628} (\bibinfo {year} {2019})}\BibitemShut {NoStop}%
		\bibitem [{\citenamefont {Persson}\ \emph {et~al.}(2000)\citenamefont
			{Persson}, \citenamefont {Rotter}, \citenamefont {St\"ockmann},\ and\
			\citenamefont {Barth}}]{Persson_IRotter_Barth_PRL2000_microwave}%
		\BibitemOpen
		\bibfield  {author} {\bibinfo {author} {\bibfnamefont {E.}~\bibnamefont
				{Persson}}, \bibinfo {author} {\bibfnamefont {I.}~\bibnamefont {Rotter}},
			\bibinfo {author} {\bibfnamefont {H.-J.}\ \bibnamefont {St\"ockmann}}, \ and\
			\bibinfo {author} {\bibfnamefont {M.}~\bibnamefont {Barth}},\ }\bibfield
		{title} {\enquote {\bibinfo {title} {Observation of resonance trapping in an
					open microwave cavity},}\ }\href {\doibase 10.1103/PhysRevLett.85.2478}
		{\bibfield  {journal} {\bibinfo  {journal} {Phys. Rev. Lett.}\ }\textbf
			{\bibinfo {volume} {85}},\ \bibinfo {pages} {2478--2481} (\bibinfo {year}
			{2000})}\BibitemShut {NoStop}%
		\bibitem [{\citenamefont {Dembowski}\ \emph {et~al.}(2001)\citenamefont
			{Dembowski}, \citenamefont {Gr\"af}, \citenamefont {Harney}, \citenamefont
			{Heine}, \citenamefont {Heiss}, \citenamefont {Rehfeld},\ and\ \citenamefont
			{Richter}}]{Dembowski_etal_PRL2001}%
		\BibitemOpen
		\bibfield  {author} {\bibinfo {author} {\bibfnamefont {C.}~\bibnamefont
				{Dembowski}}, \bibinfo {author} {\bibfnamefont {H.-D.}\ \bibnamefont
				{Gr\"af}}, \bibinfo {author} {\bibfnamefont {H.~L.}\ \bibnamefont {Harney}},
			\bibinfo {author} {\bibfnamefont {A.}~\bibnamefont {Heine}}, \bibinfo
			{author} {\bibfnamefont {W.~D.}\ \bibnamefont {Heiss}}, \bibinfo {author}
			{\bibfnamefont {H.}~\bibnamefont {Rehfeld}}, \ and\ \bibinfo {author}
			{\bibfnamefont {A.}~\bibnamefont {Richter}},\ }\bibfield  {title} {\enquote
			{\bibinfo {title} {Experimental observation of the topological structure of
					exceptional points},}\ }\href {\doibase 10.1103/PhysRevLett.86.787}
		{\bibfield  {journal} {\bibinfo  {journal} {Phys. Rev. Lett.}\ }\textbf
			{\bibinfo {volume} {86}},\ \bibinfo {pages} {787--790} (\bibinfo {year}
			{2001})}\BibitemShut {NoStop}%
		\bibitem [{\citenamefont {Dembowski}\ \emph {et~al.}(2003)\citenamefont
			{Dembowski}, \citenamefont {Dietz}, \citenamefont {Gr\"af}, \citenamefont
			{Harney}, \citenamefont {Heine}, \citenamefont {Heiss},\ and\ \citenamefont
			{Richter}}]{Dembowski_etal_PRL2003}%
		\BibitemOpen
		\bibfield  {author} {\bibinfo {author} {\bibfnamefont {C.}~\bibnamefont
				{Dembowski}}, \bibinfo {author} {\bibfnamefont {B.}~\bibnamefont {Dietz}},
			\bibinfo {author} {\bibfnamefont {H.-D.}\ \bibnamefont {Gr\"af}}, \bibinfo
			{author} {\bibfnamefont {H.~L.}\ \bibnamefont {Harney}}, \bibinfo {author}
			{\bibfnamefont {A.}~\bibnamefont {Heine}}, \bibinfo {author} {\bibfnamefont
				{W.~D.}\ \bibnamefont {Heiss}}, \ and\ \bibinfo {author} {\bibfnamefont
				{A.}~\bibnamefont {Richter}},\ }\bibfield  {title} {\enquote {\bibinfo
				{title} {Observation of a chiral state in a microwave cavity},}\ }\href
		{\doibase 10.1103/PhysRevLett.90.034101} {\bibfield  {journal} {\bibinfo
				{journal} {Phys. Rev. Lett.}\ }\textbf {\bibinfo {volume} {90}},\ \bibinfo
			{pages} {034101} (\bibinfo {year} {2003})}\BibitemShut {NoStop}%
		\bibitem [{\citenamefont {Doppler}\ \emph {et~al.}(2016)\citenamefont
			{Doppler}, \citenamefont {Mailybaev}, \citenamefont {B{\"o}hm}, \citenamefont
			{Kuhl}, \citenamefont {Girschik}, \citenamefont {Libisch}, \citenamefont
			{Milburn}, \citenamefont {Rabl}, \citenamefont {Moiseyev},\ and\
			\citenamefont {Rotter}}]{Doppler_Nature2016_encircling}%
		\BibitemOpen
		\bibfield  {author} {\bibinfo {author} {\bibfnamefont {J.}~\bibnamefont
				{Doppler}}, \bibinfo {author} {\bibfnamefont {A.A.}\ \bibnamefont
				{Mailybaev}}, \bibinfo {author} {\bibfnamefont {J.}~\bibnamefont {B{\"o}hm}},
			\bibinfo {author} {\bibfnamefont {U.}~\bibnamefont {Kuhl}}, \bibinfo {author}
			{\bibfnamefont {A.}~\bibnamefont {Girschik}}, \bibinfo {author}
			{\bibfnamefont {F.}~\bibnamefont {Libisch}}, \bibinfo {author} {\bibfnamefont
				{T.J.}\ \bibnamefont {Milburn}}, \bibinfo {author} {\bibfnamefont
				{P.}~\bibnamefont {Rabl}}, \bibinfo {author} {\bibfnamefont {N.}~\bibnamefont
				{Moiseyev}}, \ and\ \bibinfo {author} {\bibfnamefont {S.}~\bibnamefont
				{Rotter}},\ }\bibfield  {title} {\enquote {\bibinfo {title} {Dynamically
					encircling an exceptional point for asymmetric mode switching},}\ }\href
		{\doibase 10.1038/nature18605} {\bibfield  {journal} {\bibinfo  {journal}
				{Nature}\ }\textbf {\bibinfo {volume} {537}},\ \bibinfo {pages} {76}
			(\bibinfo {year} {2016})}\BibitemShut {NoStop}%
		\bibitem [{\citenamefont {Poli}\ \emph {et~al.}(2015)\citenamefont {Poli},
			\citenamefont {Bellec}, \citenamefont {Kuhl}, \citenamefont {Mortessagne},\
			and\ \citenamefont {Schomerus}}]{Schomerus_Nature2015_resonatorchain}%
		\BibitemOpen
		\bibfield  {author} {\bibinfo {author} {\bibfnamefont {C.}~\bibnamefont
				{Poli}}, \bibinfo {author} {\bibfnamefont {M.}~\bibnamefont {Bellec}},
			\bibinfo {author} {\bibfnamefont {U.}~\bibnamefont {Kuhl}}, \bibinfo {author}
			{\bibfnamefont {F.}~\bibnamefont {Mortessagne}}, \ and\ \bibinfo {author}
			{\bibfnamefont {H.}~\bibnamefont {Schomerus}},\ }\bibfield  {title} {\enquote
			{\bibinfo {title} {Selective enhancement of topologically induced interface
					states in a dielectric resonator chain},}\ }\href {\doibase
			https://doi.org/10.1038/ncomms7710} {\bibfield  {journal} {\bibinfo
				{journal} {Nature Comm.}\ }\textbf {\bibinfo {volume} {6}},\ \bibinfo {pages}
			{6710} (\bibinfo {year} {2015})}\BibitemShut {NoStop}%
		\bibitem [{\citenamefont {Chen}\ \emph {et~al.}(2017)\citenamefont {Chen},
			\citenamefont {{\"O}zdemir}, \citenamefont {Zhao}, \citenamefont {Wiersig},\
			and\ \citenamefont {Yang}}]{Chen_etal_Nature2017_microcavity}%
		\BibitemOpen
		\bibfield  {author} {\bibinfo {author} {\bibfnamefont {W.}~\bibnamefont
				{Chen}}, \bibinfo {author} {\bibfnamefont {S.K.}\ \bibnamefont
				{{\"O}zdemir}}, \bibinfo {author} {\bibfnamefont {G.}~\bibnamefont {Zhao}},
			\bibinfo {author} {\bibfnamefont {J.}~\bibnamefont {Wiersig}}, \ and\
			\bibinfo {author} {\bibfnamefont {L.}~\bibnamefont {Yang}},\ }\bibfield
		{title} {\enquote {\bibinfo {title} {Exceptional points enhance sensing in an
					optical microcavity},}\ }\href {https://doi.org/10.1038/nature23281}
		{\bibfield  {journal} {\bibinfo  {journal} {Nature}\ }\textbf {\bibinfo
				{volume} {548}},\ \bibinfo {pages} {192} (\bibinfo {year}
			{2017})}\BibitemShut {NoStop}%
		\bibitem [{\citenamefont {Yi}\ \emph {et~al.}(2018)\citenamefont {Yi},
			\citenamefont {Kullig},\ and\ \citenamefont
			{Wiersig}}]{Yi_Kullig_Wiersig_PRL2018_EPs_MicrodiskCavity}%
		\BibitemOpen
		\bibfield  {author} {\bibinfo {author} {\bibfnamefont {Chang-Hwan}\
				\bibnamefont {Yi}}, \bibinfo {author} {\bibfnamefont {Julius}\ \bibnamefont
				{Kullig}}, \ and\ \bibinfo {author} {\bibfnamefont {Jan}\ \bibnamefont
				{Wiersig}},\ }\bibfield  {title} {\enquote {\bibinfo {title} {Pair of
					exceptional points in a microdisk cavity under an extremely weak
					deformation},}\ }\href {\doibase 10.1103/PhysRevLett.120.093902} {\bibfield
			{journal} {\bibinfo  {journal} {Phys. Rev. Lett.}\ }\textbf {\bibinfo
				{volume} {120}},\ \bibinfo {pages} {093902} (\bibinfo {year}
			{2018})}\BibitemShut {NoStop}%
		\bibitem [{\citenamefont {Xu}\ \emph {et~al.}(2016)\citenamefont {Xu},
			\citenamefont {Mason}, \citenamefont {Jiang},\ and\ \citenamefont
			{Harris}}]{Xu_etal_Nature2016_optomechanical}%
		\BibitemOpen
		\bibfield  {author} {\bibinfo {author} {\bibfnamefont {H.}~\bibnamefont
				{Xu}}, \bibinfo {author} {\bibfnamefont {D.}~\bibnamefont {Mason}}, \bibinfo
			{author} {\bibfnamefont {L.}~\bibnamefont {Jiang}}, \ and\ \bibinfo {author}
			{\bibfnamefont {J.G.E.}\ \bibnamefont {Harris}},\ }\bibfield  {title}
		{\enquote {\bibinfo {title} {Topological energy transfer in an optomechanical
					system with exceptional points},}\ }\href
		{https://doi.org/10.1038/nature18604} {\bibfield  {journal} {\bibinfo
				{journal} {Nature}\ }\textbf {\bibinfo {volume} {537}},\ \bibinfo {pages}
			{80} (\bibinfo {year} {2016})}\BibitemShut {NoStop}%
		\bibitem [{\citenamefont {Zhen}\ \emph {et~al.}(2015)\citenamefont {Zhen},
			\citenamefont {Hsu}, \citenamefont {Igarashi}, \citenamefont {Lu},
			\citenamefont {Kaminer}, \citenamefont {Pick}, \citenamefont {Chua},
			\citenamefont {Joannopoulos},\ and\ \citenamefont
			{Soljacic}}]{Zhen_etal_Nature2015_EPring_from_DiracCone}%
		\BibitemOpen
		\bibfield  {author} {\bibinfo {author} {\bibfnamefont {B.}~\bibnamefont
				{Zhen}}, \bibinfo {author} {\bibfnamefont {C.~W.}\ \bibnamefont {Hsu}},
			\bibinfo {author} {\bibfnamefont {Y.}~\bibnamefont {Igarashi}}, \bibinfo
			{author} {\bibfnamefont {L.}~\bibnamefont {Lu}}, \bibinfo {author}
			{\bibfnamefont {I.}~\bibnamefont {Kaminer}}, \bibinfo {author} {\bibfnamefont
				{A.}~\bibnamefont {Pick}}, \bibinfo {author} {\bibfnamefont {S.-L.}\
				\bibnamefont {Chua}}, \bibinfo {author} {\bibfnamefont {J.D.}\ \bibnamefont
				{Joannopoulos}}, \ and\ \bibinfo {author} {\bibfnamefont {M.}~\bibnamefont
				{Soljacic}},\ }\bibfield  {title} {\enquote {\bibinfo {title} {Spawning rings
					of exceptional points out of dirac cones},}\ }\href
		{https://doi.org/10.1038/nature14889} {\bibfield  {journal} {\bibinfo
				{journal} {Nature}\ }\textbf {\bibinfo {volume} {525}},\ \bibinfo {pages}
			{354--358} (\bibinfo {year} {2015})}\BibitemShut {NoStop}%
		\bibitem [{\citenamefont {Zhou}\ \emph {et~al.}(2018)\citenamefont {Zhou},
			\citenamefont {Peng}, \citenamefont {Yoon}, \citenamefont {Hsu},
			\citenamefont {Nelson}, \citenamefont {Fu}, \citenamefont {Joannopoulos},
			\citenamefont {Solja{\v c}i{\'c}},\ and\ \citenamefont
			{Zhen}}]{Zhou_etal_Science2018_FermiArc}%
		\BibitemOpen
		\bibfield  {author} {\bibinfo {author} {\bibfnamefont {H.}~\bibnamefont
				{Zhou}}, \bibinfo {author} {\bibfnamefont {C.}~\bibnamefont {Peng}}, \bibinfo
			{author} {\bibfnamefont {Y.}~\bibnamefont {Yoon}}, \bibinfo {author}
			{\bibfnamefont {C.W.}\ \bibnamefont {Hsu}}, \bibinfo {author} {\bibfnamefont
				{K.A.}\ \bibnamefont {Nelson}}, \bibinfo {author} {\bibfnamefont
				{L.}~\bibnamefont {Fu}}, \bibinfo {author} {\bibfnamefont {J.D.}\
				\bibnamefont {Joannopoulos}}, \bibinfo {author} {\bibfnamefont
				{M.}~\bibnamefont {Solja{\v c}i{\'c}}}, \ and\ \bibinfo {author}
			{\bibfnamefont {B.}~\bibnamefont {Zhen}},\ }\bibfield  {title} {\enquote
			{\bibinfo {title} {Observation of bulk fermi arc and polarization half charge
					from paired exceptional points},}\ }\href {\doibase 10.1126/science.aap9859}
		{\bibfield  {journal} {\bibinfo  {journal} {Science}\ }\textbf {\bibinfo
				{volume} {359}},\ \bibinfo {pages} {1009--1012} (\bibinfo {year}
			{2018})}\BibitemShut {NoStop}%
		\bibitem [{\citenamefont {Zhu}\ \emph {et~al.}(2015)\citenamefont {Zhu},
			\citenamefont {Ramezani}, \citenamefont {Shi}, \citenamefont {Zhu},\ and\
			\citenamefont {Zhang}}]{Zhu_etal_JournAcousticalSoc2015}%
		\BibitemOpen
		\bibfield  {author} {\bibinfo {author} {\bibfnamefont {X.}~\bibnamefont
				{Zhu}}, \bibinfo {author} {\bibfnamefont {H.}~\bibnamefont {Ramezani}},
			\bibinfo {author} {\bibfnamefont {C.}~\bibnamefont {Shi}}, \bibinfo {author}
			{\bibfnamefont {J.}~\bibnamefont {Zhu}}, \ and\ \bibinfo {author}
			{\bibfnamefont {X.}~\bibnamefont {Zhang}},\ }\bibfield  {title} {\enquote
			{\bibinfo {title} {$\mathcal{PT}$-symmetric acoustics},}\ }\href {\doibase
			https://doi.org/10.1121/1.4920752} {\bibfield  {journal} {\bibinfo  {journal}
				{The Journal of the Acoustical Society of America}\ }\textbf {\bibinfo
				{volume} {137}},\ \bibinfo {pages} {2403--2403} (\bibinfo {year}
			{2015})}\BibitemShut {NoStop}%
		\bibitem [{\citenamefont {Fleury}\ \emph {et~al.}(2015)\citenamefont {Fleury},
			\citenamefont {Sounas},\ and\ \citenamefont
			{Alu}}]{Fleury_etal_NatComm2015_accoustic_invisible}%
		\BibitemOpen
		\bibfield  {author} {\bibinfo {author} {\bibfnamefont {R.}~\bibnamefont
				{Fleury}}, \bibinfo {author} {\bibfnamefont {D.}~\bibnamefont {Sounas}}, \
			and\ \bibinfo {author} {\bibfnamefont {A.}~\bibnamefont {Alu}},\ }\bibfield
		{title} {\enquote {\bibinfo {title} {An invisible acoustic sensor based on
					parity-time symmetry},}\ }\href {https://doi.org/10.1038/ncomms6905}
		{\bibfield  {journal} {\bibinfo  {journal} {Nature communications}\ }\textbf
			{\bibinfo {volume} {6}},\ \bibinfo {pages} {5905} (\bibinfo {year}
			{2015})}\BibitemShut {NoStop}%
		\bibitem [{\citenamefont {Shi}\ \emph {et~al.}(2016)\citenamefont {Shi},
			\citenamefont {Dubois}, \citenamefont {Chen}, \citenamefont {Cheng},
			\citenamefont {Ramezani}, \citenamefont {Wang},\ and\ \citenamefont
			{Zhang}}]{Shi_etal_NatComm2016_acoustics_unidirectional}%
		\BibitemOpen
		\bibfield  {author} {\bibinfo {author} {\bibfnamefont {C.}~\bibnamefont
				{Shi}}, \bibinfo {author} {\bibfnamefont {M.}~\bibnamefont {Dubois}},
			\bibinfo {author} {\bibfnamefont {Y.}~\bibnamefont {Chen}}, \bibinfo {author}
			{\bibfnamefont {L.}~\bibnamefont {Cheng}}, \bibinfo {author} {\bibfnamefont
				{H.}~\bibnamefont {Ramezani}}, \bibinfo {author} {\bibfnamefont
				{Y.}~\bibnamefont {Wang}}, \ and\ \bibinfo {author} {\bibfnamefont
				{X.}~\bibnamefont {Zhang}},\ }\bibfield  {title} {\enquote {\bibinfo {title}
				{Accessing the exceptional points of parity-time symmetric acoustics},}\
		}\href {\doibase https://www.nature.com/articles/ncomms11110} {\bibfield
			{journal} {\bibinfo  {journal} {Nature communications}\ }\textbf {\bibinfo
				{volume} {7}},\ \bibinfo {pages} {11110} (\bibinfo {year}
			{2016})}\BibitemShut {NoStop}%
		\bibitem [{\citenamefont {Ding}\ \emph {et~al.}(2016)\citenamefont {Ding},
			\citenamefont {Ma}, \citenamefont {Xiao}, \citenamefont {Zhang},\ and\
			\citenamefont {Chan}}]{Ding_PRX2016_acoustics_multipleEPs}%
		\BibitemOpen
		\bibfield  {author} {\bibinfo {author} {\bibfnamefont {K.}~\bibnamefont
				{Ding}}, \bibinfo {author} {\bibfnamefont {G.}~\bibnamefont {Ma}}, \bibinfo
			{author} {\bibfnamefont {M.}~\bibnamefont {Xiao}}, \bibinfo {author}
			{\bibfnamefont {Z.Q.}\ \bibnamefont {Zhang}}, \ and\ \bibinfo {author}
			{\bibfnamefont {C.T.}\ \bibnamefont {Chan}},\ }\bibfield  {title} {\enquote
			{\bibinfo {title} {Emergence, coalescence, and topological properties of
					multiple exceptional points and their experimental realization},}\ }\href
		{\doibase 10.1103/PhysRevX.6.021007} {\bibfield  {journal} {\bibinfo
				{journal} {Phys. Rev. X}\ }\textbf {\bibinfo {volume} {6}},\ \bibinfo {pages}
			{021007} (\bibinfo {year} {2016})}\BibitemShut {NoStop}%
		\bibitem [{\citenamefont {Choi}\ \emph {et~al.}(2010)\citenamefont {Choi},
			\citenamefont {Kang}, \citenamefont {Lim}, \citenamefont {Kim}, \citenamefont
			{Kim}, \citenamefont {Lee},\ and\ \citenamefont
			{An}}]{Choi_etal_PRL2010_AtomCavity}%
		\BibitemOpen
		\bibfield  {author} {\bibinfo {author} {\bibfnamefont {Y.}~\bibnamefont
				{Choi}}, \bibinfo {author} {\bibfnamefont {S.}~\bibnamefont {Kang}}, \bibinfo
			{author} {\bibfnamefont {S.}~\bibnamefont {Lim}}, \bibinfo {author}
			{\bibfnamefont {W.}~\bibnamefont {Kim}}, \bibinfo {author} {\bibfnamefont
				{J.-R.}\ \bibnamefont {Kim}}, \bibinfo {author} {\bibfnamefont {J.-H.}\
				\bibnamefont {Lee}}, \ and\ \bibinfo {author} {\bibfnamefont
				{K.}~\bibnamefont {An}},\ }\bibfield  {title} {\enquote {\bibinfo {title}
				{Quasieigenstate coalescence in an atom-cavity quantum composite},}\ }\href
		{\doibase 10.1103/PhysRevLett.104.153601} {\bibfield  {journal} {\bibinfo
				{journal} {Phys. Rev. Lett.}\ }\textbf {\bibinfo {volume} {104}},\ \bibinfo
			{pages} {153601} (\bibinfo {year} {2010})}\BibitemShut {NoStop}%
		\bibitem [{\citenamefont {Gao}\ \emph {et~al.}(2015)\citenamefont {Gao},
			\citenamefont {Estrecho}, \citenamefont {Bliokh}, \citenamefont {Liew},
			\citenamefont {Fraser}, \citenamefont {Brodbeck}, \citenamefont {Kamp},
			\citenamefont {Schneider}, \citenamefont {H{\"o}fling}, \citenamefont
			{Yamamoto}, \citenamefont {Nori}, \citenamefont {Truscott}, \citenamefont
			{Dall},\ and\ \citenamefont
			{Ostrovskaya}}]{Gao_etal_Nature2015_excitonpolariton}%
		\BibitemOpen
		\bibfield  {author} {\bibinfo {author} {\bibfnamefont {T.}~\bibnamefont
				{Gao}}, \bibinfo {author} {\bibfnamefont {E.}~\bibnamefont {Estrecho}},
			\bibinfo {author} {\bibfnamefont {K.Y.}\ \bibnamefont {Bliokh}}, \bibinfo
			{author} {\bibfnamefont {T.C.H.}\ \bibnamefont {Liew}}, \bibinfo {author}
			{\bibfnamefont {M.D.}\ \bibnamefont {Fraser}}, \bibinfo {author}
			{\bibfnamefont {S.}~\bibnamefont {Brodbeck}}, \bibinfo {author}
			{\bibfnamefont {M.}~\bibnamefont {Kamp}}, \bibinfo {author} {\bibfnamefont
				{C.}~\bibnamefont {Schneider}}, \bibinfo {author} {\bibfnamefont
				{S.}~\bibnamefont {H{\"o}fling}}, \bibinfo {author} {\bibfnamefont
				{Y.}~\bibnamefont {Yamamoto}}, \bibinfo {author} {\bibfnamefont {Y.S.}\
				\bibnamefont {Nori}, \bibfnamefont {F.~Kivshar}}, \bibinfo {author}
			{\bibfnamefont {A.}~\bibnamefont {Truscott}}, \bibinfo {author}
			{\bibfnamefont {R.}~\bibnamefont {Dall}}, \ and\ \bibinfo {author}
			{\bibfnamefont {E.A.}\ \bibnamefont {Ostrovskaya}},\ }\bibfield  {title}
		{\enquote {\bibinfo {title} {Observation of non-{H}ermitian degeneracies in a
					chaotic exciton-polariton billiard},}\ }\href {\doibase 10.1038/nature15522}
		{\bibfield  {journal} {\bibinfo  {journal} {Nature}\ }\textbf {\bibinfo
				{volume} {526}},\ \bibinfo {pages} {554} (\bibinfo {year}
			{2015})}\BibitemShut {NoStop}%
		\bibitem [{\citenamefont {Gao}\ \emph {et~al.}(2018)\citenamefont {Gao},
			\citenamefont {Li}, \citenamefont {Estrecho}, \citenamefont {Liew},
			\citenamefont {Comber-Todd}, \citenamefont {Nalitov}, \citenamefont {Steger},
			\citenamefont {West}, \citenamefont {Pfeiffer}, \citenamefont {Snoke},
			\citenamefont {Kavokin}, \citenamefont {Truscott},\ and\ \citenamefont
			{Ostrovskaya}}]{Snoke_Truscott_Ostrovskaya_PRL2018_chiral}%
		\BibitemOpen
		\bibfield  {author} {\bibinfo {author} {\bibfnamefont {T.}~\bibnamefont
				{Gao}}, \bibinfo {author} {\bibfnamefont {G.}~\bibnamefont {Li}}, \bibinfo
			{author} {\bibfnamefont {E.}~\bibnamefont {Estrecho}}, \bibinfo {author}
			{\bibfnamefont {T.~C.~H.}\ \bibnamefont {Liew}}, \bibinfo {author}
			{\bibfnamefont {D.}~\bibnamefont {Comber-Todd}}, \bibinfo {author}
			{\bibfnamefont {A.}~\bibnamefont {Nalitov}}, \bibinfo {author} {\bibfnamefont
				{M.}~\bibnamefont {Steger}}, \bibinfo {author} {\bibfnamefont
				{K.}~\bibnamefont {West}}, \bibinfo {author} {\bibfnamefont {L.}~\bibnamefont
				{Pfeiffer}}, \bibinfo {author} {\bibfnamefont {D.~W.}\ \bibnamefont {Snoke}},
			\bibinfo {author} {\bibfnamefont {A.~V.}\ \bibnamefont {Kavokin}}, \bibinfo
			{author} {\bibfnamefont {A.~G.}\ \bibnamefont {Truscott}}, \ and\ \bibinfo
			{author} {\bibfnamefont {E.~A.}\ \bibnamefont {Ostrovskaya}},\ }\bibfield
		{title} {\enquote {\bibinfo {title} {Chiral modes at exceptional points in
					exciton-polariton quantum fluids},}\ }\href {\doibase
			10.1103/PhysRevLett.120.065301} {\bibfield  {journal} {\bibinfo  {journal}
				{Phys. Rev. Lett.}\ }\textbf {\bibinfo {volume} {120}},\ \bibinfo {pages}
			{065301} (\bibinfo {year} {2018})}\BibitemShut {NoStop}%
		\bibitem [{\citenamefont {Cartarius}\ \emph {et~al.}(2007)\citenamefont
			{Cartarius}, \citenamefont {Main},\ and\ \citenamefont
			{Wunner}}]{PRL2007_hydrogenatom}%
		\BibitemOpen
		\bibfield  {author} {\bibinfo {author} {\bibfnamefont {H.}~\bibnamefont
				{Cartarius}}, \bibinfo {author} {\bibfnamefont {J.}~\bibnamefont {Main}}, \
			and\ \bibinfo {author} {\bibfnamefont {G.}~\bibnamefont {Wunner}},\
		}\bibfield  {title} {\enquote {\bibinfo {title} {Exceptional points in atomic
					spectra},}\ }\href {\doibase 10.1103/PhysRevLett.99.173003} {\bibfield
			{journal} {\bibinfo  {journal} {Phys. Rev. Lett.}\ }\textbf {\bibinfo
				{volume} {99}},\ \bibinfo {pages} {173003} (\bibinfo {year}
			{2007})}\BibitemShut {NoStop}%
		\bibitem [{\citenamefont {Lin}\ \emph {et~al.}(2011)\citenamefont {Lin},
			\citenamefont {Ramezani}, \citenamefont {Eichelkraut}, \citenamefont
			{Kottos}, \citenamefont {Cao},\ and\ \citenamefont
			{Christodoulides}}]{Kottos_Christodoulides_PRL2011_unidirectional}%
		\BibitemOpen
		\bibfield  {author} {\bibinfo {author} {\bibfnamefont {Z.}~\bibnamefont
				{Lin}}, \bibinfo {author} {\bibfnamefont {H.}~\bibnamefont {Ramezani}},
			\bibinfo {author} {\bibfnamefont {T.}~\bibnamefont {Eichelkraut}}, \bibinfo
			{author} {\bibfnamefont {T.}~\bibnamefont {Kottos}}, \bibinfo {author}
			{\bibfnamefont {H.}~\bibnamefont {Cao}}, \ and\ \bibinfo {author}
			{\bibfnamefont {D.N.}\ \bibnamefont {Christodoulides}},\ }\bibfield  {title}
		{\enquote {\bibinfo {title} {Unidirectional invisibility induced by
					$\mathcal{P}\mathcal{T}$-symmetric periodic structures},}\ }\href {\doibase
			10.1103/PhysRevLett.106.213901} {\bibfield  {journal} {\bibinfo  {journal}
				{Phys. Rev. Lett.}\ }\textbf {\bibinfo {volume} {106}},\ \bibinfo {pages}
			{213901} (\bibinfo {year} {2011})}\BibitemShut {NoStop}%
		\bibitem [{\citenamefont {Regensburger}\ \emph {et~al.}(2012)\citenamefont
			{Regensburger}, \citenamefont {Bersch}, \citenamefont {Miri}, \citenamefont
			{Onishchukov}, \citenamefont {Christodoulides},\ and\ \citenamefont
			{Peschel}}]{Regensburger_etal_NatPhys2012}%
		\BibitemOpen
		\bibfield  {author} {\bibinfo {author} {\bibfnamefont {A.}~\bibnamefont
				{Regensburger}}, \bibinfo {author} {\bibfnamefont {C.}~\bibnamefont
				{Bersch}}, \bibinfo {author} {\bibfnamefont {M.-A.}\ \bibnamefont {Miri}},
			\bibinfo {author} {\bibfnamefont {G.}~\bibnamefont {Onishchukov}}, \bibinfo
			{author} {\bibfnamefont {D.~N.}\ \bibnamefont {Christodoulides}}, \ and\
			\bibinfo {author} {\bibfnamefont {U.}~\bibnamefont {Peschel}},\ }\bibfield
		{title} {\enquote {\bibinfo {title} {Parity-time synthetic photonic
					lattices},}\ }\href {\doibase 10.1038/nature11298} {\bibfield  {journal}
			{\bibinfo  {journal} {Nature}\ }\textbf {\bibinfo {volume} {488}},\ \bibinfo
			{pages} {167--171} (\bibinfo {year} {2012})}\BibitemShut {NoStop}%
		\bibitem [{\citenamefont {Barontini}\ \emph {et~al.}(2013)\citenamefont
			{Barontini}, \citenamefont {Labouvie}, \citenamefont {Stubenrauch},
			\citenamefont {Vogler}, \citenamefont {Guarrera},\ and\ \citenamefont
			{Ott}}]{Ott_PRL2013}%
		\BibitemOpen
		\bibfield  {author} {\bibinfo {author} {\bibfnamefont {G.}~\bibnamefont
				{Barontini}}, \bibinfo {author} {\bibfnamefont {R.}~\bibnamefont {Labouvie}},
			\bibinfo {author} {\bibfnamefont {F.}~\bibnamefont {Stubenrauch}}, \bibinfo
			{author} {\bibfnamefont {A.}~\bibnamefont {Vogler}}, \bibinfo {author}
			{\bibfnamefont {V.}~\bibnamefont {Guarrera}}, \ and\ \bibinfo {author}
			{\bibfnamefont {H.}~\bibnamefont {Ott}},\ }\bibfield  {title} {\enquote
			{\bibinfo {title} {Controlling the dynamics of an open many-body quantum
					system with localized dissipation},}\ }\href {\doibase
			10.1103/PhysRevLett.110.035302} {\bibfield  {journal} {\bibinfo  {journal}
				{Phys. Rev. Lett.}\ }\textbf {\bibinfo {volume} {110}},\ \bibinfo {pages}
			{035302} (\bibinfo {year} {2013})}\BibitemShut {NoStop}%
		\bibitem [{\citenamefont {Lu}\ \emph {et~al.}(2018)\citenamefont {Lu},
			\citenamefont {Peng}, \citenamefont {Cao}, \citenamefont {Xu}, \citenamefont
			{Wiersig}, \citenamefont {Gong},\ and\ \citenamefont
			{Xiao}}]{Lu_Wiersig_etal_ScienceBulletin2018_PT_cavityQED}%
		\BibitemOpen
		\bibfield  {author} {\bibinfo {author} {\bibfnamefont {Y.-K.}\ \bibnamefont
				{Lu}}, \bibinfo {author} {\bibfnamefont {P.}~\bibnamefont {Peng}}, \bibinfo
			{author} {\bibfnamefont {Q.-T.}\ \bibnamefont {Cao}}, \bibinfo {author}
			{\bibfnamefont {D.}~\bibnamefont {Xu}}, \bibinfo {author} {\bibfnamefont
				{J.}~\bibnamefont {Wiersig}}, \bibinfo {author} {\bibfnamefont
				{Q.}~\bibnamefont {Gong}}, \ and\ \bibinfo {author} {\bibfnamefont {Y.-F.}\
				\bibnamefont {Xiao}},\ }\bibfield  {title} {\enquote {\bibinfo {title}
				{Spontaneous t-symmetry breaking and exceptional points in cavity quantum
					electrodynamics systems},}\ }\href {\doibase 10.1016/j.scib.2018.07.020}
		{\bibfield  {journal} {\bibinfo  {journal} {Science Bulletin}\ }\textbf
			{\bibinfo {volume} {63}},\ \bibinfo {pages} {1096--1100} (\bibinfo {year}
			{2018})}\BibitemShut {NoStop}%
		\bibitem [{\citenamefont {Li}\ \emph {et~al.}(2019)\citenamefont {Li},
			\citenamefont {Harter}, \citenamefont {Liu}, \citenamefont {de~Melo},
			\citenamefont {Joglekar},\ and\ \citenamefont
			{Luo}}]{Li_Harter_Joglekar_NatComm2019_Floquet_coldatom}%
		\BibitemOpen
		\bibfield  {author} {\bibinfo {author} {\bibfnamefont {J.}~\bibnamefont
				{Li}}, \bibinfo {author} {\bibfnamefont {A.K.}\ \bibnamefont {Harter}},
			\bibinfo {author} {\bibfnamefont {J.}~\bibnamefont {Liu}}, \bibinfo {author}
			{\bibfnamefont {L.}~\bibnamefont {de~Melo}}, \bibinfo {author} {\bibfnamefont
				{Y.N.}\ \bibnamefont {Joglekar}}, \ and\ \bibinfo {author} {\bibfnamefont
				{L.}~\bibnamefont {Luo}},\ }\bibfield  {title} {\enquote {\bibinfo {title}
				{Observation of parity-time symmetry breaking transitions in a dissipative
					floquet system of ultracold atoms},}\ }\href {\doibase
			10.1038/s41467-019-08596-1} {\bibfield  {journal} {\bibinfo  {journal}
				{Nature communications}\ }\textbf {\bibinfo {volume} {10}},\ \bibinfo {pages}
			{855} (\bibinfo {year} {2019})}\BibitemShut {NoStop}%
		\bibitem [{\citenamefont {Kato}(1995)}]{Kato_book1995}%
		\BibitemOpen
		\bibfield  {author} {\bibinfo {author} {\bibfnamefont {T.}~\bibnamefont
				{Kato}},\ }\href {\doibase 10.1007/978-3-642-66282-9} {\emph {\bibinfo
				{title} {Perturbation Theory for Linear Operators}}}\ (\bibinfo  {publisher}
		{Springer-Verlag Berlin Heidelberg},\ \bibinfo {year} {1995})\BibitemShut
		{NoStop}%
		\bibitem [{\citenamefont {Heiss}(2004)}]{Heiss2004}%
		\BibitemOpen
		\bibfield  {author} {\bibinfo {author} {\bibfnamefont {W.D.}\ \bibnamefont
				{Heiss}},\ }\bibfield  {title} {\enquote {\bibinfo {title} {Exceptional
					points of non-{H}ermitian operators},}\ }\href {\doibase
			10.1088/0305-4470/37/6/034} {\bibfield  {journal} {\bibinfo  {journal}
				{Journal of Physics A: Mathematical and General}\ }\textbf {\bibinfo {volume}
				{37}},\ \bibinfo {pages} {2455--2464} (\bibinfo {year} {2004})}\BibitemShut
		{NoStop}%
		\bibitem [{\citenamefont {Berry}(2004)}]{Berry2004}%
		\BibitemOpen
		\bibfield  {author} {\bibinfo {author} {\bibfnamefont {M.V.}\ \bibnamefont
				{Berry}},\ }\bibfield  {title} {\enquote {\bibinfo {title} {Physics of
					non-{H}ermitian degeneracies},}\ }\href {\doibase
			10.1023/B:CJOP.0000044002.05657.04} {\bibfield  {journal} {\bibinfo
				{journal} {Czechoslovak Journal of Physics}\ }\textbf {\bibinfo {volume}
				{54}},\ \bibinfo {pages} {1039--1047} (\bibinfo {year} {2004})}\BibitemShut
		{NoStop}%
		\bibitem [{\citenamefont {M\"uller}\ and\ \citenamefont
			{Rotter}(2008)}]{Mueller_IRotter_JPA2008_exceptional}%
		\BibitemOpen
		\bibfield  {author} {\bibinfo {author} {\bibfnamefont {M.}~\bibnamefont
				{M\"uller}}\ and\ \bibinfo {author} {\bibfnamefont {I.}~\bibnamefont
				{Rotter}},\ }\bibfield  {title} {\enquote {\bibinfo {title} {Exceptional
					points in open quantum systems},}\ }\href {\doibase
			10.1088/1751-8113/41/24/244018} {\bibfield  {journal} {\bibinfo  {journal}
				{Journal of Physics A: Mathematical and Theoretical}\ }\textbf {\bibinfo
				{volume} {41}},\ \bibinfo {pages} {244018} (\bibinfo {year}
			{2008})}\BibitemShut {NoStop}%
		\bibitem [{\citenamefont {Heiss}(2012)}]{Heiss_JPA2012}%
		\BibitemOpen
		\bibfield  {author} {\bibinfo {author} {\bibfnamefont {W.D.}\ \bibnamefont
				{Heiss}},\ }\bibfield  {title} {\enquote {\bibinfo {title} {The physics of
					exceptional points},}\ }\href {\doibase 10.1088/1751-8113/45/44/444016}
		{\bibfield  {journal} {\bibinfo  {journal} {Journal of Physics A:
					Mathematical and Theoretical}\ }\textbf {\bibinfo {volume} {45}},\ \bibinfo
			{pages} {444016} (\bibinfo {year} {2012})}\BibitemShut {NoStop}%
		\bibitem [{\citenamefont {Golub}\ and\ \citenamefont
			{Van~Loan}(1996)}]{Golub_VanLoan_MatrixComputations_book1996}%
		\BibitemOpen
		\bibfield  {author} {\bibinfo {author} {\bibfnamefont {G.H.}\ \bibnamefont
				{Golub}}\ and\ \bibinfo {author} {\bibfnamefont {C.F.}\ \bibnamefont
				{Van~Loan}},\ }\href@noop {} {\emph {\bibinfo {title} {Matrix
					Computations}}},\ Johns Hopkins Studies in the Mathematical Sciences\
		(\bibinfo  {publisher} {Johns Hopkins University Press},\ \bibinfo {year}
		{1996})\BibitemShut {NoStop}%
		\bibitem [{\citenamefont
			{Watkins}(2004)}]{Watkins_MatrixComputations_book2004}%
		\BibitemOpen
		\bibfield  {author} {\bibinfo {author} {\bibfnamefont {D.S.}\ \bibnamefont
				{Watkins}},\ }\href@noop {} {\emph {\bibinfo {title} {Fundamentals of Matrix
					Computations}}},\ Pure and Applied Mathematics: A Wiley Series of Texts,
		Monographs and Tracts\ (\bibinfo  {publisher} {Wiley},\ \bibinfo {year}
		{2004})\BibitemShut {NoStop}%
		\bibitem [{\citenamefont {Heiss}\ and\ \citenamefont
			{Harney}(2001)}]{Heiss_Harney_EPJD2001_chirality}%
		\BibitemOpen
		\bibfield  {author} {\bibinfo {author} {\bibfnamefont {W.D.}\ \bibnamefont
				{Heiss}}\ and\ \bibinfo {author} {\bibfnamefont {H.L.}\ \bibnamefont
				{Harney}},\ }\bibfield  {title} {\enquote {\bibinfo {title} {The chirality of
					exceptional points},}\ }\href {\doibase 10.1007/s100530170017} {\bibfield
			{journal} {\bibinfo  {journal} {The European Physical Journal D - Atomic,
					Molecular, Optical and Plasma Physics}\ }\textbf {\bibinfo {volume} {17}},\
			\bibinfo {pages} {149--151} (\bibinfo {year} {2001})}\BibitemShut {NoStop}%
		\bibitem [{\citenamefont {Liu}\ \emph {et~al.}(2018)\citenamefont {Liu},
			\citenamefont {Wiersig}, \citenamefont {Sun}, \citenamefont {Fan},
			\citenamefont {Ge}, \citenamefont {Yang}, \citenamefont {Xiao}, \citenamefont
			{Song},\ and\ \citenamefont
			{Cao}}]{Liu_Wiersig_etal_LaserPhotRev2018_chirality}%
		\BibitemOpen
		\bibfield  {author} {\bibinfo {author} {\bibfnamefont {S.}~\bibnamefont
				{Liu}}, \bibinfo {author} {\bibfnamefont {J.}~\bibnamefont {Wiersig}},
			\bibinfo {author} {\bibfnamefont {W.}~\bibnamefont {Sun}}, \bibinfo {author}
			{\bibfnamefont {Y.}~\bibnamefont {Fan}}, \bibinfo {author} {\bibfnamefont
				{L.}~\bibnamefont {Ge}}, \bibinfo {author} {\bibfnamefont {J.}~\bibnamefont
				{Yang}}, \bibinfo {author} {\bibfnamefont {S.}~\bibnamefont {Xiao}}, \bibinfo
			{author} {\bibfnamefont {Q.}~\bibnamefont {Song}}, \ and\ \bibinfo {author}
			{\bibfnamefont {H.}~\bibnamefont {Cao}},\ }\bibfield  {title} {\enquote
			{\bibinfo {title} {Transporting the optical chirality through the dynamical
					barriers in optical microcavities},}\ }\href {\doibase
			10.1002/lpor.201800027} {\bibfield  {journal} {\bibinfo  {journal} {Laser \&
					Photonics Reviews}\ }\textbf {\bibinfo {volume} {12}},\ \bibinfo {pages}
			{1800027} (\bibinfo {year} {2018})}\BibitemShut {NoStop}%
		\bibitem [{\citenamefont {Wiersig}(2014)}]{Wiersig_PRL2014_sensitivity}%
		\BibitemOpen
		\bibfield  {author} {\bibinfo {author} {\bibfnamefont {J.}~\bibnamefont
				{Wiersig}},\ }\bibfield  {title} {\enquote {\bibinfo {title} {Enhancing the
					sensitivity of frequency and energy splitting detection by using exceptional
					points: Application to microcavity sensors for single-particle detection},}\
		}\href {\doibase 10.1103/PhysRevLett.112.203901} {\bibfield  {journal}
			{\bibinfo  {journal} {Phys. Rev. Lett.}\ }\textbf {\bibinfo {volume} {112}},\
			\bibinfo {pages} {203901} (\bibinfo {year} {2014})}\BibitemShut {NoStop}%
		\bibitem [{\citenamefont {Wiersig}(2016)}]{Wiersig_PRA2016_Sensors}%
		\BibitemOpen
		\bibfield  {author} {\bibinfo {author} {\bibfnamefont {J.}~\bibnamefont
				{Wiersig}},\ }\bibfield  {title} {\enquote {\bibinfo {title} {Sensors
					operating at exceptional points: General theory},}\ }\href {\doibase
			10.1103/PhysRevA.93.033809} {\bibfield  {journal} {\bibinfo  {journal} {Phys.
					Rev. A}\ }\textbf {\bibinfo {volume} {93}},\ \bibinfo {pages} {033809}
			(\bibinfo {year} {2016})}\BibitemShut {NoStop}%
		\bibitem [{\citenamefont {Hodaei}\ \emph {et~al.}(2017)\citenamefont {Hodaei},
			\citenamefont {Hassan}, \citenamefont {Wittek}, \citenamefont
			{Garcia-Gracia}, \citenamefont {El-Ganainy}, \citenamefont
			{Christodoulides},\ and\ \citenamefont
			{Khajavikhan}}]{Hodei_etal_Nature2017_EnhancedSensitivity}%
		\BibitemOpen
		\bibfield  {author} {\bibinfo {author} {\bibfnamefont {H.}~\bibnamefont
				{Hodaei}}, \bibinfo {author} {\bibfnamefont {A.U.}\ \bibnamefont {Hassan}},
			\bibinfo {author} {\bibfnamefont {S.}~\bibnamefont {Wittek}}, \bibinfo
			{author} {\bibfnamefont {H.}~\bibnamefont {Garcia-Gracia}}, \bibinfo {author}
			{\bibfnamefont {R.}~\bibnamefont {El-Ganainy}}, \bibinfo {author}
			{\bibfnamefont {D.N.}\ \bibnamefont {Christodoulides}}, \ and\ \bibinfo
			{author} {\bibfnamefont {M.}~\bibnamefont {Khajavikhan}},\ }\bibfield
		{title} {\enquote {\bibinfo {title} {Enhanced sensitivity at higher-order
					exceptional points},}\ }\href {\doibase 10.1038/nature23280} {\bibfield
			{journal} {\bibinfo  {journal} {Nature}\ }\textbf {\bibinfo {volume} {548}},\
			\bibinfo {pages} {187} (\bibinfo {year} {2017})}\BibitemShut {NoStop}%
		\bibitem [{\citenamefont {Hatano}\ and\ \citenamefont
			{Nelson}(1996)}]{Hatano_Nelson_PRL1996_localization}%
		\BibitemOpen
		\bibfield  {author} {\bibinfo {author} {\bibfnamefont {N.}~\bibnamefont
				{Hatano}}\ and\ \bibinfo {author} {\bibfnamefont {D.R.}\ \bibnamefont
				{Nelson}},\ }\bibfield  {title} {\enquote {\bibinfo {title} {Localization
					transitions in non-{H}ermitian quantum mechanics},}\ }\href {\doibase
			10.1103/PhysRevLett.77.570} {\bibfield  {journal} {\bibinfo  {journal} {Phys.
					Rev. Lett.}\ }\textbf {\bibinfo {volume} {77}},\ \bibinfo {pages} {570--573}
			(\bibinfo {year} {1996})}\BibitemShut {NoStop}%
		\bibitem [{\citenamefont {Goldsheid}\ and\ \citenamefont
			{Khoruzhenko}(1998)}]{Goldsheid_PRL1998_Andersonmodel}%
		\BibitemOpen
		\bibfield  {author} {\bibinfo {author} {\bibfnamefont {I.Ya.}\ \bibnamefont
				{Goldsheid}}\ and\ \bibinfo {author} {\bibfnamefont {B.A.}\ \bibnamefont
				{Khoruzhenko}},\ }\bibfield  {title} {\enquote {\bibinfo {title}
				{Distribution of eigenvalues in non-{H}ermitian anderson models},}\ }\href
		{\doibase 10.1103/PhysRevLett.80.2897} {\bibfield  {journal} {\bibinfo
				{journal} {Phys. Rev. Lett.}\ }\textbf {\bibinfo {volume} {80}},\ \bibinfo
			{pages} {2897--2900} (\bibinfo {year} {1998})}\BibitemShut {NoStop}%
		\bibitem [{\citenamefont {Heinrichs}(2001)}]{Heinrichs_PRB2001_Andersonmodel}%
		\BibitemOpen
		\bibfield  {author} {\bibinfo {author} {\bibfnamefont {J.}~\bibnamefont
				{Heinrichs}},\ }\bibfield  {title} {\enquote {\bibinfo {title} {Eigenvalues
					in the non-{H}ermitian anderson model},}\ }\href {\doibase
			10.1103/PhysRevB.63.165108} {\bibfield  {journal} {\bibinfo  {journal} {Phys.
					Rev. B}\ }\textbf {\bibinfo {volume} {63}},\ \bibinfo {pages} {165108}
			(\bibinfo {year} {2001})}\BibitemShut {NoStop}%
		\bibitem [{\citenamefont {Barontini}\ and\ \citenamefont
			{Guarrera}(2015)}]{Barontini_PhysRevA.91.032114_Anderson}%
		\BibitemOpen
		\bibfield  {author} {\bibinfo {author} {\bibfnamefont {Giovanni}\
				\bibnamefont {Barontini}}\ and\ \bibinfo {author} {\bibfnamefont {Vera}\
				\bibnamefont {Guarrera}},\ }\bibfield  {title} {\enquote {\bibinfo {title}
				{Localization by dissipative disorder: Deterministic approach to position
					measurements},}\ }\href {\doibase 10.1103/PhysRevA.91.032114} {\bibfield
			{journal} {\bibinfo  {journal} {Phys. Rev. A}\ }\textbf {\bibinfo {volume}
				{91}},\ \bibinfo {pages} {032114} (\bibinfo {year} {2015})}\BibitemShut
		{NoStop}%
		\bibitem [{\citenamefont
			{Longhi}(2019{\natexlab{a}})}]{Longhi_PRL2019_quasicrystals}%
		\BibitemOpen
		\bibfield  {author} {\bibinfo {author} {\bibfnamefont {S.}~\bibnamefont
				{Longhi}},\ }\bibfield  {title} {\enquote {\bibinfo {title} {Topological
					phase transition in non-{H}ermitian quasicrystals},}\ }\href {\doibase
			10.1103/PhysRevLett.122.237601} {\bibfield  {journal} {\bibinfo  {journal}
				{Phys. Rev. Lett.}\ }\textbf {\bibinfo {volume} {122}},\ \bibinfo {pages}
			{237601} (\bibinfo {year} {2019}{\natexlab{a}})}\BibitemShut {NoStop}%
		\bibitem [{\citenamefont
			{Longhi}(2019{\natexlab{b}})}]{Longhi_PRB2019_AubryAndre}%
		\BibitemOpen
		\bibfield  {author} {\bibinfo {author} {\bibfnamefont {S.}~\bibnamefont
				{Longhi}},\ }\bibfield  {title} {\enquote {\bibinfo {title} {Metal-insulator
					phase transition in a non-{H}ermitian aubry-andr\'e-harper model},}\ }\href
		{\doibase 10.1103/PhysRevB.100.125157} {\bibfield  {journal} {\bibinfo
				{journal} {Phys. Rev. B}\ }\textbf {\bibinfo {volume} {100}},\ \bibinfo
			{pages} {125157} (\bibinfo {year} {2019}{\natexlab{b}})}\BibitemShut
		{NoStop}%
		\bibitem [{\citenamefont {Longhi}(2010)}]{Longhi_PRA2010_invisibility}%
		\BibitemOpen
		\bibfield  {author} {\bibinfo {author} {\bibfnamefont {S.}~\bibnamefont
				{Longhi}},\ }\bibfield  {title} {\enquote {\bibinfo {title} {Invisibility in
					non-{H}ermitian tight-binding lattices},}\ }\href {\doibase
			10.1103/PhysRevA.82.032111} {\bibfield  {journal} {\bibinfo  {journal} {Phys.
					Rev. A}\ }\textbf {\bibinfo {volume} {82}},\ \bibinfo {pages} {032111}
			(\bibinfo {year} {2010})}\BibitemShut {NoStop}%
		\bibitem [{\citenamefont {Longhi}(2017)}]{Longhi_EPL2017_Floquet}%
		\BibitemOpen
		\bibfield  {author} {\bibinfo {author} {\bibfnamefont {S.}~\bibnamefont
				{Longhi}},\ }\bibfield  {title} {\enquote {\bibinfo {title} {Non-{H}ermitian
					{F}loquet invisibility},}\ }\href {\doibase 10.1209/0295-5075/117/10005}
		{\bibfield  {journal} {\bibinfo  {journal} {{EPL} (Europhysics Letters)}\
			}\textbf {\bibinfo {volume} {117}},\ \bibinfo {pages} {10005} (\bibinfo
			{year} {2017})}\BibitemShut {NoStop}%
		\bibitem [{\citenamefont {Leykam}\ \emph {et~al.}(2017)\citenamefont {Leykam},
			\citenamefont {Flach},\ and\ \citenamefont {Chong}}]{Flach_PRB2017_flatband}%
		\BibitemOpen
		\bibfield  {author} {\bibinfo {author} {\bibfnamefont {D.}~\bibnamefont
				{Leykam}}, \bibinfo {author} {\bibfnamefont {S.}~\bibnamefont {Flach}}, \
			and\ \bibinfo {author} {\bibfnamefont {Y.D.}\ \bibnamefont {Chong}},\
		}\bibfield  {title} {\enquote {\bibinfo {title} {Flat bands in lattices with
					non-{H}ermitian coupling},}\ }\href {\doibase 10.1103/PhysRevB.96.064305}
		{\bibfield  {journal} {\bibinfo  {journal} {Phys. Rev. B}\ }\textbf {\bibinfo
				{volume} {96}},\ \bibinfo {pages} {064305} (\bibinfo {year}
			{2017})}\BibitemShut {NoStop}%
		\bibitem [{\citenamefont {Longhi}(2009)}]{Longhi_PRL2009_BlochOscil}%
		\BibitemOpen
		\bibfield  {author} {\bibinfo {author} {\bibfnamefont {S.}~\bibnamefont
				{Longhi}},\ }\bibfield  {title} {\enquote {\bibinfo {title} {Bloch
					oscillations in complex crystals with $\mathcal{P}\mathcal{T}$ symmetry},}\
		}\href {\doibase 10.1103/PhysRevLett.103.123601} {\bibfield  {journal}
			{\bibinfo  {journal} {Phys. Rev. Lett.}\ }\textbf {\bibinfo {volume} {103}},\
			\bibinfo {pages} {123601} (\bibinfo {year} {2009})}\BibitemShut {NoStop}%
		\bibitem [{\citenamefont {Bendix}\ \emph {et~al.}(2009)\citenamefont {Bendix},
			\citenamefont {Fleischmann}, \citenamefont {Kottos},\ and\ \citenamefont
			{Shapiro}}]{Bendix_Kottos_Shapiro_PRL2009_latticePT}%
		\BibitemOpen
		\bibfield  {author} {\bibinfo {author} {\bibfnamefont {O.}~\bibnamefont
				{Bendix}}, \bibinfo {author} {\bibfnamefont {R.}~\bibnamefont {Fleischmann}},
			\bibinfo {author} {\bibfnamefont {T.}~\bibnamefont {Kottos}}, \ and\ \bibinfo
			{author} {\bibfnamefont {B.}~\bibnamefont {Shapiro}},\ }\bibfield  {title}
		{\enquote {\bibinfo {title} {Exponentially fragile $\mathcal{P}\mathcal{T}$
					symmetry in lattices with localized eigenmodes},}\ }\href {\doibase
			10.1103/PhysRevLett.103.030402} {\bibfield  {journal} {\bibinfo  {journal}
				{Phys. Rev. Lett.}\ }\textbf {\bibinfo {volume} {103}},\ \bibinfo {pages}
			{030402} (\bibinfo {year} {2009})}\BibitemShut {NoStop}%
		\bibitem [{\citenamefont {Jin}\ and\ \citenamefont
			{Song}(2009)}]{Jin_Song_PRA2009_latticePT}%
		\BibitemOpen
		\bibfield  {author} {\bibinfo {author} {\bibfnamefont {L.}~\bibnamefont
				{Jin}}\ and\ \bibinfo {author} {\bibfnamefont {Z.}~\bibnamefont {Song}},\
		}\bibfield  {title} {\enquote {\bibinfo {title} {Solutions of
					$\mathcal{P}\mathcal{T}$-symmetric tight-binding chain and its equivalent
					{H}ermitian counterpart},}\ }\href {\doibase 10.1103/PhysRevA.80.052107}
		{\bibfield  {journal} {\bibinfo  {journal} {Phys. Rev. A}\ }\textbf {\bibinfo
				{volume} {80}},\ \bibinfo {pages} {052107} (\bibinfo {year}
			{2009})}\BibitemShut {NoStop}%
		\bibitem [{\citenamefont {Jin}\ and\ \citenamefont
			{Song}(2010)}]{Jin_Song_PRA2010_latticePT}%
		\BibitemOpen
		\bibfield  {author} {\bibinfo {author} {\bibfnamefont {L.}~\bibnamefont
				{Jin}}\ and\ \bibinfo {author} {\bibfnamefont {Z.}~\bibnamefont {Song}},\
		}\bibfield  {title} {\enquote {\bibinfo {title} {Physics counterpart of the
					$\mathcal{P}\mathcal{T}$ non-{H}ermitian tight-binding chain},}\ }\href
		{\doibase 10.1103/PhysRevA.81.032109} {\bibfield  {journal} {\bibinfo
				{journal} {Phys. Rev. A}\ }\textbf {\bibinfo {volume} {81}},\ \bibinfo
			{pages} {032109} (\bibinfo {year} {2010})}\BibitemShut {NoStop}%
		\bibitem [{\citenamefont {Joglekar}\ \emph {et~al.}(2010)\citenamefont
			{Joglekar}, \citenamefont {Scott}, \citenamefont {Babbey},\ and\
			\citenamefont
			{Saxena}}]{Joglekar_Scott_Babbey_Saxena_PRA2010_tightbindingPT}%
		\BibitemOpen
		\bibfield  {author} {\bibinfo {author} {\bibfnamefont {Y.~N.}\ \bibnamefont
				{Joglekar}}, \bibinfo {author} {\bibfnamefont {D.}~\bibnamefont {Scott}},
			\bibinfo {author} {\bibfnamefont {M.}~\bibnamefont {Babbey}}, \ and\ \bibinfo
			{author} {\bibfnamefont {A.}~\bibnamefont {Saxena}},\ }\bibfield  {title}
		{\enquote {\bibinfo {title} {Robust and fragile $\mathcal{PT}$-symmetric
					phases in a tight-binding chain},}\ }\href {\doibase
			10.1103/PhysRevA.82.030103} {\bibfield  {journal} {\bibinfo  {journal} {Phys.
					Rev. A}\ }\textbf {\bibinfo {volume} {82}},\ \bibinfo {pages} {030103}
			(\bibinfo {year} {2010})}\BibitemShut {NoStop}%
		\bibitem [{\citenamefont {Jin}\ and\ \citenamefont
			{Song}(2016)}]{Jin_Song_PRA2016_PTflux}%
		\BibitemOpen
		\bibfield  {author} {\bibinfo {author} {\bibfnamefont {L.}~\bibnamefont
				{Jin}}\ and\ \bibinfo {author} {\bibfnamefont {Z.}~\bibnamefont {Song}},\
		}\bibfield  {title} {\enquote {\bibinfo {title} {Parity-time symmetry under
					magnetic flux},}\ }\href {\doibase 10.1103/PhysRevA.93.062110} {\bibfield
			{journal} {\bibinfo  {journal} {Phys. Rev. A}\ }\textbf {\bibinfo {volume}
				{93}},\ \bibinfo {pages} {062110} (\bibinfo {year} {2016})}\BibitemShut
		{NoStop}%
		\bibitem [{\citenamefont {Zhu}\ \emph {et~al.}(2016)\citenamefont {Zhu},
			\citenamefont {L\"u},\ and\ \citenamefont
			{Chen}}]{Zhu_etal_PRA2016_latticePT_scattering}%
		\BibitemOpen
		\bibfield  {author} {\bibinfo {author} {\bibfnamefont {B.}~\bibnamefont
				{Zhu}}, \bibinfo {author} {\bibfnamefont {R.}~\bibnamefont {L\"u}}, \ and\
			\bibinfo {author} {\bibfnamefont {S.}~\bibnamefont {Chen}},\ }\bibfield
		{title} {\enquote {\bibinfo {title} {$\mathcal{PT}$-symmetry breaking for the
					scattering problem in a one-dimensional non-{H}ermitian lattice model},}\
		}\href {\doibase 10.1103/PhysRevA.93.032129} {\bibfield  {journal} {\bibinfo
				{journal} {Phys. Rev. A}\ }\textbf {\bibinfo {volume} {93}},\ \bibinfo
			{pages} {032129} (\bibinfo {year} {2016})}\BibitemShut {NoStop}%
		\bibitem [{\citenamefont {Ortega}\ \emph {et~al.}(2019)\citenamefont {Ortega},
			\citenamefont {Stegmann}, \citenamefont {L.},\ and\ \citenamefont
			{Larralde}}]{Ortega_etal_arXiV2019_latticePT}%
		\BibitemOpen
		\bibfield  {author} {\bibinfo {author} {\bibfnamefont {A.}~\bibnamefont
				{Ortega}}, \bibinfo {author} {\bibfnamefont {T.}~\bibnamefont {Stegmann}},
			\bibinfo {author} {\bibfnamefont {Benet.}\ \bibnamefont {L.}}, \ and\
			\bibinfo {author} {\bibfnamefont {H.}~\bibnamefont {Larralde}},\ }\href@noop
		{} {\enquote {\bibinfo {title} {$\mathcal{PT}$-symmetric tight-binding chain
					with gain and loss: A completely solvable model},}\ } (\bibinfo {year}
		{2019}),\ \Eprint {http://arxiv.org/abs/1906.10116v1} {arXiv:1906.10116v1
			[quant-ph]} \BibitemShut {NoStop}%
		\bibitem [{\citenamefont {Ozawa}\ \emph {et~al.}(2019)\citenamefont {Ozawa},
			\citenamefont {Price}, \citenamefont {Amo}, \citenamefont {Goldman},
			\citenamefont {Hafezi}, \citenamefont {Lu}, \citenamefont {Rechtsman},
			\citenamefont {Schuster}, \citenamefont {Simon}, \citenamefont {Zilberberg},\
			and\ \citenamefont
			{Carusotto}}]{NGoldman_Zilberberg_Carusotto_RMP2018_topologicalphotonics}%
		\BibitemOpen
		\bibfield  {author} {\bibinfo {author} {\bibfnamefont {T.}~\bibnamefont
				{Ozawa}}, \bibinfo {author} {\bibfnamefont {H.~M.}\ \bibnamefont {Price}},
			\bibinfo {author} {\bibfnamefont {A.}~\bibnamefont {Amo}}, \bibinfo {author}
			{\bibfnamefont {N.}~\bibnamefont {Goldman}}, \bibinfo {author} {\bibfnamefont
				{M.}~\bibnamefont {Hafezi}}, \bibinfo {author} {\bibfnamefont
				{L.}~\bibnamefont {Lu}}, \bibinfo {author} {\bibfnamefont {M.C.}\
				\bibnamefont {Rechtsman}}, \bibinfo {author} {\bibfnamefont {D.}~\bibnamefont
				{Schuster}}, \bibinfo {author} {\bibfnamefont {J.}~\bibnamefont {Simon}},
			\bibinfo {author} {\bibfnamefont {O.}~\bibnamefont {Zilberberg}}, \ and\
			\bibinfo {author} {\bibfnamefont {I.}~\bibnamefont {Carusotto}},\ }\bibfield
		{title} {\enquote {\bibinfo {title} {Topological photonics},}\ }\href
		{\doibase 10.1103/RevModPhys.91.015006} {\bibfield  {journal} {\bibinfo
				{journal} {Rev. Mod. Phys.}\ }\textbf {\bibinfo {volume} {91}},\ \bibinfo
			{pages} {015006} (\bibinfo {year} {2019})}\BibitemShut {NoStop}%
		\bibitem [{\citenamefont {Martinez~Alvarez}\ \emph {et~al.}(2018)\citenamefont
			{Martinez~Alvarez}, \citenamefont {Barrios~Vargas}, \citenamefont
			{Berdakin},\ and\ \citenamefont
			{Foa~Torres}}]{MartinezAlvarez_etal_EPJST2018_topological_review}%
		\BibitemOpen
		\bibfield  {author} {\bibinfo {author} {\bibfnamefont {V.~M.}\ \bibnamefont
				{Martinez~Alvarez}}, \bibinfo {author} {\bibfnamefont {J.~E.}\ \bibnamefont
				{Barrios~Vargas}}, \bibinfo {author} {\bibfnamefont {M.}~\bibnamefont
				{Berdakin}}, \ and\ \bibinfo {author} {\bibfnamefont {L.~E.~F.}\ \bibnamefont
				{Foa~Torres}},\ }\bibfield  {title} {\enquote {\bibinfo {title} {Topological
					states of non-{H}ermitian systems},}\ }\href {\doibase
			10.1140/epjst/e2018-800091-5} {\bibfield  {journal} {\bibinfo  {journal} {The
					European Physical Journal Special Topics}\ }\textbf {\bibinfo {volume}
				{227}},\ \bibinfo {pages} {1295--1308} (\bibinfo {year} {2018})}\BibitemShut
		{NoStop}%
		\bibitem [{\citenamefont {Torres}(2019)}]{torres_arXiv2019_review_topological}%
		\BibitemOpen
		\bibfield  {author} {\bibinfo {author} {\bibfnamefont {Luis E F~Foa}\
				\bibnamefont {Torres}},\ }\bibfield  {title} {\enquote {\bibinfo {title}
				{Perspective on topological states of non-hermitian lattices},}\ }\href
		{\doibase 10.1088/2515-7639/ab4092} {\bibfield  {journal} {\bibinfo
				{journal} {Journal of Physics: Materials}\ }\textbf {\bibinfo {volume} {3}},\
			\bibinfo {pages} {014002} (\bibinfo {year} {2019})}\BibitemShut {NoStop}%
		\bibitem [{\citenamefont {McClarty}\ and\ \citenamefont
			{Rau}(2019)}]{McClarty_Rau_PRB2019_magnon}%
		\BibitemOpen
		\bibfield  {author} {\bibinfo {author} {\bibfnamefont {P.~A.}\ \bibnamefont
				{McClarty}}\ and\ \bibinfo {author} {\bibfnamefont {J.~G.}\ \bibnamefont
				{Rau}},\ }\bibfield  {title} {\enquote {\bibinfo {title} {Non-{H}ermitian
					topology of spontaneous magnon decay},}\ }\href {\doibase
			10.1103/PhysRevB.100.100405} {\bibfield  {journal} {\bibinfo  {journal}
				{Phys. Rev. B}\ }\textbf {\bibinfo {volume} {100}},\ \bibinfo {pages}
			{100405} (\bibinfo {year} {2019})}\BibitemShut {NoStop}%
		\bibitem [{\citenamefont {Luitz}\ and\ \citenamefont
			{Piazza}(2019)}]{LuitzPiazza_2019_manybody}%
		\BibitemOpen
		\bibfield  {author} {\bibinfo {author} {\bibfnamefont {David~J.}\
				\bibnamefont {Luitz}}\ and\ \bibinfo {author} {\bibfnamefont {Francesco}\
				\bibnamefont {Piazza}},\ }\bibfield  {title} {\enquote {\bibinfo {title}
				{Exceptional points and the topology of quantum many-body spectra},}\ }\href
		{\doibase 10.1103/PhysRevResearch.1.033051} {\bibfield  {journal} {\bibinfo
				{journal} {Phys. Rev. Research}\ }\textbf {\bibinfo {volume} {1}},\ \bibinfo
			{pages} {033051} (\bibinfo {year} {2019})}\BibitemShut {NoStop}%
		\bibitem [{\citenamefont
			{Allcock}(1969{\natexlab{a}})}]{Allcock_AnnPhys1969_1}%
		\BibitemOpen
		\bibfield  {author} {\bibinfo {author} {\bibfnamefont {G.R.}\ \bibnamefont
				{Allcock}},\ }\bibfield  {title} {\enquote {\bibinfo {title} {The time of
					arrival in quantum mechanics {I}. formal considerations},}\ }\href {\doibase
			https://doi.org/10.1016/0003-4916(69)90251-6} {\bibfield  {journal} {\bibinfo
				{journal} {Annals of Physics}\ }\textbf {\bibinfo {volume} {53}},\ \bibinfo
			{pages} {253 -- 285} (\bibinfo {year} {1969}{\natexlab{a}})}\BibitemShut
		{NoStop}%
		\bibitem [{\citenamefont
			{Allcock}(1969{\natexlab{b}})}]{Allcock_AnnPhys1969_2}%
		\BibitemOpen
		\bibfield  {author} {\bibinfo {author} {\bibfnamefont {G.R.}\ \bibnamefont
				{Allcock}},\ }\bibfield  {title} {\enquote {\bibinfo {title} {The time of
					arrival in quantum mechanics {II}. the individual measurement},}\ }\href
		{\doibase https://doi.org/10.1016/0003-4916(69)90252-8} {\bibfield  {journal}
			{\bibinfo  {journal} {Annals of Physics}\ }\textbf {\bibinfo {volume} {53}},\
			\bibinfo {pages} {286 -- 310} (\bibinfo {year}
			{1969}{\natexlab{b}})}\BibitemShut {NoStop}%
		\bibitem [{\citenamefont
			{Allcock}(1969{\natexlab{c}})}]{Allcock_AnnPhys1969_3}%
		\BibitemOpen
		\bibfield  {author} {\bibinfo {author} {\bibfnamefont {G.R.}\ \bibnamefont
				{Allcock}},\ }\bibfield  {title} {\enquote {\bibinfo {title} {The time of
					arrival in quantum mechanics {III}: The measurement ensemble},}\ }\href
		{\doibase 10.1016/0003-4916(69)90253-X} {\bibfield  {journal} {\bibinfo
				{journal} {Annals of Physics}\ }\textbf {\bibinfo {volume} {53}},\ \bibinfo
			{pages} {311 -- 348} (\bibinfo {year} {1969}{\natexlab{c}})}\BibitemShut
		{NoStop}%
		\bibitem [{\citenamefont {Misra}\ and\ \citenamefont
			{Sudarshan}(1977)}]{Misra_Sudarshan_Zeno_JMathPhys1977}%
		\BibitemOpen
		\bibfield  {author} {\bibinfo {author} {\bibfnamefont {B.}~\bibnamefont
				{Misra}}\ and\ \bibinfo {author} {\bibfnamefont {E.C.G.}\ \bibnamefont
				{Sudarshan}},\ }\bibfield  {title} {\enquote {\bibinfo {title} {The zeno’s
					paradox in quantum theory},}\ }\href {\doibase 10.1063/1.523304} {\bibfield
			{journal} {\bibinfo  {journal} {Journal of Mathematical Physics}\ }\textbf
			{\bibinfo {volume} {18}},\ \bibinfo {pages} {756--763} (\bibinfo {year}
			{1977})}\BibitemShut {NoStop}%
		\bibitem [{\citenamefont {Facchi}\ and\ \citenamefont
			{Pascazio}(2008)}]{Facchi_Pascazio_JPA2008}%
		\BibitemOpen
		\bibfield  {author} {\bibinfo {author} {\bibfnamefont {P.}~\bibnamefont
				{Facchi}}\ and\ \bibinfo {author} {\bibfnamefont {S.}~\bibnamefont
				{Pascazio}},\ }\bibfield  {title} {\enquote {\bibinfo {title} {Quantum zeno
					dynamics: mathematical and physical aspects},}\ }\href {\doibase
			10.1088/1751-8113/41/49/493001} {\bibfield  {journal} {\bibinfo  {journal}
				{Journal of Physics A: Mathematical and Theoretical}\ }\textbf {\bibinfo
				{volume} {41}},\ \bibinfo {pages} {493001} (\bibinfo {year}
			{2008})}\BibitemShut {NoStop}%
		\bibitem [{\citenamefont {Zezyulin}\ \emph {et~al.}(2012)\citenamefont
			{Zezyulin}, \citenamefont {Konotop}, \citenamefont {Barontini},\ and\
			\citenamefont {Ott}}]{Zezyulin_Barontini_Ott_MacroscopicZeno_PRL2012}%
		\BibitemOpen
		\bibfield  {author} {\bibinfo {author} {\bibfnamefont {D.~A.}\ \bibnamefont
				{Zezyulin}}, \bibinfo {author} {\bibfnamefont {V.~V.}\ \bibnamefont
				{Konotop}}, \bibinfo {author} {\bibfnamefont {G.}~\bibnamefont {Barontini}},
			\ and\ \bibinfo {author} {\bibfnamefont {H.}~\bibnamefont {Ott}},\ }\bibfield
		{title} {\enquote {\bibinfo {title} {Macroscopic zeno effect and stationary
					flows in nonlinear waveguides with localized dissipation},}\ }\href {\doibase
			10.1103/PhysRevLett.109.020405} {\bibfield  {journal} {\bibinfo  {journal}
				{Phys. Rev. Lett.}\ }\textbf {\bibinfo {volume} {109}},\ \bibinfo {pages}
			{020405} (\bibinfo {year} {2012})}\BibitemShut {NoStop}%
		\bibitem [{\citenamefont {Krapivsky}\ \emph {et~al.}(2014)\citenamefont
			{Krapivsky}, \citenamefont {Luck},\ and\ \citenamefont
			{Mallick}}]{Krapivsky_Luck_Mallick_JSP2014_survival}%
		\BibitemOpen
		\bibfield  {author} {\bibinfo {author} {\bibfnamefont {P.~L.}\ \bibnamefont
				{Krapivsky}}, \bibinfo {author} {\bibfnamefont {J.~M.}\ \bibnamefont {Luck}},
			\ and\ \bibinfo {author} {\bibfnamefont {K.}~\bibnamefont {Mallick}},\
		}\bibfield  {title} {\enquote {\bibinfo {title} {Survival of classical and
					quantum particles in the presence of traps},}\ }\href {\doibase
			10.1007/s10955-014-0936-8} {\bibfield  {journal} {\bibinfo  {journal}
				{Journal of Statistical Physics}\ }\textbf {\bibinfo {volume} {154}},\
			\bibinfo {pages} {1430--1460} (\bibinfo {year} {2014})}\BibitemShut {NoStop}%
		\bibitem [{\citenamefont {Dhar}\ \emph
			{et~al.}(2015{\natexlab{a}})\citenamefont {Dhar}, \citenamefont {Dasgupta},
			\citenamefont {Dhar},\ and\ \citenamefont {Sen}}]{Dhar_PRA2015_measurements}%
		\BibitemOpen
		\bibfield  {author} {\bibinfo {author} {\bibfnamefont {S.}~\bibnamefont
				{Dhar}}, \bibinfo {author} {\bibfnamefont {S.}~\bibnamefont {Dasgupta}},
			\bibinfo {author} {\bibfnamefont {A.}~\bibnamefont {Dhar}}, \ and\ \bibinfo
			{author} {\bibfnamefont {D.}~\bibnamefont {Sen}},\ }\bibfield  {title}
		{\enquote {\bibinfo {title} {Detection of a quantum particle on a lattice
					under repeated projective measurements},}\ }\href {\doibase
			10.1103/PhysRevA.91.062115} {\bibfield  {journal} {\bibinfo  {journal} {Phys.
					Rev. A}\ }\textbf {\bibinfo {volume} {91}},\ \bibinfo {pages} {062115}
			(\bibinfo {year} {2015}{\natexlab{a}})}\BibitemShut {NoStop}%
		\bibitem [{\citenamefont {Dhar}\ \emph
			{et~al.}(2015{\natexlab{b}})\citenamefont {Dhar}, \citenamefont {Dasgupta},\
			and\ \citenamefont {Dhar}}]{Dhar_JPA2015_arrival}%
		\BibitemOpen
		\bibfield  {author} {\bibinfo {author} {\bibfnamefont {S.}~\bibnamefont
				{Dhar}}, \bibinfo {author} {\bibfnamefont {S.}~\bibnamefont {Dasgupta}}, \
			and\ \bibinfo {author} {\bibfnamefont {A.}~\bibnamefont {Dhar}},\ }\bibfield
		{title} {\enquote {\bibinfo {title} {Quantum time of arrival distribution in
					a simple lattice model},}\ }\href {\doibase 10.1088/1751-8113/48/11/115304}
		{\bibfield  {journal} {\bibinfo  {journal} {Journal of Physics A:
					Mathematical and Theoretical}\ }\textbf {\bibinfo {volume} {48}},\ \bibinfo
			{pages} {115304} (\bibinfo {year} {2015}{\natexlab{b}})}\BibitemShut
		{NoStop}%
		\bibitem [{\citenamefont {Kozlowski}\ \emph {et~al.}(2016)\citenamefont
			{Kozlowski}, \citenamefont {Caballero-Benitez},\ and\ \citenamefont
			{Mekhov}}]{Kozlowski_Mekhov_PRA2016_Zeno}%
		\BibitemOpen
		\bibfield  {author} {\bibinfo {author} {\bibfnamefont {W.}~\bibnamefont
				{Kozlowski}}, \bibinfo {author} {\bibfnamefont {S.~F.}\ \bibnamefont
				{Caballero-Benitez}}, \ and\ \bibinfo {author} {\bibfnamefont {I.~B.}\
				\bibnamefont {Mekhov}},\ }\bibfield  {title} {\enquote {\bibinfo {title}
				{Non-{H}ermitian dynamics in the quantum zeno limit},}\ }\href {\doibase
			10.1103/PhysRevA.94.012123} {\bibfield  {journal} {\bibinfo  {journal} {Phys.
					Rev. A}\ }\textbf {\bibinfo {volume} {94}},\ \bibinfo {pages} {012123}
			(\bibinfo {year} {2016})}\BibitemShut {NoStop}%
		\bibitem [{\citenamefont {Fr\"oml}\ \emph {et~al.}(2019)\citenamefont
			{Fr\"oml}, \citenamefont {Chiocchetta}, \citenamefont {Kollath},\ and\
			\citenamefont {Diehl}}]{Chiocchetta_Kollath_Diehl_PRL2019_QuantumZeno}%
		\BibitemOpen
		\bibfield  {author} {\bibinfo {author} {\bibfnamefont {H.}~\bibnamefont
				{Fr\"oml}}, \bibinfo {author} {\bibfnamefont {A.}~\bibnamefont
				{Chiocchetta}}, \bibinfo {author} {\bibfnamefont {C.}~\bibnamefont
				{Kollath}}, \ and\ \bibinfo {author} {\bibfnamefont {S.}~\bibnamefont
				{Diehl}},\ }\bibfield  {title} {\enquote {\bibinfo {title}
				{Fluctuation-induced quantum zeno effect},}\ }\href {\doibase
			10.1103/PhysRevLett.122.040402} {\bibfield  {journal} {\bibinfo  {journal}
				{Phys. Rev. Lett.}\ }\textbf {\bibinfo {volume} {122}},\ \bibinfo {pages}
			{040402} (\bibinfo {year} {2019})}\BibitemShut {NoStop}%
		\bibitem [{\citenamefont {Muga}\ \emph {et~al.}(2004)\citenamefont {Muga},
			\citenamefont {Palao}, \citenamefont {Navarro},\ and\ \citenamefont
			{Egusquiza}}]{MUGA2004357}%
		\BibitemOpen
		\bibfield  {author} {\bibinfo {author} {\bibfnamefont {J.G.}\ \bibnamefont
				{Muga}}, \bibinfo {author} {\bibfnamefont {J.P.}\ \bibnamefont {Palao}},
			\bibinfo {author} {\bibfnamefont {B.}~\bibnamefont {Navarro}}, \ and\
			\bibinfo {author} {\bibfnamefont {I.L.}\ \bibnamefont {Egusquiza}},\
		}\bibfield  {title} {\enquote {\bibinfo {title} {Complex absorbing
					potentials},}\ }\href {\doibase
			https://doi.org/10.1016/j.physrep.2004.03.002} {\bibfield  {journal}
			{\bibinfo  {journal} {Physics Reports}\ }\textbf {\bibinfo {volume} {395}},\
			\bibinfo {pages} {357 -- 426} (\bibinfo {year} {2004})}\BibitemShut {NoStop}%
		\bibitem [{\citenamefont {Griffiths}(2017)}]{Griffiths_QM_book2017}%
		\BibitemOpen
		\bibfield  {author} {\bibinfo {author} {\bibfnamefont {D.J.}\ \bibnamefont
				{Griffiths}},\ }\href@noop {} {\emph {\bibinfo {title} {Introduction to
					Quantum Mechanics}}}\ (\bibinfo  {publisher} {Cambridge University Press},\
		\bibinfo {year} {2017})\BibitemShut {NoStop}%
		\bibitem [{\citenamefont {Graefe}\ \emph {et~al.}(2008)\citenamefont {Graefe},
			\citenamefont {G{\"u}nther}, \citenamefont {Korsch},\ and\ \citenamefont
			{Niederle}}]{GraefeKorsch_JPA2008_higherorder}%
		\BibitemOpen
		\bibfield  {author} {\bibinfo {author} {\bibfnamefont {E.-M.}\ \bibnamefont
				{Graefe}}, \bibinfo {author} {\bibfnamefont {U.}~\bibnamefont {G{\"u}nther}},
			\bibinfo {author} {\bibfnamefont {H.J.}\ \bibnamefont {Korsch}}, \ and\
			\bibinfo {author} {\bibfnamefont {A.E.}\ \bibnamefont {Niederle}},\
		}\bibfield  {title} {\enquote {\bibinfo {title} {A non-{H}ermitian symmetric
					bose--hubbard model: eigenvalue rings from unfolding higher-order exceptional
					points},}\ }\href {\doibase 10.1088/1751-8113/41/25/255206} {\bibfield
			{journal} {\bibinfo  {journal} {Journal of Physics A: Mathematical and
					Theoretical}\ }\textbf {\bibinfo {volume} {41}},\ \bibinfo {pages} {255206}
			(\bibinfo {year} {2008})}\BibitemShut {NoStop}%
		\bibitem [{\citenamefont {Heiss}(2008)}]{Heiss_JPA2008_higherorder_chirality}%
		\BibitemOpen
		\bibfield  {author} {\bibinfo {author} {\bibfnamefont {W.D.}\ \bibnamefont
				{Heiss}},\ }\bibfield  {title} {\enquote {\bibinfo {title} {Chirality of
					wavefunctions for three coalescing levels},}\ }\href {\doibase
			10.1088/1751-8113/41/24/244010} {\bibfield  {journal} {\bibinfo  {journal}
				{Journal of Physics A: Mathematical and Theoretical}\ }\textbf {\bibinfo
				{volume} {41}},\ \bibinfo {pages} {244010} (\bibinfo {year}
			{2008})}\BibitemShut {NoStop}%
		\bibitem [{\citenamefont {Demange}\ and\ \citenamefont
			{Graefe}(2011)}]{Demange_Graefe_JPA2011_higherorder}%
		\BibitemOpen
		\bibfield  {author} {\bibinfo {author} {\bibfnamefont {G.}~\bibnamefont
				{Demange}}\ and\ \bibinfo {author} {\bibfnamefont {E.-M.}\ \bibnamefont
				{Graefe}},\ }\bibfield  {title} {\enquote {\bibinfo {title} {Signatures of
					three coalescing eigenfunctions},}\ }\href {\doibase
			10.1088/1751-8113/45/2/025303} {\bibfield  {journal} {\bibinfo  {journal}
				{Journal of Physics A: Mathematical and Theoretical}\ }\textbf {\bibinfo
				{volume} {45}},\ \bibinfo {pages} {025303} (\bibinfo {year}
			{2011})}\BibitemShut {NoStop}%
		\bibitem [{\citenamefont {Wiersig}(2018)}]{Wiersig_PRA2018}%
		\BibitemOpen
		\bibfield  {author} {\bibinfo {author} {\bibfnamefont {J.}~\bibnamefont
				{Wiersig}},\ }\bibfield  {title} {\enquote {\bibinfo {title} {Role of
					nonorthogonality of energy eigenstates in quantum systems with localized
					losses},}\ }\href {\doibase 10.1103/PhysRevA.98.052105} {\bibfield  {journal}
			{\bibinfo  {journal} {Phys. Rev. A}\ }\textbf {\bibinfo {volume} {98}},\
			\bibinfo {pages} {052105} (\bibinfo {year} {2018})}\BibitemShut {NoStop}%
		\bibitem [{\citenamefont {Fr\"oml}\ \emph {et~al.}(2020)\citenamefont
			{Fr\"oml}, \citenamefont {Muckel}, \citenamefont {Kollath}, \citenamefont
			{Chiocchetta},\ and\ \citenamefont
			{Diehl}}]{Chiocchetta_Kollath_Diehl_arXiv2019_manybodyQuantumZeno}%
		\BibitemOpen
		\bibfield  {author} {\bibinfo {author} {\bibfnamefont {Heinrich}\
				\bibnamefont {Fr\"oml}}, \bibinfo {author} {\bibfnamefont {Christopher}\
				\bibnamefont {Muckel}}, \bibinfo {author} {\bibfnamefont {Corinna}\
				\bibnamefont {Kollath}}, \bibinfo {author} {\bibfnamefont {Alessio}\
				\bibnamefont {Chiocchetta}}, \ and\ \bibinfo {author} {\bibfnamefont
				{Sebastian}\ \bibnamefont {Diehl}},\ }\bibfield  {title} {\enquote {\bibinfo
				{title} {Ultracold quantum wires with localized losses: {M}any-body quantum
					{Z}eno effect},}\ }\href {\doibase 10.1103/PhysRevB.101.144301} {\bibfield
			{journal} {\bibinfo  {journal} {Phys. Rev. B}\ }\textbf {\bibinfo {volume}
				{101}},\ \bibinfo {pages} {144301} (\bibinfo {year} {2020})}\BibitemShut
		{NoStop}%
		\bibitem [{\citenamefont {Labouvie}\ \emph {et~al.}(2015)\citenamefont
			{Labouvie}, \citenamefont {Santra}, \citenamefont {Heun}, \citenamefont
			{Wimberger},\ and\ \citenamefont {Ott}}]{Labouvie_PhysRevLett.115.050601}%
		\BibitemOpen
		\bibfield  {author} {\bibinfo {author} {\bibfnamefont {R.}~\bibnamefont
				{Labouvie}}, \bibinfo {author} {\bibfnamefont {B.}~\bibnamefont {Santra}},
			\bibinfo {author} {\bibfnamefont {S.}~\bibnamefont {Heun}}, \bibinfo {author}
			{\bibfnamefont {S.}~\bibnamefont {Wimberger}}, \ and\ \bibinfo {author}
			{\bibfnamefont {H.}~\bibnamefont {Ott}},\ }\bibfield  {title} {\enquote
			{\bibinfo {title} {Negative differential conductivity in an interacting
					quantum gas},}\ }\href {\doibase 10.1103/PhysRevLett.115.050601} {\bibfield
			{journal} {\bibinfo  {journal} {Phys. Rev. Lett.}\ }\textbf {\bibinfo
				{volume} {115}},\ \bibinfo {pages} {050601} (\bibinfo {year}
			{2015})}\BibitemShut {NoStop}%
		\bibitem [{\citenamefont {Labouvie}\ \emph {et~al.}(2016)\citenamefont
			{Labouvie}, \citenamefont {Santra}, \citenamefont {Heun},\ and\ \citenamefont
			{Ott}}]{Labouvie_PhysRevLett.116.235302}%
		\BibitemOpen
		\bibfield  {author} {\bibinfo {author} {\bibfnamefont {R.}~\bibnamefont
				{Labouvie}}, \bibinfo {author} {\bibfnamefont {B.}~\bibnamefont {Santra}},
			\bibinfo {author} {\bibfnamefont {S.}~\bibnamefont {Heun}}, \ and\ \bibinfo
			{author} {\bibfnamefont {H.}~\bibnamefont {Ott}},\ }\bibfield  {title}
		{\enquote {\bibinfo {title} {Bistability in a driven-dissipative
					superfluid},}\ }\href {\doibase 10.1103/PhysRevLett.116.235302} {\bibfield
			{journal} {\bibinfo  {journal} {Phys. Rev. Lett.}\ }\textbf {\bibinfo
				{volume} {116}},\ \bibinfo {pages} {235302} (\bibinfo {year}
			{2016})}\BibitemShut {NoStop}%
		\bibitem [{\citenamefont {Heinrich}\ \emph
			{et~al.}(2014{\natexlab{a}})\citenamefont {Heinrich}, \citenamefont {Miri},
			\citenamefont {St{\"u}zer}, \citenamefont {El-Ganainy}, \citenamefont
			{Nolte}, \citenamefont {Szameit},\ and\ \citenamefont
			{Christodoulides}}]{Szameit_nature_2014_susy_mode_conv}%
		\BibitemOpen
		\bibfield  {author} {\bibinfo {author} {\bibfnamefont {M.}~\bibnamefont
				{Heinrich}}, \bibinfo {author} {\bibfnamefont {M.}~\bibnamefont {Miri}},
			\bibinfo {author} {\bibfnamefont {S.}~\bibnamefont {St{\"u}zer}}, \bibinfo
			{author} {\bibfnamefont {R.}~\bibnamefont {El-Ganainy}}, \bibinfo {author}
			{\bibfnamefont {S.}~\bibnamefont {Nolte}}, \bibinfo {author} {\bibfnamefont
				{A.}~\bibnamefont {Szameit}}, \ and\ \bibinfo {author} {\bibfnamefont
				{D.~N.}\ \bibnamefont {Christodoulides}},\ }\bibfield  {title} {\enquote
			{\bibinfo {title} {Supersymmetric mode converters},}\ }\href {\doibase
			10.1038/ncomms4698} {\bibfield  {journal} {\bibinfo  {journal} {Nature
					Communications}\ }\textbf {\bibinfo {volume} {5}},\ \bibinfo {pages} {3698}
			(\bibinfo {year} {2014}{\natexlab{a}})}\BibitemShut {NoStop}%
		\bibitem [{\citenamefont {Eichelkraut}\ \emph {et~al.}(2018)\citenamefont
			{Eichelkraut}, \citenamefont {Kremer}, \citenamefont {Ornigotti},\ and\
			\citenamefont {Szameit}}]{Eichelkraut2018}%
		\BibitemOpen
		\bibfield  {author} {\bibinfo {author} {\bibfnamefont {S.}~\bibnamefont
				{Eichelkraut}, \bibfnamefont {T.and~Weimann}}, \bibinfo {author}
			{\bibfnamefont {M.}~\bibnamefont {Kremer}}, \bibinfo {author} {\bibfnamefont
				{M.}~\bibnamefont {Ornigotti}}, \ and\ \bibinfo {author} {\bibfnamefont
				{A.}~\bibnamefont {Szameit}},\ }\enquote {\bibinfo {title} {Passive {$\cal
					PT$}-symmetry in laser-written optical waveguide structures},}\ in\ \href
		{\doibase 10.1007/978-981-13-1247-2_5} {\emph {\bibinfo {booktitle}
				{Parity-time Symmetry and Its Applications}}},\ \bibinfo {editor} {edited by\
			\bibinfo {editor} {\bibfnamefont {Demetrios}\ \bibnamefont
				{Christodoulides}}\ and\ \bibinfo {editor} {\bibfnamefont {Jianke}\
				\bibnamefont {Yang}}}\ (\bibinfo  {publisher} {Springer Singapore},\ \bibinfo
		{address} {Singapore},\ \bibinfo {year} {2018})\ pp.\ \bibinfo {pages}
		{123--153}\BibitemShut {NoStop}%
		\bibitem [{\citenamefont {Morandotti}\ \emph {et~al.}(1999)\citenamefont
			{Morandotti}, \citenamefont {Peschel}, \citenamefont {Aitchison},
			\citenamefont {Eisenberg},\ and\ \citenamefont
			{Silberberg}}]{Silberberg_photoniclattice_BlochOscil_PRL1999}%
		\BibitemOpen
		\bibfield  {author} {\bibinfo {author} {\bibfnamefont {R.}~\bibnamefont
				{Morandotti}}, \bibinfo {author} {\bibfnamefont {U.}~\bibnamefont {Peschel}},
			\bibinfo {author} {\bibfnamefont {J.~S.}\ \bibnamefont {Aitchison}}, \bibinfo
			{author} {\bibfnamefont {H.~S.}\ \bibnamefont {Eisenberg}}, \ and\ \bibinfo
			{author} {\bibfnamefont {Y.}~\bibnamefont {Silberberg}},\ }\bibfield  {title}
		{\enquote {\bibinfo {title} {Experimental observation of linear and nonlinear
					optical {B}loch oscillations},}\ }\href {\doibase
			10.1103/PhysRevLett.83.4756} {\bibfield  {journal} {\bibinfo  {journal}
				{Phys. Rev. Lett.}\ }\textbf {\bibinfo {volume} {83}},\ \bibinfo {pages}
			{4756--4759} (\bibinfo {year} {1999})}\BibitemShut {NoStop}%
		\bibitem [{\citenamefont {Schwartz}\ \emph {et~al.}(2007)\citenamefont
			{Schwartz}, \citenamefont {Bartal}, \citenamefont {Fishman},\ and\
			\citenamefont {Segev}}]{Segev_photoniclattice_andersonloc_Nature2007}%
		\BibitemOpen
		\bibfield  {author} {\bibinfo {author} {\bibfnamefont {T.}~\bibnamefont
				{Schwartz}}, \bibinfo {author} {\bibfnamefont {G.}~\bibnamefont {Bartal}},
			\bibinfo {author} {\bibfnamefont {S.}~\bibnamefont {Fishman}}, \ and\
			\bibinfo {author} {\bibfnamefont {M.}~\bibnamefont {Segev}},\ }\bibfield
		{title} {\enquote {\bibinfo {title} {Transport and anderson localization in
					disordered two-dimensional photonic lattices},}\ }\href {\doibase
			10.1038/nature05623} {\bibfield  {journal} {\bibinfo  {journal} {Nature}\
			}\textbf {\bibinfo {volume} {446}},\ \bibinfo {pages} {52--55} (\bibinfo
			{year} {2007})}\BibitemShut {NoStop}%
		\bibitem [{\citenamefont {Guzman-Silva}\ \emph {et~al.}(2020)\citenamefont
			{Guzman-Silva}, \citenamefont {Heinrich}, \citenamefont {Biesenthal},
			\citenamefont {Kartashov},\ and\ \citenamefont
			{Szameit}}]{Szameit_dynamicloc_Andersonloc_OptExpr2020}%
		\BibitemOpen
		\bibfield  {author} {\bibinfo {author} {\bibfnamefont {D.}~\bibnamefont
				{Guzman-Silva}}, \bibinfo {author} {\bibfnamefont {M.}~\bibnamefont
				{Heinrich}}, \bibinfo {author} {\bibfnamefont {T.}~\bibnamefont
				{Biesenthal}}, \bibinfo {author} {\bibfnamefont {Y.V.}\ \bibnamefont
				{Kartashov}}, \ and\ \bibinfo {author} {\bibfnamefont {A.}~\bibnamefont
				{Szameit}},\ }\bibfield  {title} {\enquote {\bibinfo {title} {Experimental
					study of the interplay between dynamic localization and {A}nderson
					localization},}\ }\href {\doibase 10.1364/OL.380399} {\bibfield  {journal}
			{\bibinfo  {journal} {Opt. Lett.}\ }\textbf {\bibinfo {volume} {45}},\
			\bibinfo {pages} {415--418} (\bibinfo {year} {2020})}\BibitemShut {NoStop}%
		\bibitem [{\citenamefont {Szameit}\ \emph {et~al.}(2011)\citenamefont
			{Szameit}, \citenamefont {Rechtsman}, \citenamefont {Bahat-Treidel},\ and\
			\citenamefont {Segev}}]{Szameit_Segev_PRA2011}%
		\BibitemOpen
		\bibfield  {author} {\bibinfo {author} {\bibfnamefont {A.}~\bibnamefont
				{Szameit}}, \bibinfo {author} {\bibfnamefont {M.~C.}\ \bibnamefont
				{Rechtsman}}, \bibinfo {author} {\bibfnamefont {O.}~\bibnamefont
				{Bahat-Treidel}}, \ and\ \bibinfo {author} {\bibfnamefont {M.}~\bibnamefont
				{Segev}},\ }\bibfield  {title} {\enquote {\bibinfo {title}
				{$\mathcal{P}\mathcal{T}$-symmetry in honeycomb photonic lattices},}\ }\href
		{\doibase 10.1103/PhysRevA.84.021806} {\bibfield  {journal} {\bibinfo
				{journal} {Phys. Rev. A}\ }\textbf {\bibinfo {volume} {84}},\ \bibinfo
			{pages} {021806} (\bibinfo {year} {2011})}\BibitemShut {NoStop}%
		\bibitem [{\citenamefont {Heinrich}\ \emph
			{et~al.}(2014{\natexlab{b}})\citenamefont {Heinrich}, \citenamefont {Miri},
			\citenamefont {St{\"u}tzer}, \citenamefont {Nolte}, \citenamefont
			{Christodoulides},\ and\ \citenamefont
			{Szameit}}]{Szameit_supersymmetric_OptLett2014}%
		\BibitemOpen
		\bibfield  {author} {\bibinfo {author} {\bibfnamefont {M.}~\bibnamefont
				{Heinrich}}, \bibinfo {author} {\bibfnamefont {M.-A.}\ \bibnamefont {Miri}},
			\bibinfo {author} {\bibfnamefont {S.}~\bibnamefont {St{\"u}tzer}}, \bibinfo
			{author} {\bibfnamefont {S.}~\bibnamefont {Nolte}}, \bibinfo {author}
			{\bibfnamefont {D.N.}\ \bibnamefont {Christodoulides}}, \ and\ \bibinfo
			{author} {\bibfnamefont {A.}~\bibnamefont {Szameit}},\ }\bibfield  {title}
		{\enquote {\bibinfo {title} {Observation of supersymmetric scattering in
					photonic lattices},}\ }\href {\doibase 10.1364/OL.39.006130} {\bibfield
			{journal} {\bibinfo  {journal} {Opt. Lett.}\ }\textbf {\bibinfo {volume}
				{39}},\ \bibinfo {pages} {6130--6133} (\bibinfo {year}
			{2014}{\natexlab{b}})}\BibitemShut {NoStop}%
		\bibitem [{\citenamefont {Brazhnyi}\ \emph {et~al.}(2009)\citenamefont
			{Brazhnyi}, \citenamefont {Konotop}, \citenamefont {P\'erez-Garc\'{\i}a},\
			and\ \citenamefont {Ott}}]{Konotop_Ott_PRL2009}%
		\BibitemOpen
		\bibfield  {author} {\bibinfo {author} {\bibfnamefont {V.~A.}\ \bibnamefont
				{Brazhnyi}}, \bibinfo {author} {\bibfnamefont {V.~V.}\ \bibnamefont
				{Konotop}}, \bibinfo {author} {\bibfnamefont {V.~M.}\ \bibnamefont
				{P\'erez-Garc\'{\i}a}}, \ and\ \bibinfo {author} {\bibfnamefont
				{H.}~\bibnamefont {Ott}},\ }\bibfield  {title} {\enquote {\bibinfo {title}
				{Dissipation-induced coherent structures in bose-einstein condensates},}\
		}\href {\doibase 10.1103/PhysRevLett.102.144101} {\bibfield  {journal}
			{\bibinfo  {journal} {Phys. Rev. Lett.}\ }\textbf {\bibinfo {volume} {102}},\
			\bibinfo {pages} {144101} (\bibinfo {year} {2009})}\BibitemShut {NoStop}%
	\end{thebibliography}
\end{document}